\documentclass[showpacs,preprintnumbers,amsmath,amssymb,prd,superscriptaddress]{revtex4}
\usepackage{amssymb}
\usepackage{epsfig,amsmath,graphics}
\usepackage{epstopdf}
\usepackage{subfigure}
\usepackage{verbatim}

\newcommand{\OO}{\mathcal{O}}

\newcommand{\eff}{\text{eff}}
\newcommand{\keff}{k_\text{eff}}

\newcommand{\half}{\frac{1}{2}}

\newcommand{\SNR}{\text{SNR}}
\renewcommand{\v}[1]{\ensuremath{\mathbf{#1}}}
\newcommand{\ket}[1]{\ensuremath{\left|#1\right>}}

\newcommand{\glavg}{g_{L}^{avg}}
\newcommand{\rlavg}{R_{L}^{avg}}
\newcommand{\vlavg}{v_{L}^{avg}}

\newcommand{\Isat}{I_\text{sat}}
\newcommand{\EarthShotNoise}{10^{-15} \frac{g}{\sqrt{\text{Hz}}}}
\newcommand{\spshotnoise}{10^{-19} g( \frac{\omega}{10^{-2}\text{ Hz}})^2}
\newcommand{\degree}{^{\circ}}
\newcommand{\rthz}{\sqrt{\text{Hz}}}
\newcommand{\sqna}{\frac{1}{\sqrt{N_a}}}
\newcommand{\sqnasubs}{\sqrt{\frac{N_a}{10^8}}}

\begin{document}


\title{An Atomic Gravitational Wave Interferometric Sensor (AGIS)}

\author{Savas Dimopoulos}
\email{savas@stanford.edu}
\affiliation{Department of Physics, Stanford University, Stanford, California 94305}

\author{Peter W. Graham}
\email{pwgraham@stanford.edu}
\affiliation{SLAC, Stanford University, Menlo Park, California 94025}

\author{Jason M. Hogan}
\email{hogan@stanford.edu}
\affiliation{Department of Physics, Stanford University, Stanford, California 94305}

\author{Mark A. Kasevich}
\email{kasevich@stanford.edu}
\affiliation{Department of Physics, Stanford University, Stanford, California 94305}

\author{Surjeet Rajendran}
\email{surjeet@stanford.edu}
\affiliation{SLAC, Stanford University, Menlo Park, California 94025}
\affiliation{Department of Physics, Stanford University, Stanford, California 94305}

\date{\today}

\begin{abstract}
We propose two distinct atom interferometer gravitational wave detectors, one terrestrial and another satellite-based, utilizing the core technology of the Stanford $10\,\text{m}$ atom interferometer presently under construction.  Each configuration compares two widely separated atom interferometers run using common lasers.  The signal scales with the distance between the interferometers, which can be large since only the light travels over this distance, not the atoms. The terrestrial experiment with two $\sim 10 \text{ m}$ atom interferometers separated by a $\sim 1 \text{ km}$ baseline can operate with strain sensitivity $ \sim \frac{10^{-19}}{\sqrt{\text{Hz}}}$ in the 1 Hz - 10 Hz band, inaccessible to LIGO,  and can  detect gravitational waves from solar mass binaries out to megaparsec distances. The satellite experiment with two atom interferometers separated by a  $\sim 1000 \text{ km}$ baseline can probe the same frequency spectrum as LISA with comparable strain sensitivity $ \sim \frac{10^{-20}}{\sqrt{\text{Hz}}}$.  The use of ballistic atoms  (instead of mirrors) as inertial test masses improves systematics coming from vibrations, acceleration noise, and significantly reduces spacecraft control requirements.  We analyze the backgrounds in this configuration and discuss methods for controlling them to the required levels.
\end{abstract}

\maketitle

\tableofcontents

\section{Introduction}

Gravitational waves offer a rich, unexplored source of information about the universe \cite{Schutz:1999xj, Cutler:2002me}.  Many phenomena can only be explored with gravitational, not electromagnetic, radiation.  These include accepted sources such as white dwarf, neutron star, or black hole binaries whose observation could provide useful data on astrophysics and general relativity.  It has even been proposed that these compact binaries could be used as standard sirens to determine astronomical distances and possibly the expansion rate of the universe more precisely \cite{Schutz:1986gp}.  Gravitational waves could also be one of the only ways to learn about the early universe before the surface of last scattering.  There are many speculative cosmological sources including inflation and reheating, early universe phase transitions, and cosmic strings.  For all these applications, it is important to be able to observe gravitational waves as broadly and over as large a range of frequencies and amplitudes as possible.


In this article we expand on a previous article \cite{Dimopoulos:2007cj}, giving the details of our proposal for an Atomic Gravitational wave Interferometric Sensor (AGIS).  We develop proposals for two experiments, one terrestrial, the other satellite-based.  We will see that, at least in the configurations proposed here, it is primarily useful for observing gravitational waves with frequencies between about $10^{-3} \text{~Hz}$ and $10 \text{~Hz}$.  In particular, the terrestrial experiment is sensitive to gravitational waves with frequencies $\sim 1 - 10 \text{~Hz}$, below the range of any other terrestrial gravitational wave detector.  This arises in part from the vast reduction in systematics available with atom interferometers but impossible with laser interferometers.  The satellite-based experiment will have peak sensitivity to gravitational waves in the $\sim 10^{-3} - 1 \text{~Hz}$ band.  The use of atomic interferometry also leads to a natural reduction in many systematic backgrounds, allowing such an experiment to reach sensitivities comparable to and perhaps better than LISA's with reduced engineering requirements.

The ability to detect gravitational waves in such a low frequency band greatly affects the potential sources.  Binary stars live much longer and are more numerous at these frequencies than in the higher band around 100 Hz where they are about to merge.  Stochastic gravitational waves from cosmological sources can also be easier to detect at these low frequencies.  These sources are usually best described in terms of the fractional energy density, $\Omega_\text{GW}$, that they produce in gravitational waves.  The energy density of a gravitational wave scales quadratically with frequency (as $\rho_\text{GW} \sim h^2 f^2 M^2_\text{pl}$).  Thus, a type of source that produces a given energy density is easier to detect at lower frequencies because the amplitude of the gravitational wave is higher.  This makes low frequency experiments particularly useful for observing cosmological sources of gravitational waves.

There are several exciting proposed and existing experiments to search for gravitational waves including broad-band laser interferometers such as LIGO and LISA, resonant bar detectors \cite{Astone:2006ut, Zendri:2002em}, and microwave cavity detectors \cite{Ballantini:2005am}.  Searching for gravitational waves with atomic interferometry is motivated by the rapid advance of this technology in recent years.  Atom interferometers have been used for many high precision applications including atomic clocks \cite{Oskay}, metrology, gyroscopes \cite{PhysRevLett.78.2046}, gradiometers \cite{PhysRevLett.81.971}, and gravimeters \cite{0026-1394-38-1-4}.  We consider the use of such previously demonstrated technology to achieve the sensitivity needed to observe gravitational waves.  Further, we consider technological advances in atom interferometry that are currently being explored, including in the Stanford 10 m interferometer, and the possible impact these will have on the search for gravitational waves.


The different sections of this paper are as independent as possible.  In section \ref{Sec:AI} we give a brief description of atom interferometry.  In Section \ref{Sec: GW Signal} we calculate the gravitational wave signal in an atom interferometer.  In section \ref{Sec: Earth} we discuss an experimental configuration for observing this signal on the earth and the relevant backgrounds.  In section \ref{Sec: Space} we give the setup and backgrounds for a satellite-based experiment.  In section \ref{Sec: Sources} we give a brief summary of the astrophysical and cosmological sources of gravitational waves that are relevant for such experiments.  In section \ref{Sec: Sensitivities} we give a description of the projected sensitivities of the earth- and space-based experiments.  In Section \ref{Sec: Conclusion} we compare this work with previous ideas on atom interferometry and gravitational waves and summarize our findings.

\section{Atom Interferometry}
\label{Sec:AI}

We propose to search for gravitational waves using light pulse atom interferometry.  In a light pulse atom interferometer, an atom is forced to follow a superposition of two spatially and temporally separated free-fall paths.  This is accomplished by coherently splitting the atom wavefunction with a pulse of light that transfers momentum to a part of the atom.  When the atom is later recombined, the resulting interference pattern depends on the relative phase accumulated along the two paths.  This phase shift results from both the free-fall evolution of the quantum state along each path as well as from the local phase of the laser which is imprinted on the atom during each of the light pulses.  Consequently, the phase shift is exquisitely sensitive to inertial forces present during the interferometer sequence, since it precisely compares the motion of the atom to the reference frame defined by the laser phase fronts.  Equivalently, the atom interferometer phase shift can be viewed as a clock comparison between the time kept by the laser's phase evolution and the atom's own internal clock.  Sensitivity to gravitational waves may be understood as arising from this time comparison, since the presence of space-time strain changes the light travel time between the atom and the laser.

A single phase measurement in an atom interferometer consists of three steps: atom cloud preparation, interferometer pulse sequence, and detection.  In the first step, the cold atom cloud is prepared.  Using laser cooling and perhaps evaporative cooling techniques \cite{atomicsources}, a sub-microkelvin cloud of atoms is formed.  Cold atom clouds are needed so that as many atoms as possible will travel along the desired trajectory and contribute to the signal.  In addition, many potential systematic errors (see Sections \ref{Sec: Earth Backgrounds} and \ref{Sec: space backgrounds}) are sensitive to the atom's initial conditions, so cooling can mitigate these unwanted effects.
At the end of the cooling procedure, the final cloud has a typical density which is low enough so that atom-atom interactions within the cloud are negligible (for example see Section V D of  \cite{GR Atom}).  This dilute ensemble of cold atoms is then launched with velocity $v_L$ by transferring momentum from laser light.  To avoid heating the cloud during launch, the photon recoil momenta are transferred to the atoms coherently, and spontaneous emission is avoided \cite{Phillips2002:JPhysB}.

\begin{figure}
\begin{center}
\includegraphics[width=4.0 in]{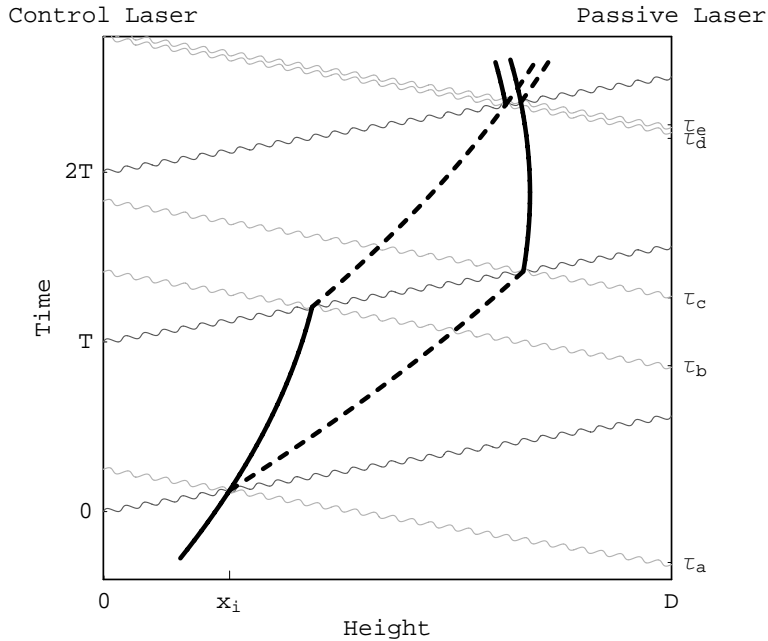}
\caption{ \label{Fig:AI-SingleInterferometer} A space-time diagram of a light pulse atom interferometer. The black lines indicate the motion of a single atom. Laser light used to manipulate the atom is incident from above (light gray) and below (dark gray) and travels along null geodesics.  The finite speed of the light has been exaggerated. }
\end{center}
\end{figure}

In the second phase of the measurement, the atoms follow free-fall trajectories and the interferometry is performed.  A sequence of laser pulses serve as beamsplitters and mirrors that coherently divide each atom's wavepacket and then later recombine it to produce the interference.  Figure \ref{Fig:AI-SingleInterferometer} is a space-time diagram illustrating this process for a single atom.  The atom beamsplitter is implemented using a stimulated two-photon  transition.  In this process, laser light incident from the right of Fig. \ref{Fig:AI-SingleInterferometer} with wavevector $\v{k_1}$ is initially absorbed by the atom.  Subsequently, laser light with wavevector $\v{k_2}$ incident from the left stimulates the emission of a $\v{k_2}$-photon from the atom, resulting in a net momentum transfer of $\v{k_{\eff}}=\v{k_2}-\v{k_1}\approx 2\v{k_2}$.  These two-photon atom optics are represented in Fig. \ref{Fig:AI-SingleInterferometer} by the intersection of two counter-propagating photon paths at each interaction node.

\begin{figure}
\begin{center}
\subfigure[ ]{\label{Fig:Raman}}
\includegraphics[width=150pt]{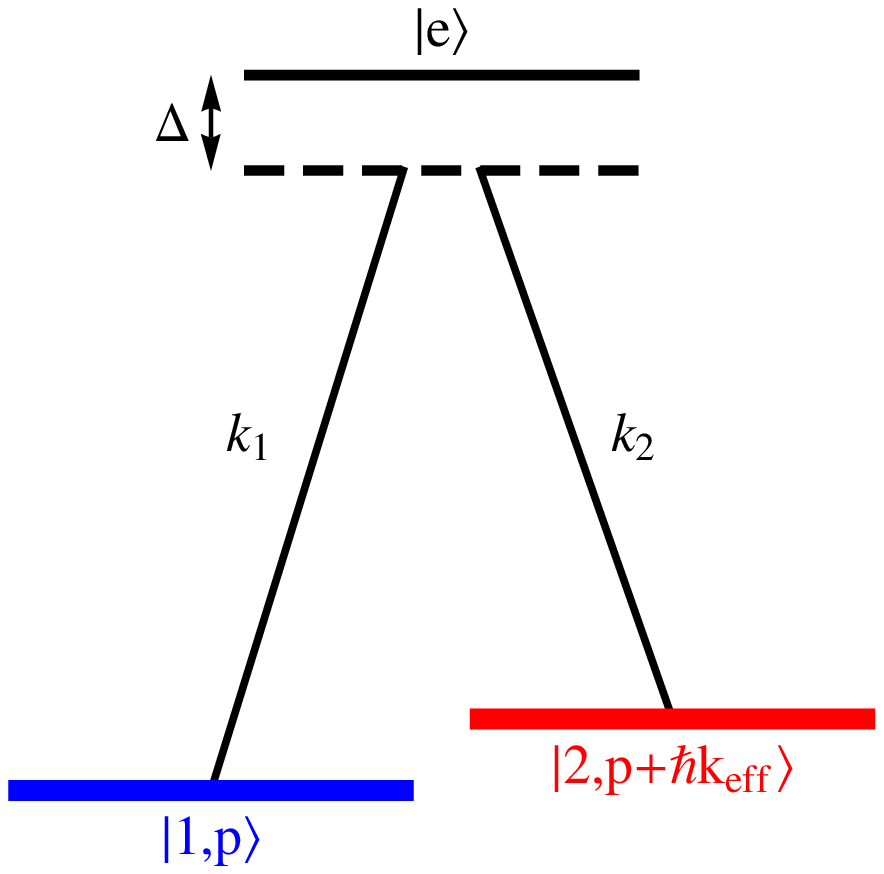}
\hspace{1 in}
\subfigure[ ]{\label{Fig:RabiPlot}}
\includegraphics[width=200pt]{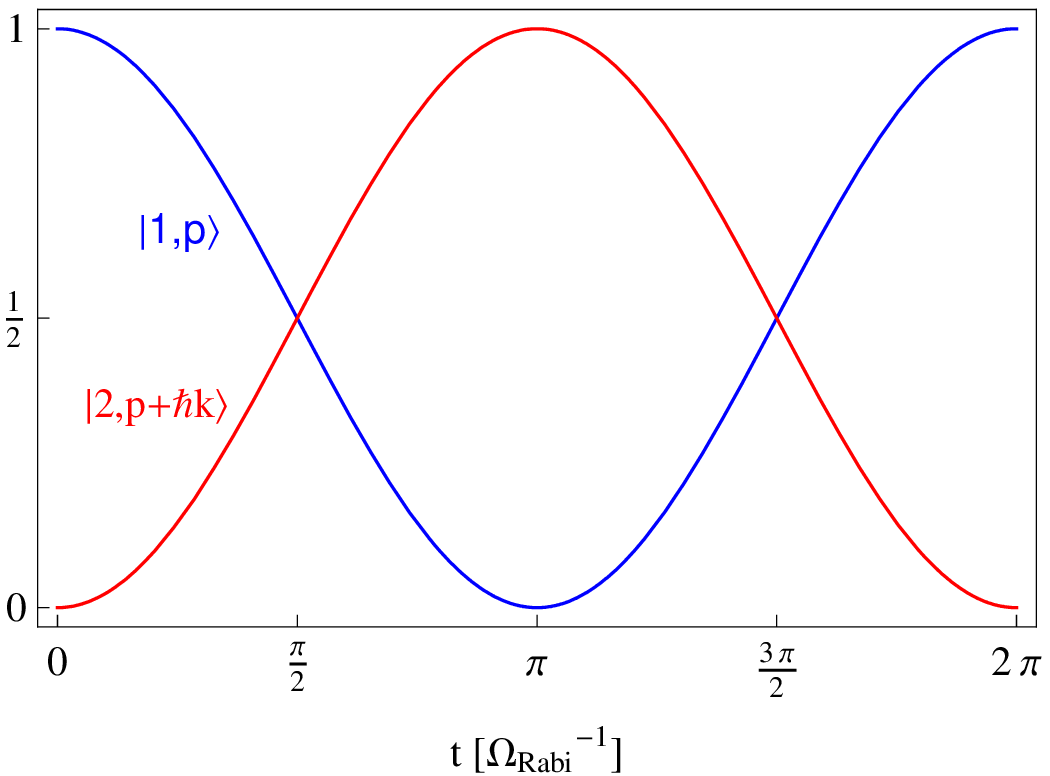}
\caption{(Color online) Figure \ref{Fig:Raman} shows an energy level diagram for a stimulated Raman transition between atomic states $\ket{1}$ and $\ket{2}$ through a virtual excited state using lasers of frequency $k_1$ and $k_2$.  Figure \ref{Fig:RabiPlot} shows the probability that the atom is in states $\ket{1}$ and $\ket{2}$ in the presence of these lasers as a function of the time the lasers are on.  A $\frac{\pi}{2}$ pulse is a beamsplitter since the atom ends up in a superposition of states $\ket{1}$ and $\ket{2}$ while a $\pi$ pulse is a mirror since the atom's state is changed completely.}
\end{center}
\end{figure}

\begin{figure}
\begin{center}
\includegraphics[width=200pt]{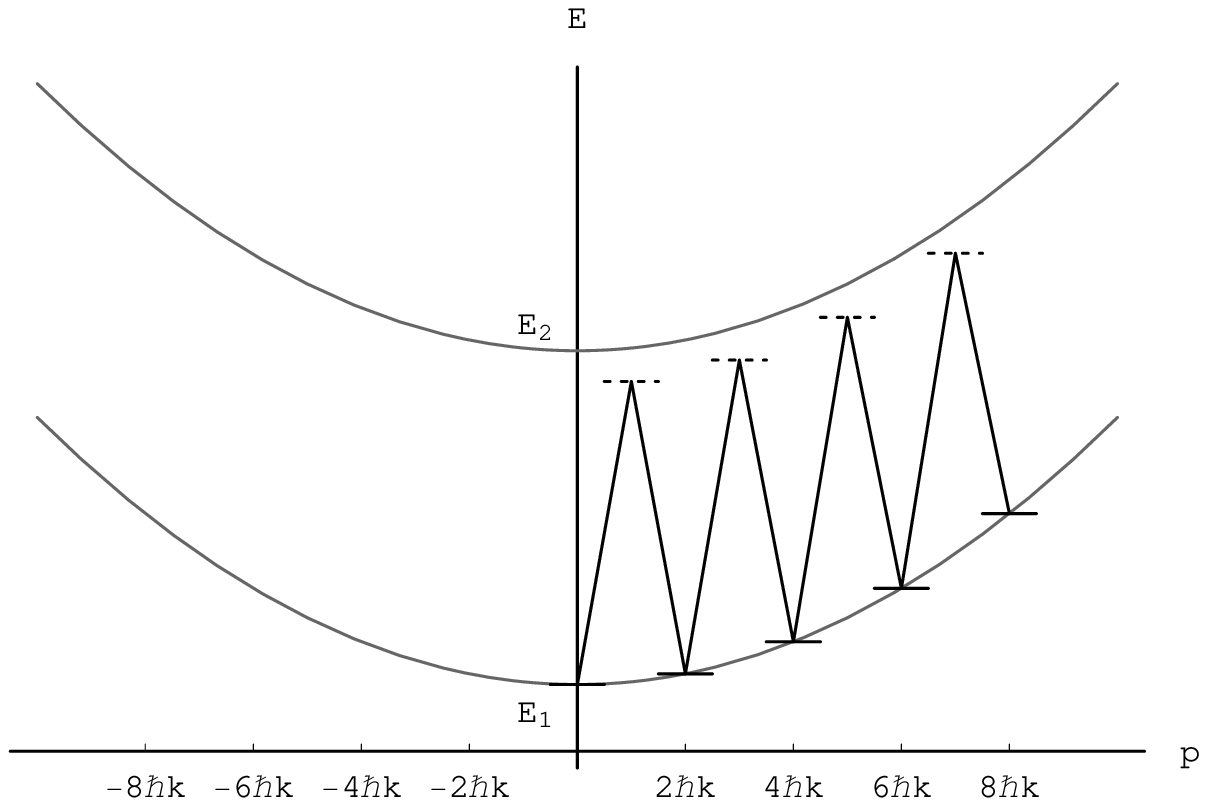}
\caption{ \label{Fig:BraggEnergyLevels} The atomic energy level diagram for a Bragg process plotted as energy versus momentum.  The horizontal lines indicate the states through which the atom is transitioned.  By sweeping the laser frequencies the atom can be given a large momentum.}
\end{center}
\end{figure}

There are several schemes for exchanging momentum between the atoms and the lasers.  Figure \ref{Fig:Raman} shows the case of a Raman transition in which the initial and final states are different internal atomic energy levels.  The light fields entangle the internal and external degrees of freedom of the atom, resulting in an energy level change and a momentum kick.  As an alternative to this, it is also possible to use Bragg transitions in which momentum is transferred to the atom while the internal atomic energy level stays fixed (see Fig. \ref{Fig:BraggEnergyLevels}).  In both the Raman and Bragg scheme, the two lasers are far detuned from the optical transitions, resulting in a negligibly small occupancy of the excited state $\ket{e}$.  This avoids spontaneous emission from the short-lived excited state.  To satisfy the resonance condition for the desired two-photon process, the frequency difference between the two lasers is set equal to the atom's recoil kinetic energy (Bragg) plus any internal energy shift (Raman).  While the laser light is on, the atom undergoes Rabi oscillations between states $\ket{\v{p}}$ and $\ket{\v{p}+\v{k_{\eff}}}$ (see Fig. \ref{Fig:RabiPlot}). A beamslitter results when the laser pulse time is equal to a quarter of a Rabi period ($\frac{\pi}{2}$ pulse), and a mirror requires half a Rabi period ($\pi$ pulse).

After the initial beamsplitter ($\frac{\pi}{2}$) pulse, the atom is in a superposition of states which differ in velocity by $\v{k_{\eff}}/m$.  The resulting spatial separation of the halves of the atom is proportional to the interferometer's sensitivity to gravitational wave-induced strain along the direction of $\v{k_{\eff}}$.  After a time $T$, a mirror ($\pi$) pulse reverses the relative velocity of the two components of the atom, eventually leading to spatial overlap.  To complete the sequence, a final beamsplitter pulse applied at time $2 T$ interferes these overlapping components at the intersection point of the two paths.  In this work we primarily consider this beamsplitter-mirror-beamsplitter $(\frac{\pi}{2}-\pi-\frac{\pi}{2})$ sequence \cite{PhysRevLett.67.181}, the simplest implementation of an accelerometer and the matter-wave analog of a Mach-Zender interferometer.

The third and final step of each measurement is atom detection.  At the end of the interferometer sequence, each atom is in a superposition of the two output velocity states, as shown by the diverging paths at the top of Fig. \ref{Fig:AI-SingleInterferometer}. Since these two states differ in velocity by $\sim k_{\eff}/m$, they spatially separate.  After an appropriate drift time, the two paths can be separately resolved, and the populations are then measured by absorption imaging. These two final velocity states are directly analogous to the two output ports of a Mach-Zehnder light interferometer after the final recombining beamsplitter.  As with a light interferometer, the probability that an atom will be found in a particular output port depends on the relative phase acquired along the two paths of the atom interferometer.

To explore the potential reach of AI-based gravitational wave detectors, we consider progressive phase sensitivities that are likely to be feasible in the near future.  Recent atom interferometers have already demonstrated sensor noise levels limited only by the quantum projection noise of the atoms (atom shot noise) \cite{clocks}.  For a typical time--average atom flux of $n=10^{6}~\text{atoms}/s$, the resulting phase sensitivity is $\sim 1/\sqrt{n}=10^{-3}~\text{rad}/\sqrt{\text{Hz}}$.  For example, modern light pulse atom interferometers of the type considered here achieve an atom flux at this level by periodically launching $\sim 10^6$ atoms per shot at a repetition rate of $\sim 1~\text{Hz}$.  For our most aggressive terrestrial proposal, we assume quantum projection noise--limited detection of $10^{8}~\text{atoms}$ per shot at a repetition rate of $10~\text{Hz}$, implying a phase sensitivity of $3\times 10^{-5}~\text{rad}/\sqrt{\text{Hz}}$.  In the satellite-based proposal we assume $10^{8}~\text{atoms}$ per shot with a $1~\text{Hz}$ repetition rate, yielding $10^{-4}~\text{rad}/\sqrt{\text{Hz}}$.

Cold atom clouds with $10^8$ to $10^{10}$ atoms are readily produced using modern laser cooling techniques \cite{Metcalf}.  However, the challenges in this application are to cool to the required narrow velocity distribution and to do so in a short enough time to support a high repetition rate.  As discussed in Sections \ref{Sec: Earth Backgrounds} and \ref{Sec: space backgrounds}, suppression of velocity-dependent backgrounds requires RMS velocity widths as small as $\sim 100~\mu\text{m}/\text{s}$, corresponding to 1D cloud temperatures of $\sim 100~\text{pK}$.  The required $\sim 100~\mu\text{m}/\text{s}$ wide cloud could conceivably be extracted from a very large ($\gtrsim 10^{10}$ atoms) $\mu \text{K}$-temperature thermal cloud by applying a highly velocity-selective cut\footnote{Such a cut may be implemented using a Doppler sensitive two--photon transition.  This technique results in a narrow velocity distribution along the longitudinal direction without reducing the velocity width in the directions transverse to the cut.  Ensembles such as these which are only cold along a single dimension can still be useful for suppressing certain systematics (e.g. gravity gradients).  See Sections \ref{Sec: Earth Backgrounds} and \ref{Sec: space backgrounds} for specific velocity width requirements.}, or by using evaporative cooling techniques.  In either case, low densities are desirable to mitigate possible systematic noise sources associated with cold collisions.



The repetition rate required for each proposal is a function of the gravitational wave signal frequency range that the experiment probes.  On earth, a $10~\text{Hz}$ repetition rate is necessary to avoid under-sampling signals in the target frequency band of $\sim 1 - 10 \text{~Hz}$.  The satellite experiment we consider is sensitive to the $\sim 10^{-3} - 1 \text{~Hz}$ band, so a $1~\text{Hz}$ rate is sufficient.  However, in both cases, multiple interferometers must be overlapped in time since the duration of a single interferometer sequence ($T\sim 1~\text{s}$ for earth, $\sim 100~\text{s}$ for space) exceeds the time between shots.  Section \ref{Sec:space fd limit} discusses the logistics of simultaneously manipulating a series of temporally overlapping interferometers and describes the implications for atom detection.

Sensor noise performance can potentially be improved by using squeezed atom states instead of uncorrelated thermal atom ensembles \cite{spin_squeezing}.  For a suitably entangled source, the Heisenberg limit is $\SNR \sim n$, a factor of $\sqrt{n}$ improvement.  For $n\sim 10^{6}$ entangled atoms, the potential sensitivity improvement is $10^3$.  Recent progress using these techniques may soon make improvements in $\SNR$ on the order of 10 to 100 realistic \cite{Tuchman_PRA}.  Even squeezing by factor of $10$ can potentially relax the atom number requirements by $10^2$.


Another sensitivity improvement involves the use of more sophisticated atom optics.  The phase sensitivity to gravitational waves is proportional to the effective momentum $\hbar \keff$ transferred to the atom during interactions with the laser.  Both the Bragg and Raman schemes described above rely on a two--photon process for which \cite{McGuirk} $\hbar \keff=2\hbar k$, but large momentum transfer (LMT) beamsplitters with up to $10\hbar k$ or perhaps $100\hbar k$ are possible \cite{HolgerLMT}.  Promising LMT beamsplitter candidates include optical lattice manipulations \cite{Phillips2002:JPhysB}, sequences of Raman pulses \cite{McGuirk} and multiphoton Bragg diffraction \cite{HolgerLMT}.  Figure \ref{Fig:BraggEnergyLevels} illustrates an example of an LMT process consisting of a series of sequential two--photon Bragg transitions as may be realized in an optical lattice.  As the atom accelerates, the resonance condition is maintained by increasing the frequency difference between the lasers.

Finally, we consider the acceleration sensitivity of the atom interferometer gravitational wave detectors proposed here.  The intrinsic sensitivity of the atom interferometer to inertial forces makes it necessary to tightly constrain many time-dependent perturbing accelerations, since background acceleration inputs in the relevant frequency band cannot be distinguished from the gravitational wave signal of interest.  The theoretical maximum acceleration sensitivity of the apparatus follows from the shot-noise limited phase sensitivity discussed above, combined with the well-known acceleration response of the atom interferometer, $\phi=\keff a T^2$:
\begin{equation}
\frac{\delta a}{a}=\frac{\delta \phi}{\phi}\sim \frac{1/\text{SNR}}{\keff a T^2}=\left(\frac{1}{\keff a T^2\sqrt{n}}\right)\tau^{-1/2}
\end{equation}
where the total signal-to-noise ratio is $\text{SNR}\sim \sqrt{n\tau}$ for a detected atom flux of $n$ atoms per second during an averaging time $\tau$.  For the terrestrial apparatus we propose, the resulting sensitivity in terms of the gravitational acceleration $g$ of the earth is $4\times 10^{-16} \left( \frac{1~\text{s}}{T} \right)^2 \left( \frac{1000 k}{\keff} \right) \left(\frac{10^9\text{atoms}/\text{s}}{n}\right)^{\half} g/\sqrt{\text{Hz}}$.  Likewise, in the satellite experiment the acceleration sensitivity is $1\times 10^{-18} \left( \frac{100~\text{s}}{T} \right)^2 \left( \frac{100 k}{\keff} \right)\left(\frac{10^8\text{atoms}/\text{s}}{n}\right)^{\half} g/\sqrt{\text{Hz}}$.  The sum of all perturbing acceleration noise sources must be kept below these levels in order for the apparatus to reach its theoretical noise limit.  Sections \ref{Sec: Earth Backgrounds} and \ref{Sec: space backgrounds} identify many of these potential backgrounds and discusses the requirements necessary to control them.


\section{Gravitational Wave Signal}
\label{Sec: GW Signal}

In this section we will discuss the details of the calculation of the phase shift in an atom interferometer due to a passing gravitational wave.  This calculation follows the method for a relativistic calculation discussed in \cite{Dimopoulos:2006nk, GR Atom}.  The method itself will not be reviewed here, only its application to a gravitational wave and the properties of the resultant phase shift will be discussed.  For the rest of the paper, only the answer from this calculation is necessary.  We will see that the signal of a gravitational wave in the interferometer is an oscillatory phase shift with frequency equal to the gravitational wave's frequency that scales with the length between the laser and the atom interferometer.

Intuitively the atom interferometer can be thought of as precisely comparing the time kept by the laser's clock (the laser's phase), and the time kept by the atom's clock (the atom's phase).  A passing gravitational wave changes the normal flat space relation between these two clocks by a factor proportional to the distance between them.  This change oscillates in time with the frequency of the gravitational wave.  This is the signal that can be looked for with an atom interferometer.  Equivalently, the atom interferometer can be thought of as a way of laser ranging the atom's motion to precisely measure its acceleration.  Calculating the acceleration that would be seen by laser ranging a test mass some distance away in the metric of the gravitational wave \eqref{Eqn: GW metric TT gauge} shows a similar oscillatory acceleration in time, and this is the signal of a gravitational wave in an atom interferometer.  This radar ranging calculation gives essentially the same answer as the full atom interferometer calculation in this case.

\subsection{Phase Shift Calculation}

For the full atom interferometer calculation we will consider the following metric for a plane gravitational wave traveling in the $z$-direction
\begin{equation}
\label{Eqn: GW metric TT gauge}
ds^2 = dt^2 - \left( 1+h \sin \left(\omega (t-z) + \phi_0 \right) \right) dx^2 - \left( 1-h \sin \left(\omega (t-z) + \phi_0 \right) \right) dy^2 - dz^2
\end{equation}
where $\omega$ is the frequency of the wave, $h$ is its dimensionless strain, and $\phi_0$ is an arbitrary initial phase.  Note that this metric is only approximate, valid to linear order in $h$.  This choice of coordinates for the gravitational wave is known as the ``+" polarization in the TT gauge.  For simplicity, we will consider a 1-dimensional atom interferometer with its axis along the $x$-direction.  An orientation for the interferometer along the $x$- or $y$-axes gives a maximal signal amplitude, while along the $z$-axis gives zero signal.

We will work throughout only to linear order in $h$ and up to quadratic order in all velocities.  These approximations are easily good enough since even the largest gravitational waves we will consider have $h \sim 10^{-18}$ and the atomic velocities in our experiment are at most $v \sim 10^{-7}$.  For simplicity we take $\hbar = c = 1$.

The total phase shift in the interferometer is the sum of three parts: the propagation phase, the laser interaction phase, and the final wavepacket separation phase. The usual formulae for these must be modified in GR to be coordinate invariants.  Our calculation has been discussed in detail in \cite{GR Atom}.  Here we will only briefly summarize how to apply that formalism to a gravitational wave metric.  The space-time paths of the atoms and lasers are geodesics of Eqn. \ref{Eqn: GW metric TT gauge}.  The propagation phase is
\begin{equation}
\phi_\text{propagation} = \int L dt = \int m ds
\end{equation}
where $L$ is the Lagrangian and the integral is along the atom's geodesic.  The separation
phase is taken as
\begin{equation}
\phi_\text{separation} = \int \overline{p}_\mu dx^\mu \sim
\overline{E} \Delta t - \vec{\overline{p}} \cdot \Delta \vec{x}
\end{equation}
where, for coordinate independence, the integral is over the null geodesic connecting the classical endpoints of the two arms of the interferometer, and $\overline{p}$ is the average of the classical 4-momenta of the two arms after the third pulse.  The laser phase shift due to interaction with the light is the constant phase of the light along its null geodesic, which is its phase at the time it leaves the laser.

We will make use of the fact that the laser phase in the atom interferometer comes entirely from the second laser, the `passive laser', which is taken to be always on so it does not affect the timing of the interferometer.  Instead the first laser, the `control laser', defines the time at which the atom-light interaction vertices occur.  For a more complete discussion of this point, see Section 3 of \cite{GR Atom}.

In practice, we will consider the atom interferometer to be 1-dimensional so that the atoms and light move only in the $x$-direction and remain at a constant $y=z=0$.  This is not an exact solution of the geodesic equation for metric \eqref{Eqn: GW metric TT gauge}.  In the full solution the atoms and light are forced to move slightly in the $z$-direction because of the $z$ in $g_{xx}$.  However the amplitude of this motion is proportional to $h$ which will mean that it only has effects on the calculation at $\OO(h^2)$.  This was shown by a full interferometer calculation in two dimensions.  It can also be understood intuitively since all displacements, velocities, and accelerations in the second dimension are $\OO(h)$.  The separation phase is then $\phi_\text{separation} \sim p_z \Delta x_z \sim \OO(h^2)$.  The extra $z$ piece of propagation phase is $\sim g_{zz} \Delta z^2 \sim \OO(h^2)$.  Changes to the calculated $x$ and $t$ coordinates and to $g_{xx}$ will also be $\OO(h^2)$ so propagation phase is only affected at this level by the motion in the $z$ direction.  Finally, if the laser phase fronts are flat in the $z$ dimension as they travel in the $x$ direction then there will be no affect of the displacement in the $z$ direction.  However the phase fronts cannot be made perfectly flat and so there will be an $\OO(h)$ effect of the displacement in the $z$ direction times the amount of bending of the laser phase fronts.  This is clearly much smaller than the leading $\OO(h)$ signal from the gravitational wave and so we will ignore it since we are primarily interested in calculating the signal.

The lasers will be taken to be at the origin and at spatial position ($D$,0,0).  We can make this choice because a fixed spatial coordinate location is a geodesic of metric \eqref{Eqn: GW metric TT gauge}.  Without loss of generality, we take the initial position of the atom to lie on the null geodesic originating from the origin (the first beamsplitter pulse) with $x=x_A$.  Of course, this means that $x_A$ is not a physically measurable variable but is instead a coordinate dependent choice.  The results are still fully correct in terms of $x_A$ and the interpretation will also be clear because the coordinate dependence only enters at $\OO(h)$.  Since, as we will see, the leading order piece of the signal being computed is $\OO(h)$, this ambiguity can only have an $\OO(h^2)$ effect on the answer.  We can ignore this and consider $x_A$ to be the physical length between the atom's initial position and the laser, by any reasonable definition of this length.  Similar reasoning allows us to define the initial launch velocity of the atoms at $x_A$ as the coordinate velocity $v_L \equiv \left. \frac{dx}{d\tau} \right|_\text{initial}$.  The ability to ignore the $\OO(h)$ corrections to the coordinate expressions for quantities such as the initial position and velocity of the atoms relies on the fact that this is a null experiment so the leading order piece of the phase shift is proportional to $h$.

With the choices above we find the geodesics of metric \eqref{Eqn: GW metric TT gauge} are given as functions of the proper time $\tau$ by
\begin{eqnarray}
x(\tau) = x_0+{v_x}_0 \tau +h \left(-\frac{{v_x}_0 \cos \left(\phi_0+t_0 \omega \right)}{\sqrt{\eta+{v_x}_0^2} \omega
}+\frac{{v_x}_0 \cos\left(\phi_0+t_0 \omega +\sqrt{\eta+{v_x}_0^2} \tau  \omega \right)}{\sqrt{\eta+{v_x}_0^2} \omega }+{v_x}_0 \tau  \sin \left(\phi_0+t_0 \omega \right)\right) \\ 
t(\tau) = t_0+\sqrt{\eta+{v_x}_0^2} \tau + \frac{h {v_x}_0^2}{\left(\eta+{v_x}_0^2\right)} \left( \frac{ \cos\left(\phi_0+t_0 \omega +\sqrt{\eta+{v_x}_0^2} \tau  \omega \right) - \cos \left(\phi_0+t_0 \omega \right)}{2 \omega }
+ \tau \sqrt{\eta+{v_x}_0^2} \sin\left(\phi_0+t_0 \omega \right) \right)
\end{eqnarray}
to linear order in $h$ where $\eta = g_{\mu\nu} \frac{dx^\mu}{d\tau} \frac{dx^\nu}{d\tau}$ is 0 for null geodesics and 1 for time-like geodesics.  The leading order pieces of these are just the normal trajectories in flat space.

Using these trajectories, the intersection points can be calculated and the final phase shift found as in the general method laid out in \cite{GR Atom}.  Here the relevant equations are made solvable by expanding always to first order in $h$.  Note that here the use of the local Lorentz frame to calculate the atom-light interactions is unnecessary.  The interaction rules are applied at one space-time point so the local curvature of the space is irrelevant.  Further, the choice of boost (the velocity of the frame) can only make corrections of $\OO(v^2)$ to the transferred momentum, giving a $\OO(v^3)$ correction to the overall velocity of the atom, which is negligible.

To define the lasers' frequencies in a physically meaningful way as in \cite{GR Atom}, we take each laser to have a frequency, $k$, given in terms of the coordinate momenta of the light by
\begin{equation}
\left. \left( g_{\mu \nu} U^\mu \frac{dx_\text{light}^\nu}{d\lambda} \right) \right|_{x_L} = k
\end{equation}
where $U^\mu = \frac{dx_\text{obs}^\mu}{d\tau}$ is the four-velocity of an observer at the position of the laser.  The momenta of the light is then changed by the gravitational wave as it propagates and the kick it gives to the atom is given by its momenta at the point of interaction. 



\subsection{Results}
\label{Sec: Signal Results}

Following the method above, the phase difference seen in the atom interferometer in metric \eqref{Eqn: GW metric TT gauge} is shown in Table \ref{Tab: phases}.  As in Figure \ref{Fig:AI-SingleInterferometer}, the lasers are taken to be a distance $D$ apart, with the atom initially a distance $x_i$ from the left laser and moving with initial velocity $v_L$.  The left laser is the control laser and the right is the passive laser, as defined above.  The atom's rest mass in the lower ground state is $m$, the atomic energy level splitting is $\omega_a$, the laser frequencies are $k_1$ and $k_2$, and $h$, $\omega$, and $\phi_0$ are respectively the amplitude, frequency and initial phase of the gravitational wave. We will be considering a situation in which $D \sim 1 \text{ km}$ is much larger than the size of the interferometer region $v_r T \sim 1 \text{ m}$.

\begin{table}
\begin{center}
\begin{math}
\begin{array}{|c|c|}
\hline
\text{Phase Shift} & \text{Size (rad)} \\
\hline
4 \frac{h k_2}{\omega} \sin^2 \left( \frac{\omega T}{2} \right) \sin \left( \omega \left( x_i - \frac{D}{2} \right) \right) \sin \left( \phi_0 + \omega \left( x_i - \frac{D}{2} \right) + \omega T \right) & 3 \times 10^{-7} \\
-4 h k_2 v_L T \sin \left( \frac{\omega T}{2} \right) \cos \left( \phi_0 + 2 \omega \left( x_i - \frac{D}{2} \right) + \frac{3 \omega T}{2} \right) & 3 \times 10^{-9} \\
4 \frac{h \omega_A}{\omega} \sin^2 \left( \frac{\omega T}{2} \right) \sin \left( \frac{\omega x_i}{2} \right) \sin \left( \phi_0 + \frac{\omega x_i}{2} + \omega T \right) & 10^{-12} \\
8 \frac{h k_2 v_L}{\omega} \sin^2 \left( \frac{\omega T}{2} \right) \cos \left( \frac{\omega x_i}{2} \right) \cos \left( \phi_0 + \frac{\omega x_i}{2} + \omega T \right) & 10^{-14} \\
\hline
\end{array}
\end{math}
\caption{\label{Tab: phases} A size ordered list of the largest terms in the calculated phase shift due to a gravitational wave.  The sizes are given assuming $D \sim x_i \sim 1 \text{~km}$, $h \sim 10^{-17}$, $\omega \sim 1 \text{~rad} / \text{s}$, $k_2 \sim 10^{7} \text{~m}^{-1}$, $v_L \sim 3 \times 10^{-8}$, and $T \sim 1 \text{~s}$.}
\end{center}
\end{table}

The first term in Table \ref{Tab: phases} is the largest phase shift and the source of the gravitational wave effect we will look for in our proposed experiment:
\begin{equation}
\label{Eqn:GW phase shift}
\Delta \phi_\text{tot} = 4 \frac{h k_2}{\omega} \sin^2 \left( \frac{\omega T}{2} \right) \sin \left( \omega \left( x_i - \frac{D}{2} \right) \right) \sin \left( \phi_0 + \omega \left( x_i - \frac{D}{2} \right) + \omega T \right) + ...
\end{equation}
This is proportional to $k_2$, just as in the phase shift from Newtonian gravity \cite{GR Atom}.  This arises from the choice of laser 2 as the passive laser which is always on and laser 1 as the control laser which defines the timing of the beamsplitter and mirror interactions.  As shown in \cite{GR Atom}, the laser phase from laser 1 is zero and the main effect then arises from the laser phase of laser 2 and so is proportional to $k_2$.  Under certain conditions, this term is proportional to the baseline length, $D$, between the lasers (see Eqn. \ref{Eqn:GW phase expand}).

The effect of a gravitational wave is always proportional to a length scale.  The second term in Table \ref{Tab: phases} is not proportional to the distance $D$ between the lasers, but is proportional to the distance the atom travels during the interferometer $\sim v_L T$.  Thus it cannot be increased by scaling the laser baseline.  Instead it depends on the size of the region available for the atomic fountain, which is more difficult to increase experimentally.  Thus this term will likely be several orders of magnitude smaller than the first term in a practical experimental setup, as seen in Table \ref{Tab: phases}.  This $v_L T$ term is essentially the same term that has been found by previous authors (e.g. \cite{Tino:2007hs}).  We will not use this term for the signal in our proposed experiment as it is smaller than the first term.

The second term can be loosely understood from the intuition that the type of atom interferometer being considered is an accelerometer.  In the frame in which the lasers are stationary, this atom interferometer configuration is usually said to respond to the acceleration $a$ of the atom with a phase shift $\sim k a T^2$.  In this frame, the motion of the atom in the presence of a gravitational wave (metric \eqref{Eqn: GW metric TT gauge}) appears to have a coordinate acceleration $a \sim h \omega v$.  This would then give a phase shift $k a T^2 \sim k h \omega v T^2$ which is approximately the second term in Table \ref{Tab: phases}.  Of course, this is clearly coordinate-dependent intuition and will not work in a different coordinate system.  Nevertheless, it is interesting that this `accelerometer' term arises in a fully relativistic, coordinate-invariant calculation.

The third term in Table \ref{Tab: phases} is essentially the same as the first term, but with $k_2$ replaced by $\omega_a$ since it arises from the difference in rest masses between the two atomic states.  We are considering a Raman transition between two nearly degenerate ground states so $\omega_a \ll k_2$.  However for an atom interferometer made with a single laser driving a transition directly between two atomic states, the $k_2$ terms would be gone and the terms proportional to $\omega_a$ would be the leading order phase shift.  In this case, $\omega_a$ would be the same size as the $k$ of the laser in order to make the atomic transition possible.  Such a configuration may be difficult to achieve experimentally.

To understand the answer for the gravitational wave phase shift in Eqn. \eqref{Eqn:GW phase shift}, consider the limit where the period of the gravitational wave is longer than the interrogation time of the interferometer.  Expanding Eqn. \eqref{Eqn:GW phase shift} in the small quantities $\omega T$, $\omega D$ and $\omega x_i$ gives
\begin{eqnarray}
\label{Eqn:GW phase expand}
\Delta \phi_\text{tot} = h k_2 \omega^2 T^2 \left( x_i - \frac{D}{2} \right) \left( \sin\left(\phi_0 \right) + \omega T \cos\left(\phi_0 \right) - \frac{7}{12} \omega^2 T^2 \sin \left(\phi_0 \right) + \OO(\omega^3 T^3) \right) + ...
\end{eqnarray}
The phase shift is proportional to the distance of the atom from the midpoint between the two lasers.  This had to be the case because the leading order phase shift does not depend on the atom's velocity, resulting in a parity symmetry about the midpoint.  The signal increases with the size of the interferometer and the interrogation time $T$.  Of course, as we see from Eqn. \eqref{Eqn:GW phase shift} this increase stops when the size and time of the interferometer become comparable to the wavelength and period, $\frac{1}{\omega}$, of the gravitational wave.  Note that in the intermediate regime where $T > \frac{1}{\omega} > D$ then we can expand in $\omega$ times the distances so
\begin{eqnarray}
\label{Eqn:GW phase length expand}
\Delta \phi_\text{tot} = 4 h k_2 \left( x_i - \frac{D}{2} \right) \sin^2 \left( \frac{\omega T}{2} \right) \sin \left( \phi_0 + \omega \left( x_i - \frac{D}{2} \right) + \omega T \right) + ...
\end{eqnarray}
When $\omega T \sim 1$, this is very similar to the phase shift in LIGO which goes as $h k \ell$.

Although we will not go through the details of the whole calculation here, we will motivate the origin of the main effect, $\Delta \phi_\text{tot} \propto h k_2 (x_i - \frac{D}{2})$.  In other words, we work in the limit of Eqn. \ref{Eqn:GW phase length expand} when $\omega T \approx 1$.  We will be interested in the case where the length of the atom's paths are small compared to the distance between the lasers, $v_L T \ll D$ so the interferometer essentially takes place entirely at position $x_i$.  The main effect comes from laser phase from the passive laser, hence from the timing of these laser pulses.  The control laser's pulses are always at $0$, $T$, and $2T$.  As an example, the first beamsplitter pulse from the control laser then would reach the atom at time $x_i$ in flat space and so the passive laser pulse then originates at $2 x_i - D$.  However if the gravitational wave is causing an expansion of space the control pulse is `delayed' and actually reaches the atom at time $\sim x_i (1 + h)$.  Then the passive laser pulse originated at $\sim (2 x_i - D) (1+h)$.  Thus the laser phase from the passive laser pulse has been changed by the gravitational wave by an amount $k_2 h (2 x_i - D)$.  This is our signal.  Although this motivation is coordinate dependent, it provides intuition for the result of the full gauge invariant calculation.

Eqn. \eqref{Eqn:GW phase shift} is the main effect of a gravitational wave in an atom interferometer.  Therefore, the signal we are searching for is a phase shift in the interferometer that oscillates in time with the frequency of the gravitational wave.  Note that Eqn. \eqref{Eqn:GW phase shift} and all terms in Table \ref{Tab: phases} are oscillatory because $\phi_0$, the phase of the gravitational wave at the time the atom interferometer begins (the time of the first beamsplitter pulse), oscillates in time.  In other words, the phase shift measured by the atom interferometer changes from shot to shot because the phase of the gravitational wave changes.

One way to enhance this signal is to use large momentum transfer (LMT) beamsplitters as described in Section \ref{Sec:AI}.  This can be thought of as giving a large number of photon kicks to the atom, transferring momentum $N \hbar k$.  This enhances the signal by a factor of $N$ since laser phase is enhanced by $N$. The phase shift calculation is then exactly as if $k$ of the laser is replaced by $N k$.

As usual for a gravitational wave detector, this answer would be modulated by the angle between the direction of propagation of the gravitational wave and the orientation of the detector.  If the gravitational wave is propagating in the same direction as the lasers in the interferometer there will be no signal (for the results above we assumed a gravitational wave propagating perpendicularly to the laser axis).  This is clear since a gravitational wave is transverse, so space is not stretched in the direction of propagation.



\section{Terrestrial Experiment}
\label{Sec: Earth}

\subsection{Experimental Setup}
\label{Sec: Earth Setup}
We have shown that there is an oscillatory gravitational  wave signal in an atom interferometer.  To determine whether this signal is detectable requires examining the backgrounds in a possible experiment.  Two of the most important backgrounds are vibrations and laser phase noise.  As experience with LIGO would suggest, vibrational noise can be orders of magnitude larger than a gravitational  wave signal.  For an atom interferometer, laser phase noise can also directly affect the measurement and can be larger than the signal.  Reducing these backgrounds must therefore dictate the experimental configuration.

After the atom clouds are launched, they are inertial and  do not feel vibrations.  The vibrations they feel while in the atomic trap do not directly affect the final measured phase shift because the first beamsplitter pulse has not been applied yet.  Both vibrational and laser phase noise arise only from the lasers which run the atom interferometer.  We propose a differential measurement between two simultaneous atom interferometers run with the same laser pulses to greatly reduce these backgrounds.  In order to maximize a gravitational  wave signal, these atom interferometers should be separated by a distance $L$ which is as large as experimentally achievable.

\begin{figure}%
\begin{center}
\subfigure[ ]{\label{Fig:space-time} 
\includegraphics[width=3.5 in]{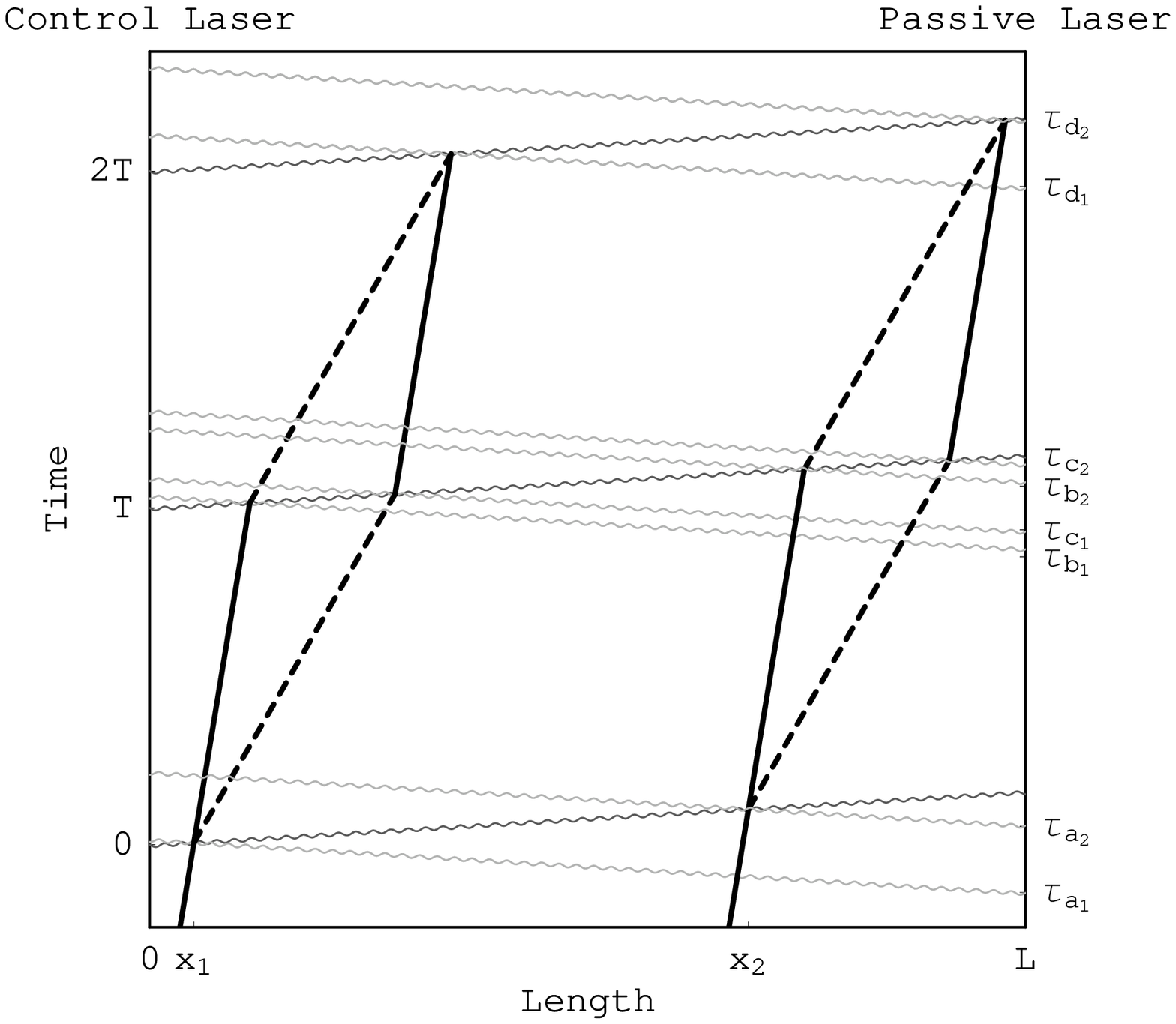}}
\hspace{1 in}
\subfigure[ ]{\label{Fig:earthsetup} 
\includegraphics[height=3.5 in]{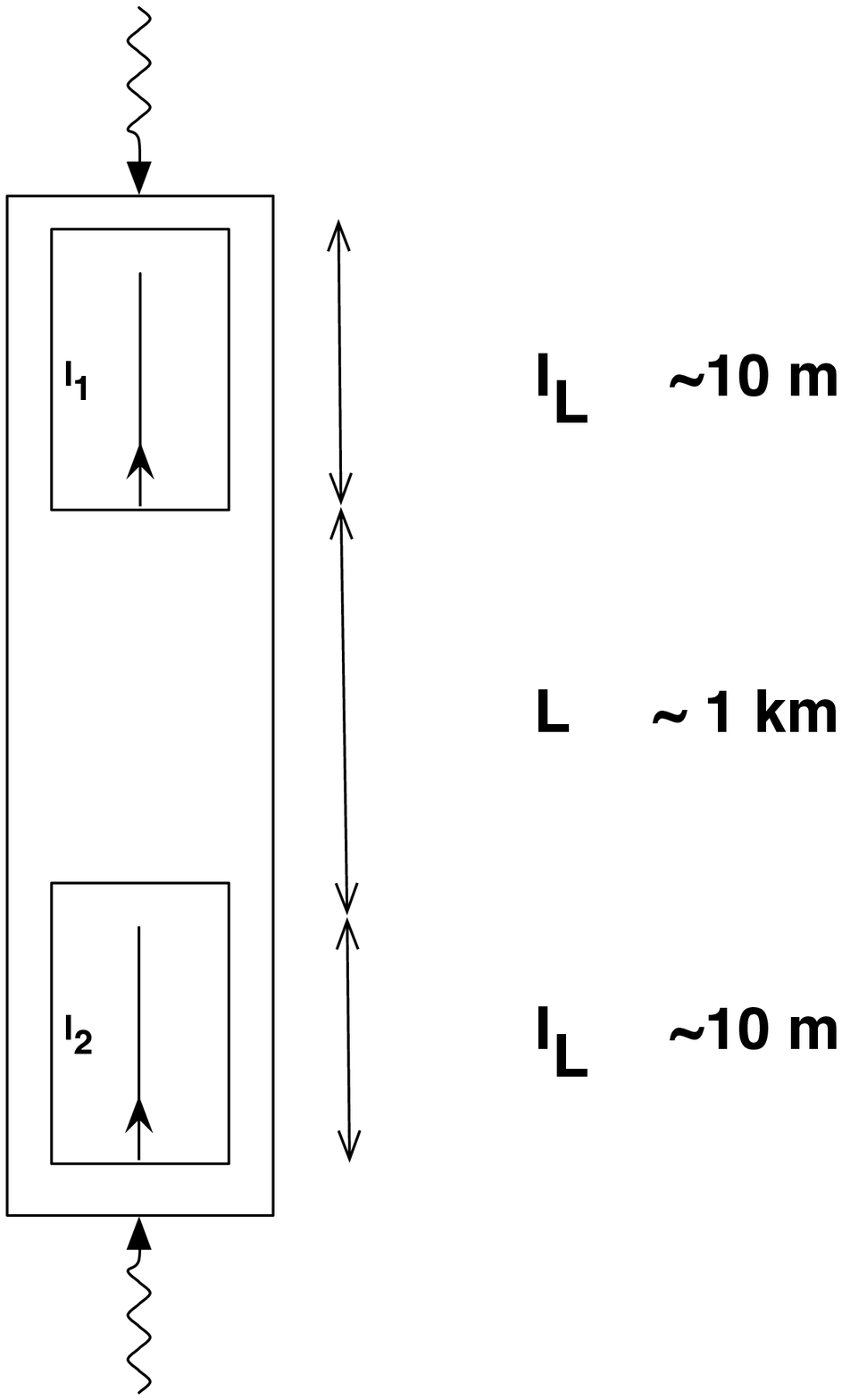}}\\ 

\caption{\label{Fig: Earth diff} Figure \ref{Fig:space-time} is a space-time diagram of two light pulse interferometers in the proposed differential configuration, as in Figure \ref{Fig:AI-SingleInterferometer}. \\ \\  Figure \ref{Fig:earthsetup} is a diagram of the proposed setup for a terrestrial experiment. The straight lines represent the path of the atoms in the two $I_L \sim 10$ m  interferometers $I_{1}$ and $I_{2}$ separated vertically by $L \sim 1$ km.   The wavy lines represent the paths of the lasers.}
\end{center}

\end{figure}

On the earth, one possible experimental configuration is to have a long, vertical shaft with one interferometer near the top and the other near the bottom of the shaft.  The atom interferometers would be run vertically along the axis  defined by the common laser pulses applied from the bottom and top of the shaft, as shown in Fig. \ref{Fig:earthsetup}.  For reference we will consider a $L \sim 1 \text{ km}$ long shaft, with two $I_{L} \sim 10 \text{ m}$ long atom interferometers $I_1$ and $I_2$.  Each atom interferometer then has $T \sim 1 \text{ s}$ of interrogation time, so such a setup will have maximal sensitivity to a gravitational  wave of frequency around $1 \text{ Hz}$.  Because the two atom interferometers are separated by a distance $L$, the gravitational  wave signal in each will not have the same magnitude but will differ by $\sim h L \omega^2 T^2$ as shown in the previous section.  Such a differential measurement can  reduce  backgrounds without reducing the signal.

One reason to use only 10 m at the top and bottom of the shaft for the atom interferometers themselves is to reduce the cost scaling with length.  There are more stringent requirements on the interferometer regions than on the region between them.  The interferometer regions should have a constant bias magnetic field applied vertically in order to fix the atomic spins to the vertical axis.  This bias field must be larger than any ambient magnetic field, and this, along with the desire to reduce phase shifts from this ambient field (as will be seen later), requires magnetically shielding the interferometer regions.  Further, the regions must be in ultra-high vacuum $\sim 10^{-10} \, \text{Torr}$ in order to avoid destroying the cold atom cloud.  At this pressure and room temperature the vacuum contains $n \approx 3 \times 10^{12} \frac{1}{\text{m}^3}$ particles at an average velocity of $v \approx 500 \, \frac{\text{m}}{\text{s}}$.  The $\text{N}_2$-Rb cross section is $\sigma \approx 4 \times 10^{-18} \text{ m}^2$ \cite{Rapol}.  The average time between collisions is $\frac{1}{n \sigma v} \approx 200 \text{ s}$, so the cold atom clouds can last the required 1 to 10 s. 

As discussed in section \ref{Sec: GW Signal}, the signal in the interferometer arises from the time dependent variation of the distance between the interferometers as sensed by the laser pulses executing the interferometry. The success of this measurement strategy requires the optical path length in the region between the interferometers to be stable in the measurement band. In order to detect a gravitational wave of strain $h_{rms} \sim  \frac{10^{-19}}{\sqrt{\text{Hz}}}$ (see  section \ref{Sec: Sensitivities}), time variations in the index of refraction $\eta$ of the region between the interferometers, in the 1 Hz band, must be smaller than $h_{rms}$. The index of refraction of air is $\eta \sim 1 + 10^{-4} \left(\frac{P}{\text{760 Torr}}\right) \left(\frac{\text{300 K}}{\tau}\right)$ where $P$ is the pressure and $\tau$ is the temperature. Time variations $\delta \tau$ of the temperature cause time variations in the index of refraction $\delta \eta \sim 10^{-4} \left(\frac{P}{\text{760 Torr}}\right) \left(\frac{\text{300 K}}{\tau}\right) \left( \frac{\delta \tau}{\tau} \right)$. The required stability in $\eta$ can be achieved if the region between the interferometers is evacuated to pressures $P \sim 10^{-7} \text{ Torr} \left( \frac{\text{0.01 K}}{\delta \tau} \right)$  with temperature variations $\delta \tau$ over time scales $\sim 1$ second. 

If the entire length $L$ of the shaft can be evacuated to $\sim 10^{-10} \text{ torr}$ and magnetically shielded then the atom interferometers can be run over a much larger length $I_L \sim L$, yielding a larger interrogation time and greater sensitivity to low frequency gravitational  waves.  The signal sensitivity is $\propto (L - \frac{1}{2} g T^2) (\omega T)^2$ for a gravitational  wave of frequency $\omega \leq T^{-1}$. For such a low frequency gravitational  wave, this is maximized when $I_L = \frac{1}{2} g T^2 = \frac{1}{2} L$, so the length of  each interferometer should be chosen to be equal to half the distance between the lasers.  In the case of a $1 \text{ km}$ long shaft this would give an interrogation time of $T \approx 10 \text{ s}$ so a peak sensitivity to $0.1 \text{ Hz}$ gravitational  waves.

In order to run an atom interferometer over such a long baseline, it is necessary to have the laser power to drive the stimulated 2-photon transitions (Raman or Bragg) used to make beamsplitter and mirror pulses from that distance.   It is possible to obtain a sufficiently rapid Rabi oscillation frequency using a $\sim 1 \, \text{W}$ laser with a Rayleigh range $\sim 10 \, \text{km}$ which is easy to achieve with a waist of $\sim 10 \, \text{cm}$.  This will be a more restrictive requirement for the satellite based experiment and so will be considered in greater detail in Section \ref{Sec:space baseline limit}.

The atom interferometer configurations discussed here are maximally sensitive to frequencies as low as $\frac{1}{T}$ and lose sensitivity at lower frequencies.  However, the sensitivity is also limited at high frequencies by the data-taking rate, $f_d$, the frequency of running cold atom clouds through the interferometer.  Gravitational wave frequencies higher than the Nyquist frequency, $\frac{f_d}{2}$, will be aliased to lower frequencies which is undesirable since we wish to measure the frequency of the gravitational wave.  We will cut off our sensitivity curves at the Nyquist frequency.  If only one cloud of atoms is run through the interferometer at a time, the Nyquist frequency will be below $\frac{1}{T}$.  Thus it is important to be able to simultaneously run more than one cloud of atoms through the same interferometer concurrently.  Using the same spatial paths for all the cold atom clouds is useful since otherwise DC systematic offsets in the phase of the different interferometers could give a spurious signal at a frequency $\sim f_d$.

\begin{figure}
\begin{center}
\includegraphics[width=1.0 in]{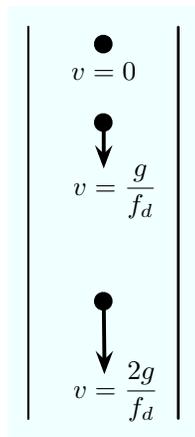}
\caption{ \label{Fig:many shots} A diagram of several clouds of atoms being run through the atom interferometer sequence concurrently. The arrows indicate the velocity of each cloud of atoms at a single instant in time.  Earlier shots will be moving with more downward velocity, allowing the clouds to be individually addressed with Doppler detuned laser frequencies.}
\end{center}
\end{figure}

It is then necessary to estimate how high a data-taking rate is achievable.  We will show that it is possible to have multiple atom clouds running concurrent atom interferometers in each of the two interferometer regions.  This is possible because the atom clouds are dilute and so pass through each other and also because it is possible for the beamsplitter and mirror laser pulses to interact only with a particular atom cloud even though all the atom clouds are along the laser propagation axis.  This is accomplished by Doppler detuning the required laser frequencies of each atomic transition by having all the atom clouds moving with different velocities at any instant of time.  We imagine having different clouds shot sequentially with the same launch position and velocity, with a time difference $\frac{1}{f_d} < T$ (see Figure \ref{Fig:many shots}).  The atom clouds accelerate under gravity and so each successive atom cloud always has a velocity difference from the preceding one of $\frac{g}{f_d}$.  When a beamsplitter or mirror pulse is applied along the axis, it must be tuned to the Doppler shifted atomic transition frequency.  The width of the two-photon transitions that make the beamsplitters and mirrors is set by the Rabi frequency and so can be roughly $\Omega^{-1} \sim 10^{4} \text{~Hz}$.  Taking a laser frequency of $3 \times 10^{14} \text{~Hz}$ implies that the clouds must have velocities that differ by at least $\Delta v \sim \frac{10^4}{3 \times 10^{14}} \approx 3 \times 10^{-11} \approx 1 \frac{\text{cm}}{\text{s}}$ in order for the Doppler shift to be larger than the width of the transition.  In practice, every cloud besides the one being acted upon should be many line-widths away from resonance which can be accomplished if $\frac{g}{f_d} \gg  3 \times 10^{-11}$.  

While doppler detuning prevents unwanted stimulated transitions, the laser field can drive  spontaneous 2-photon transitions as discussed in subsection \ref{Sec:space baseline limit}. A significant fraction of the atoms should not undergo spontaneous transitions in order for the interferometer to operate with the desired sensitivity. Using the formalism discussed in subsection \ref{Sec:space baseline limit},  the spontaneous emission rate $R$  is given by $R \sim \frac{2 \Omega_{\text{st}}^2}{\Gamma \frac{I}{I_{sat}}}$ where $\Omega_\text{st}$ is the Rabi frequency of the stimulated 2-photon transition,  $\Gamma$ is the decay rate of the excited state,  $I$  the intensity of the lasers at the location of the atoms and $I_{sat}$ the saturation intensity of the chosen atomic states. With  $\Omega_\text{st} \sim 2 \pi \left( 10^4 \, \text{Hz} \right)$, lasers of waist $\sim 3 \, \text{cm}$ and power $\sim 1 \, \text{W}$,  atomic parameters (e.g. for Rb or Cs) $\Isat \approx 2.5 \, \frac{\text{mW}}{\text{cm}^2}$, and $\Gamma  \approx 3 \times 10^7 \, \frac{\text{rad}}{\text{s}}$ \cite{Metcalf, Steck} the spontaneous emission rate is $R \sim \left(5 \text{ s}\right)^{-1}$. These parameters will allow for the operation of up to $\sim 5$ concurrent interferometers using up to $N \sim 1000$ LMT beamsplitters. This implies a data-taking rate of $f_d \sim 5 \text{~Hz}$.  This number of course depends on the particular atomic species being used and the laser intensity. A more judicious choice of atom species or increased laser power will directly increase the data-taking rate. Our only desire here is to show that it should be possible to have a data-taking rate of $f_d \sim 10 \text{~Hz}$.  This is the number we will use for our sensitivity plots.  In the actual experiment, the data-taking rate will depend on many complex details, including the atom cooling mechanism.  There is a tradeoff between the rate of cooling and the number of atoms in the cloud.  Here we only assume that cooling can be done at this rate, if necessary with several different atomic traps, since this rate is not drastically higher than presently achievable rates. 

In order to detect a gravitational wave it is only necessary to have one such pair of atom interferometers.  However, it may be desirable to have several such devices operating simultaneously. Detecting a stochastic background of gravitational waves requires cross-correlating the output from two independent gravitational wave detectors.  Even for a single, periodic source, correlated measurements would increase the confidence of a detection. Furthermore, cross-correlating the outputs of independent gravitational wave detectors will help reduce the effects of backgrounds with long coherence times.  In addition, with three such single-axis gravitational wave detectors whose axes point in different directions, it is possible to determine information on the direction of the gravitational wave source.  Such independent experiments could be oriented vertically in different locations on the earth, giving different axes.  It is also possible to consider orienting the laser axis of such a pair of atom interferometers horizontally, though to maintain sensitivity to $\sim 1 \text{ Hz}$ gravitational waves the atom interferometers themselves would still have to be 10 m long vertically.  Such a configuration would still have the same signal, proportional to the length between the interferometers, though some backgrounds could be different. 

It may be desirable to operate two, non-parallel atom interferometer baselines that share a common passive laser in a LIGO-like configuration. For example, one baseline could be vertical with the other horizontal, or both baselines can be horizontal. Each baseline consists of two interferometers. The interferometers along each baseline are operated by a common control laser. The passive laser is placed at the intersection of the two baselines with appropriate optical beamsplitters so that the beam from the passive laser is shared by both baselines. As discussed in sub section \ref{Sec: Earth Laser Phase Noise}, laser phase noise in this configuration is significantly suppressed.

\subsection{Backgrounds}
\label{Sec: Earth Backgrounds}


We consider the terrestrial setup discussed in the above section with two atom interferometers separated vertically by a $L \sim 1$ km long baseline. The interferometers will be operated by common lasers and the experiment will measure the differential phase shift between the two atom interferometers. This strategy mitigates the effects of vibration and laser phase noise. Based upon realistic extrapolations from current performance levels, atom interferometers  could conceivably reach a  per shot phase sensitivity $\sim 10^{-5}$ rad. This will make the interferometer sensitive to accelerations  $\sim \EarthShotNoise \left(\frac{1 \text{s}}{T}\right)^2$ (Section \ref{Sec: Sensitivities}). We will assume a range of sensitivities. In what follows, we will show that backgrounds can be controlled to better than the most optimistic sensitivity $\sim 10^{-5}$ rad. 

A gravitational  wave of amplitude $h$ and frequency $\omega$ produces an acceleration $\sim h L \omega^2$. With an acceleration sensitivity of $\EarthShotNoise$, the experiment will have a gravitational wave strain sensitivity $\sim \frac{10^{-18}}{\sqrt{\text{Hz}}} (\frac{1 \text{ km}}{L})$. This sensitivity will allow  the detection of gravitational waves of amplitude $h \sim 10^{-22} (\frac{4 \text{ km}}{L})$ after $\sim 10^6$ s of integration time (Section \ref{Sec: Sensitivities}). The detection of gravitational  waves at these sensitivities requires time varying differential phase shifts in the interferometer to be smaller than the per shot phase sensitivity $\sim 10^{-5}$. In particular, time varying differential acceleration backgrounds must be smaller than the target acceleration sensitivity $\EarthShotNoise$. 

In addition to stochastic noise, there might be backgrounds with long coherence times in a given detector.  Since these backgrounds will not efficiently integrate down, the sensitivity of any single detector will be limited by the floor set by these backgrounds. However, as discussed in sub section \ref{Sec: Earth Setup},  it may be desirable to build and simultaneously operate a network of several such gravitational wave detectors. The gravitational wave signal in a given detector depends upon the orientation of the detector relative to the incident direction, the polarization and the arrival time of the gravitational wave at the detector. If the detectors are sufficiently far away, then the gravitational wave signal in the detectors are in a well defined relationship which is different from the contribution of backgrounds with long coherence time. If the output of these detectors are cross-correlated, then the sensitivity of the network will be limited by the stochastic noise floor.  In the following,  we will assume that such a network of independent gravitational wave detectors can be constructed  with their cross-correlated sensitivity limited by stochastic noise. We discuss these stochastic backgrounds  and strategies to suppress them to the required level. 

\subsubsection{Vibration Noise}

The phase shift in the interferometer is accrued by the atom during the time between the initial and final beamsplitters. In this period, the atoms are in free fall and are  coupled to ambient vibrations only through gravity. In addition to this coupling, vibrations of the trap used to confine the atoms before their launch will lead to fluctuations in the launch velocity of the atom cloud. These fluctuations do not directly cause a phase shift since the initial beamsplitter is applied to the atoms after their launch. However, variations in the launch velocity will make the atoms move along different trajectories. In a non-uniform gravitational field, different trajectories will see different gravitational fields thereby producing time dependent phase shifts. But, since these effects arise from gravitational interactions, their impact on the experiment is significantly reduced. A detailed discussion of these gravitational backgrounds is contained in  subsection  \ref{SubSec: EarthNewtonianGravity}.

The vibrations of the lasers contribute directly to the phase shift through the laser pulses used to execute the interferometry. The pulses from the control laser (Section \ref{Sec: GW Signal}) at times $0$, $T$ and $2T$ (Figure \ref{Fig:space-time}) are common to both interferometers and contributions from the vibrations of this laser to the differential phase shift are completely cancelled. The vibrations of the passive laser (Section \ref{Sec: GW Signal}) are not completely common. The pulses from the passive laser that hit one interferometer ($\tau_{a_1}, \tau_{b_1}, \tau_{c_1}, \tau_{d_1}$) are displaced in time by $L$  from the pulses  ($\tau_{a_2}, \tau_{b_2}, \tau_{c_2}, \tau_{d_2}$) that hit the other interferometer due to the spatial separation $L$ between interferometers (Figure  \ref{Fig:space-time} ). 

The proposed experiment relies on using LMT beamsplitters to boost the sensitivity of the interferometer. The effect of a LMT pulse on the atom can be understood by modeling the LMT pulse as being composed of $N$ ($\sim 1000$)  regular laser pulses. If the time duration of each regular pulse is greater than the light travel time $L$ between the two interferometers, then all but the beginning and end of each LMT pulse will be common to the interferometers. The time duration of the pulses can be modified by changing the Rabi frequency of the transition of interest by manipulating the detuning and intensity of the lasers from the intermediate state used to facilitate the 2-photon Raman transitions. With Rabi frequencies $\sim 3 \times 10^5 \text{ Hz } (\frac{\text{1 km}}{L})$, the duration of a regular pulse is equal to the distance between the interferometers. The beginning and end of the LMT pulse from the passive laser that hits one interferometer is displaced in time by $L$ from the pulse that hits the other interferometer. Vibrations $\delta x$  of the passive laser position in this time interval are uncommon and result in a phase shift $\sim k \delta x$ instead of $\keff \delta x$. Contributions to the phase shift from vibrations of the passive laser at frequencies smaller than $\frac{1}{L}$ are common to both interferometers and are absent in the differential phase shift.  

The net phase shift $k  \delta x$ is smaller than $10^{-5}$ if $\delta x \lessapprox 10^{-12} \text{ m } (\frac{10^{7} \text{m}^{-1}}{k})$. Here $\delta x$ is the amount by which the passive laser moves in the light travel time $L$ between the two interferometers.  A vibration at frequency $\nu$ with amplitude $a$ contributes to the displacement $\delta x$ of the laser in a time $L$ by an amount $a \nu L$. This displacement is smaller than $10^{-12} \text{ m}$ if $a < \left(\frac{10^{-12} \text{ m}}{\nu L}\right) = 3 \times 10^{-7} \text{ m} \left(\frac{\text{1 Hz}}{\nu}\right) \left( \frac{1 \text{ km}}{L}\right)$. This can be achieved by placing the passive laser on vibration isolation stacks that damp its motion below  $10^{-7} \frac{\text{ m}}{\sqrt{\text{Hz}}} \left(\frac{1 \text{ Hz}}{\nu}\right)^{\frac{3}{2}} \left(\frac{1 \text{ km}}{L}\right) $. It is only necessary to damp the motion of the lasers below this value in the frequency band  $3 \times 10^5 \text{ Hz}  \left(\frac{1 \text{km}}{L}\right) \gtrsim \nu \gtrsim 1 \text{ Hz}$. The high frequency cutoff is established since contributions to the phase shift from vibrations at frequencies above $3 \times 10^5 \text{ Hz}  \left(\frac{1 \text{km}}{L}\right)$ are suppressed by the size of the Rabi pulse.  The low frequency cutoff arises since vibrations at frequencies below $1 \text{ Hz}$ are irrelevant to the detection of gravitational waves at $1 \text{ Hz}$. 

While these vibrations may have long coherence times in a single gravitational wave detector, cross correlating multiple detectors with different vibrational noise should allow these vibrations to be reduced to the stochastic floor. 

\subsubsection{Laser Phase Noise}
\label{Sec: Earth Laser Phase Noise}

The gravitational  wave signal in the interferometer arises from an asymmetry in the time durations between the first and second and between the second and third laser pulses. The interferometer is operated by pulsing the control laser at equal time intervals. The corresponding pulses from the passive laser that interact with the atom must then have been emitted at unequal time intervals since in the presence of a gravitational  wave,  pulses emitted at different times travel along different trajectories \footnote{Note that the passive laser is turned on well before the control laser is pulsed, and then we are only referring to the part of the passive laser pulse which triggers the atomic transitions.}.  The phase of the passive laser reflects this temporal asymmetry. This phase is impinged on the atom during the interaction between the atom and the laser field  producing a phase shift in the interferometer (see Section \ref{Sec: GW Signal}). Noise in the evolution of the laser's phase will mimic a temporal variation and is a background to the experiment. The pulses from the control laser are common to both interferometers. Noise in the phase of this laser does not contribute to the differential phase shift as these contributions are completely cancelled.

The pulses from the passive laser that hit one interferometer ($\tau_{a_1}, \tau_{b_1}, \tau_{c_1}, \tau_{d_1}$) are displaced in time by $L$  from the pulses  ($\tau_{a_2}, \tau_{b_2}, \tau_{c_2}, \tau_{d_2}$) that hit the other interferometer (Figure  \ref{Fig:space-time}). Since the pulses are not completely common, phase noise in the passive laser will contribute to the differential phase shift. Phase noise in a laser operating at a central frequency $k$ during a time interval $\delta T$  can be characterized as the difference $\delta \phi = \phi_m - k \delta T$ where $\phi_m$ is the phase measured after $\delta T$. The pulses from the passive laser that hit the two interferometers are separated in time by $L$ and phase noise of the laser during this period will contribute to the differential phase shift. An additional contribution arises from the drift of the central frequency of the laser in the time $T$ between pulses. A drift, $\delta k$, in the central frequency of the laser between pulses changes the evolution of the laser phase  mimicking a change in the time of emission of the laser pulse. The pulses from the passive laser that interact with the two interferometers are separated by a time $L \ll T$. The contribution to the phase of these pulses from the frequency drift are common except for the additional phase accrued by the laser during the time $L$. This additional phase $\delta k \; L$  contributes to the differential phase shift in the interferometer.

\begin{figure}
\begin{center}
\includegraphics[width=4.5 in]{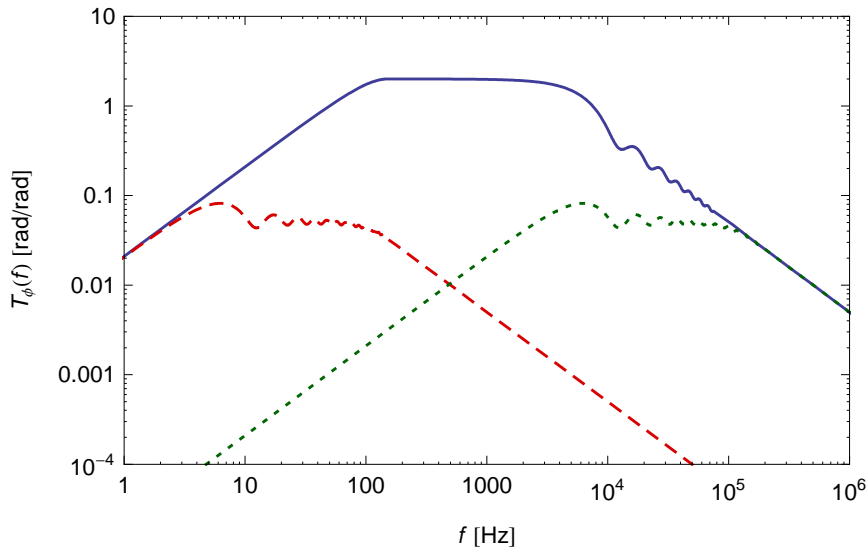}
\caption{ \label{Fig:phasenoise}(Color online) Interferometer Response to laser phase noise in a single $\frac{\pi}{2}$ pulse.  The dotted (green) curve represents a differential measurement strategy with $L=1 \text{ km}$ and a Rabi period of $10^{-4} \text{ s}$.  The solid (blue) curve is the same Rabi period but $L= 1000 \text{ km}$.  The dashed (red) curve is $L= 1000 \text{ km}$ and a Rabi period of $10^{-1} \text{ s}$.  Sharp spikes in the response curves above the Rabi frequency have been enveloped.
}
\end{center}
\end{figure}

The proposed experiment relies on using LMT beamsplitters to boost the sensitivity of the interferometer. As a proof of principle, we will model the LMT pulse as a sequential Bragg process.  In other words, it is composed of $N$ ($\sim 1000$) of the regular laser pulses (those used to drive a single 2-photon Bragg transition) run consecutively with no time delay between them. Each regular laser pulse transfers a momentum $2 k$ to the atom resulting in an overall momentum transfer $\keff = 2 N k$. The phase of every laser pulse is registered on the atom, amplifying the phase noise transferred to the atoms to $\sqrt{N} \delta \phi \oplus N \delta k \; L$ ($\oplus$ means add in quadrature). However, if the time duration of each regular pulse is greater than the light travel time $L$ between the two interferometers, then all but the beginning and end of each LMT pulse will be common to the interferometers. This can be achieved by setting the Rabi frequency of the transition to be $\lesssim \frac{1}{L} \sim 3 \times 10^5 \text{ Hz } (\frac{\text{1 km}}{L})$. The Rabi frequency can be tuned by manipulating the detuning of the lasers  from the intermediate state used to facilitate the 2-photon Bragg transitions. The differential phase shift will then receive contributions from the phase noise in the beginning and end of each LMT pulse alone.  For example, if the light travel time $L$ is equal to the 2-photon Rabi period, then only the first and last of the regular laser pulses making up the LMT pulse will have uncommon phase noise, all the rest will give common phase noise to the two interferometers which will cancel.  In this case the laser phase noise would be reduced to $\sqrt{2} \delta \phi \oplus 2 \delta k \; L$, independent of $N$.  This method for reducing the laser phase noise from an LMT pulse was discussed for a sequential Bragg process as a demonstration, but similar ways of reducing the phase noise may exist for other LMT methods \footnote{For example, a sequential Raman process with reduced phase noise might be realizable in a similar manner if the light travel time $L$ is set equal to twice the Rabi period and both lasers alternate between the two frequencies needed to run the Raman process with no intervening time.}.


Low frequency phase noise is suppressed by the differential measurement strategy outlined above at frequencies below $\sim \frac{1}{L}$.  High frequency phase noise is reduced by averaging over the finite time length of the pulse, and will be suppressed above the Rabi frequency.  To see these reductions, the calculated atom response to phase noise is shown in Fig. \ref{Fig:phasenoise} for several different configurations (for a description of a similar calculation see \cite{phasenoisecalc}, here we have also added in a time delay due to the finite speed of light). With Rabi frequencies $\sim 3 \times 10^5 \text{ Hz } (\frac{\text{1 km}}{L})$, the contribution of the laser phase noise to the differential phase shift is $\sim \delta \phi \oplus \delta k \; L$. This phase shift must be smaller than $10^{-5}$.

$\delta \phi$ is the phase noise in the laser at frequencies $\sim 3 \times10^{5} \text{ Hz } (\frac{\text{1 km}}{L})$. This is smaller than $10^{-5}$ if the phase noise of  the laser is smaller than  $-140 \frac{\text{dBc}}{\text{Hz}}$ at a  $\sim 3 \times 10^{5} \text{ Hz } (\frac{\text{1 km}}{L})$ offset. The $\delta k \; L$ term is smaller than $10^{-5}$ if $\delta k \lessapprox 10^{-9} \text{ m}^{-1} (\frac{\text{1 km}}{L})$ which requires fractional stability in the laser frequency $\sim 10^{-15} (\frac{10^7 \text{m}^{-1}}{k})$ over time scales $\sim T$. These requirements can be met using lasers locked to high finesse cavities \cite{LudlowStableLaser}.

Another scheme that could be employed to deal with laser phase noise is to operate interferometers along two, non parallel baselines that share a common passive laser in a LIGO-like configuration. For example, one baseline could be vertical with the other horizontal, or both baselines can be horizontal. Each baseline consists of two interferometers. The interferometers along each baseline are operated by a common control laser. The passive laser is placed at the intersection of the two baselines with appropriate optical beamsplitters so that the beam from the passive laser is shared by both baselines. The same pulses from the passive laser can trigger transitions along the interferometers in both baselines if the control lasers along the two baselines are simultaneously triggered. The laser phase noise in the difference of the differential phase shift along each baseline is greatly suppressed since phase noise from the control laser is common to the interferometers along each baseline and the phase noise from the passive laser is common to the baselines. The gravitational wave signal is retained in this measurement strategy since the gravitational wave will have different components along the two non parallel baselines. As discussed in sub section \ref{Subsec:SpaceVibrations}, laser phase noise along the two arms can be cancelled up to knowledge of the arm lengths of each baseline. With $\sim 10$ cm knowledge of the arm lengths, these contributions are smaller than shot noise if the frequency drift $\delta k$ of the laser is controlled to better than $\sim 10^{4} \frac{\text{Hz}}{\rthz}$ at frequencies $\omega \sim 1$ Hz.

\subsubsection{ Newtonian Gravity Backgrounds }

\label{SubSec: EarthNewtonianGravity}

\begin{figure}
\begin{center}
\includegraphics[width=4.5 in]{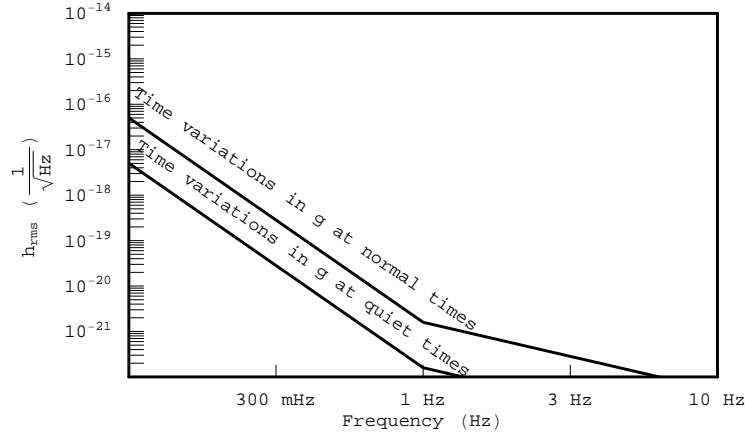}
\caption{ \label{Fig:timevargrav10Km} Interferometer response in strain per $\sqrt{\text{Hz}}$ to a time varying g with a 10 km baseline setup.  See text for more details.
}
\end{center}
\end{figure}

\begin{figure}
\begin{center}
\includegraphics[width=4.5 in]{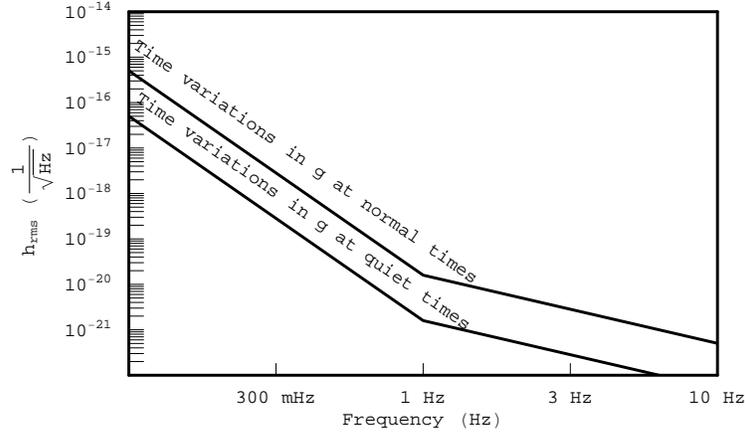}
\caption{ \label{Fig:timevargrav1Km} Interferometer response in strain per $\sqrt{\text{Hz}}$ to a time varying g with a 1 km baseline setup.  See text for more details.
}
\end{center}
\end{figure}

The average gravitational field $g_L$ along the space-time trajectory of each atom contributes to the phase shift in the interferometer. Each shot of the experiment measures the average phase shift of all the atoms in the cloud and is hence sensitive to the average value ($\glavg$) of $g_L$ over all the atoms in the cloud. Time variations in $\glavg$ are a background to the experiment. Seismic and atmospheric activity are the dominant natural causes for time variations in $\glavg$. The gravitational effects of these phenomena were studied in \cite{Saulson:1984yg} and \cite{Hughes:1998pe}. Using the interferometer transfer functions evaluated in these papers, we find that time varying gravitational accelerations will not limit the detection of gravitational  waves at sensitivities $\sim \frac{10^{-17}}{\sqrt{\text{Hz}}} (\frac{\text{1 km}}{L})$ at frequencies above 300 mHz (Figures \ref{Fig:timevargrav10Km}, \ref{Fig:timevargrav1Km}).  Human activity can also cause time variations in $\glavg$.  Any object whose motion has a significant overlap with the 1 Hz band is a background to the experiment. This background is smaller than $\EarthShotNoise$ if such objects (of mass $M$) are at distances larger than 1 km $\sqrt{\frac{M}{\text{1000 \text{ kg}}}}$.

The trajectory of the atom is determined by its initial position ($R_L$)  and velocity ($v_L$). Variations in $R_L$ and $v_L$ will change the trajectory of the atom. In a non-uniform gravitational field, different trajectories will have different values of $g_L$. The interferometer has to run several shots during the period of the gravitational  wave source in order to detect the time varying phase shift from the gravitational  wave.  The average launch position and velocity of the atoms may change from shot to shot thereby changing the average gravitational field sensed by the interferometer. These variations cause time dependent phase shifts.  We estimate the size of these effects by writing $g_L$ in terms of the length $I_L \sim v_L T$ of the interferometer as: 
\begin{equation}
g_L = g(R_L) + \nabla g(R_L) v_L T + \dots
\end{equation}
where $g(R_L)$ and $\nabla g(R_L)$ are the gravitational field and its gradient at the initial position $R_L$ of the atom. $\glavg$ can then be expressed in terms of the average initial position ($\rlavg$) and velocity ($\vlavg$) of the atom cloud as: 
\begin{equation}
\glavg = g(\rlavg) + \nabla g(\rlavg) \vlavg T + \dots
\end{equation}
Shot to shot variations $\delta \rlavg$ and $\delta \vlavg$ in the average position and velocity of the atom cloud will result in accelerations $\sim \nabla g \delta \rlavg + \nabla g \delta \vlavg T$. These accelerations must be made smaller than $\EarthShotNoise$. 

The gradient of the Earth's gravitational field in a vertical interferometer is $\nabla g \sim \frac{G M_E}{R_E^3}$. The corresponding accelerations $\frac{G M_E}{R_E^2} \frac{\delta \rlavg}{R_E}$ and  $\frac{G M_E}{R_E^2} \frac{\delta \vlavg T}{R_E}$ are smaller than $\EarthShotNoise$ if $\delta \rlavg$ and $\delta \vlavg$ are smaller than $10 \frac{\text{nm}}{\sqrt{\text{Hz}}}$ and $ 10 \frac{\text{nm/s}}{\sqrt{\text{Hz}}}$ respectively. $\delta \rlavg$ and $\delta \vlavg$ are caused by vibrations of the atom traps used to confine the atoms and thermal effects in the atom cloud. 

Vibrations of the atom traps are caused by seismic motion and fluctuations in the magnetic fields used to confine the atoms. Seismic vibrations in the 1 Hz band have been measured to be $\sim 10 \frac{\text{nm}}{\sqrt{\text{Hz}}}$ \cite{Fix} at an average site on the Earth.  During noisier times, these vibrations may be as large as  $\sim 100 \frac{\text{nm}}{\sqrt{\text{Hz}}}$ \cite{Fix}. Seismic vibrations of the trap are therefore only marginally bigger than the $10 \frac{\text{nm}}{\sqrt{\text{Hz}}}$ control required by this experiment and hence these vibrations can be sufficiently damped by vibration isolation systems. 

The magnetic fields used to trap the atom will fluctuate due to variations in the currents used to produce these fields. The trap used in this experiment can be modelled as a harmonic oscillator with frequency $\omega_T =  \sqrt{\frac{\kappa}{M_{A}}} \sim 100 \text{ Hz}$ where $M_{A}$ is the mass of the atom and the ``spring constant" $\kappa$ is proportional to the (curvature of) applied magnetic field. Fluctuations in the equillibrium position of this oscillator due to variations in $\kappa$ are $\sim \frac{g}{\omega_T^2} \frac{\delta \kappa}{\kappa}$ and these are smaller than $10 \frac{\text{nm}}{\sqrt{\text{Hz}}}$ if $\frac{\delta \kappa}{\kappa} \lessapprox \frac{10^{-5}}{\sqrt{\text{Hz}}}$. Since $\kappa$ is proportional to the applied current, fractional stability $\sim \frac{10^{-5}}{\sqrt{\text{Hz}}}$ in the current source will adequately stabilize the equillibrium position of the trap. 

The requirements on the control over the atom traps can be ameliorated by using a common optical lattice to launch the atoms in both interferometers. The vibrations of the lattice will then be common to both interferometers and the first non-zero contribution to the differential phase shift from the Earth's gravitational field arises from the quadratic gradient of this field. These contributions are smaller than $\EarthShotNoise$ if the vibrations of the lattice are $\lessapprox \frac{10^{-4} \text{m}}{\sqrt{\text{Hz}}}$ in the 1 Hz band.



The average velocity of the atom clouds will change from shot to shot due to the random fluctuations in the thermal velocities of the atoms.  These variations are smaller than 10 nm/s if the average thermal velocity of each atom cloud is smaller than 10 nm/s. In an atom cloud with $\sim 10^8$ atoms, the average thermal velocity is smaller than 10 nm/s if the thermal velocities of the atoms are  $\sim 100 \mu \text{m/s}$. Such thermal velocities can be achieved by cooling the cloud to $\sim 100$ picokelvin temperatures. Similarly, the average position of the atom cloud changes from shot to shot due to thermal effects. These fluctuations can be made smaller than 10 nm by confining the atoms within a region of size 100 $\mu \text{m}$.


The control required over the launch parameters of the atom cloud is directly proportional to the gravity gradient $\nabla g$.  Thus these requirements may be ameliorated by reducing the local gravity gradient. We estimated these controls using the natural value of $\nabla g \sim \frac{G M_E}{R_E^3}$ on the surface of the Earth. However for a $\sim 10 \text{ m}$ atom interferometer, it may be possible to reduce gravity gradients to  $\sim 1 \% $ their natural value by shimming the local gravitational field using a suitably chosen local mass density. The density of the Earth varies significantly with distance below its surface. The average density of the Earth is $\bar{\rho} \sim  \text{5.5 gm/cm}^3$ while the average density of its crust is  $\rho_{c} \sim \text{3 gm/cm}^3$. The earth's  gravitational field in a vertical interferometer  inside the Earth's crust can be modeled as arising from a sphere of radius $R_E$ with average density equal to $\rho_c$ and a point object of mass $(\frac{4 \pi}{3}) (\bar{\rho} - \rho_{c}) R_E^3$ located at the center of the Earth. The gradient of this field is $\sim G (\frac{4 \pi}{3})  ( 2 (\bar{\rho}-\rho_c) - \rho_c)$ and the effect of this gradient can be cancelled by surrounding the upper end of the interferometer by a sphere of radius $\sim 1$ m and average density $\sim 2 (\bar{\rho}-\rho_c) \sim \text{ 5 gm/cm}^3$ (see also Section V of \cite{GR Atom}). The demands on the launch parameters of the experiment can be relaxed to the extent to which the local gravity gradient can be reduced. For example, if the gravity gradient in each interferometer is reduced to a percent of its natural value, the experiment can reach the target sensitivity with $1 \frac{\mu \text{m}}{\sqrt{\text{Hz}}}$ control over the average position and $1 \frac{\mu \text{m/s}}{\sqrt{\text{Hz}}}$ control over the average velocity of the atom clouds in the 1 Hz frequency band.

\begin{figure}
\begin{center}
\includegraphics[width=3.5 in]{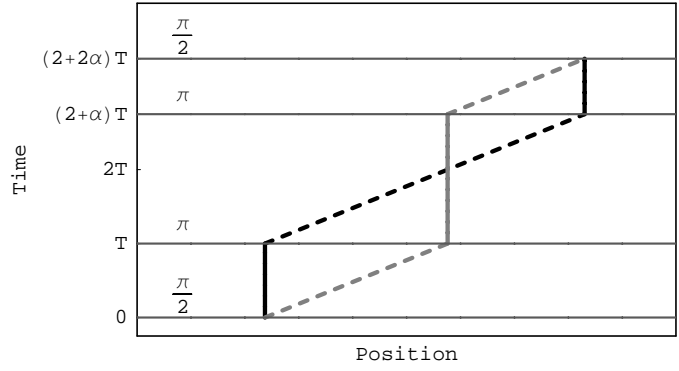}
\caption{ \label{Fig:double-diamond} A space-time diagram of the double loop interferometer. The black and gray lines indicate the two halves of the wave function after the initial beamsplitter. The dashed and solid lines represent the two internal states of the atom. The laser light used to manipulate the atom is shown as horizontal dark gray lines. The speed of light has been exaggerated. 
}
\end{center}
\end{figure}

The interferometer configuration discussed above executes its control pulses in the Mach-Zender sequence $\frac{\pi}{2} - \pi - \frac{\pi}{2}$ with equal time between pulses.  The interferometer can also be run in the $\frac{\pi}{2} - \pi - \pi - \frac{\pi}{2}$ double loop configuration with the atom spending a time $2 T$ in the lower loop and a time $(\sqrt{5} - 1) T$ in the upper loop (Figure \ref{Fig:double-diamond} with $\alpha = \frac{\sqrt{5}-1}{2}$). In this configuration, the phase shift from constant accelerations is retained while the contribution from linear acceleration gradients is identically cancelled \cite{McGuirk, BorisDiamonds}.  In this case the only velocity-dependent contributions come from second gradients of the gravitational field.  Phase shift variations from shot to shot variations in the average velocity of the atom cloud are then smaller than the requirement if this average velocity is controlled to better than 1 $\frac{\text{cm/s}}{\rthz}$. Since this configuration does not cancel constant accelerations, the average position of the atom clouds must still be controlled to 10 nm in order to achieve target sensitivity. The gravitational  wave signal in this configuration is $\sim \keff h L (\omega T)^2$ just like the Mach Zender interferometer. However, this interferometer has to run for a time $(1 + \sqrt{5}) T$ instead of $2 T$ in order to resonantly couple to gravitational  waves of frequency $\omega \sim \frac{1}{T}$. With fixed total interferometer time, the Mach Zender configuration can probe lower frequencies than the double loop. It is therefore preferrable to run the interferometer in the Mach-Zender configuration. The double loop configuration can however be used if the stringent control over the average launch velocity of the atom cloud proves to be technically challenging. The double loop configuration can also be run with the atom spending equal times in both loops (Figure \ref{Fig:double-diamond} with $\alpha = 1$) . Constant accelerations do not contribute to the phase shift in this configuration \cite{McGuirk, BorisDiamonds}.  This sequence relaxes the control required over the average launch position of the cloud but does not alleviate the control required over the average launch velocity of the cloud. 

\begin{figure}
\begin{center}
\includegraphics[width=4.5 in]{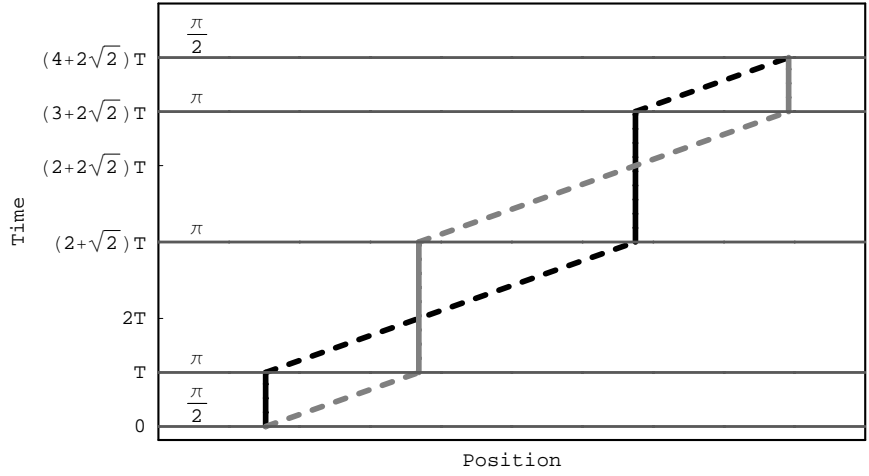}
\caption{ \label{Fig:triple-diamond} A space-time diagram of the triple loop interferometer. The black and gray lines indicate the two halves of the wave function after the initial beamsplitter. The dashed and solid lines represent the two internal states of the atom. The laser light used to manipulate the atom is shown as horizontal dark gray lines. The speed of light has been exaggerated. 
}
\end{center}
\end{figure}

In addition to the double loop configuration, the interferometer can also be operated with the pulse sequence  $\frac{\pi}{2} - \pi  - \pi - \pi - \frac{\pi}{2}$ with the time between the $\frac{\pi}{2} - \pi$ and $\pi - \pi$ pulses in the ratio  $\frac{1}{1 + \sqrt{2}}$ (Figure \ref{Fig:triple-diamond}). In this configuration, constant accelerations and time independent linear acceleration gradients do not contribute to the phase shift \cite{McGuirk, BorisDiamonds}. The first non-zero phase shift in such an interferometer comes from the quadratic gradient  $\nabla(\nabla g)$ which produces an acceleration $\sim \frac{G M_E}{R_E^2} (\frac{\vlavg T}{R_E})^2$ in a vertical terrestrial interferometer. This acceleration is orders of magnitude smaller than g and its linear gradient $\nabla g$. Fluctuations of this acceleration due to variations in the launch position and velocity of the atom clouds can be made smaller than $\EarthShotNoise$ with minimal control over these parameters. For instance, the contribution from shot to shot variations in the average velocity of the atom cloud are smaller than $\EarthShotNoise$ if these variations are smaller than $1 \frac{\text{cm/s}}{\sqrt{\text{Hz}}}$. The gravitational  wave signal in this multiloop configuration is $\sim \keff h L (\omega T)^4$. This interferometer is equally sensitive to gravitational  waves at the interferometer's resonant frequency ($T \sim \frac{1}{\omega}$) as the double loop configuration considered earlier but its bandwidth is suppressed by $\sim (\omega T)^2$ relative to the double loop interferometer. Furthermore, this interferometer needs to run for a time  $ (4 + 2 \sqrt{2}) T$ in order to resonantly couple to a gravitational  wave of frequency $\omega \sim \frac{1}{T}$.  The triple loop can  be used if control over both the average launch position and velocity of the atom clouds becomes difficult. 

The effects of position and velocity noise may be amplified due to the presence of local mass anomalies near the interferometer. A local anomaly is a mass distribution  near the interferometer whose field changes by $\mathcal{O}(1)$ over the length of the interferometer. The phase shift from such an anomaly of mass $M$ at a distance  $R \lessapprox v_R T$ from the interferometer can be calculated using the methods of \cite{GR Atom} and was found to be
\begin{equation}
\Delta \phi \sim  \keff \left( \frac{G M}{ (R \;  v_R T)} \right) \left(1   -    \left(\frac{v_L T}{R} \right) \right)T^2 + \dots
\end{equation}
where $v_L$ and  $v_R = \frac{\keff}{m_\text{atom}}$ are the launch and recoil velocity of the atoms with $v_L T \ll R$. Time varying accelerations from shot to shot variations in the average position or velocity of the atom cloud with respect to this anomaly are smaller than $\EarthShotNoise$  if $\delta \rlavg \lessapprox 1 \frac{\mu m}{\sqrt{\text{Hz}}} \left(\frac{\text{1000 kg}}{M}\right) \left(\frac{R}{\text{1 m}}\right)^2 $ and $\delta \vlavg \lessapprox 1 \frac{\mu \text{m/s}}{\sqrt{\text{Hz}}}  \left(\frac{\text{1000 kg}}{M}\right) \left(\frac{R}{\text{1 m}}\right)^2$. The constraints on launch position and velocity demanded by local mass anomalies are less stringent than the demands imposed by the Earth's gravity gradient. 

The time varying signal caused by a local mass anomaly is due to fluctuations in the relative position and velocity of the atom cloud with respect to the anomaly. Since the fractional fluctuations in these quantities can be controlled to $\sim \frac{10^{-6}}{\sqrt{\text{Hz}}}$ relatively easily, the anomaly must produce a relatively large gravitational field ($\sim 10^{-9} g$) inside the interferometer in order for these position and velocity fluctuations to cause accelerations $\sim \EarthShotNoise$. The gravitational field in the interferometer can be measured to $10^{-9}g$ using conventional gravimeters enabling the detection of mass anomalies of interest. The effects of these anomalies can then be minimized by strategically positioning mass sources that shim the gravitational field in the detector. 

The measurement of $\glavg$ can also fluctuate from shot to shot due to fluctuations $\delta \keff$ in the frequency of the lasers over the time scale of a second. The differential phase shift caused by this effect is $\delta \keff \nabla g L T^2$.   Fractional stability in the laser frequency $\sim 10^{-11} \left(\frac{\text{1 km}}{L}\right)$ in the 1 Hz band is required to push this background below shot noise.  The experiment will employ lasers with fractional stability $\sim 10^{-15}$  to tackle laser phase noise. Hence this background will be smaller than shot noise.

\subsubsection{Timing Errors}

The interferometer is initiated by the first $\frac{\pi}{2}$ pulse  which hits the atom causing it to split into two arms, one moving with the original launch velocity $v_L$ and the other with  velocity $v_L + v_R$. The $\pi$ pulse switches the velocities of the arms after which the final $\frac{\pi}{2}$ pulse interferes the arms. An asymmetry $\delta T$ in the time between the $\frac{\pi}{2} - \pi$ and $\pi - \frac{\pi}{2}$ stages of the interferometer results in the arms spending unequal times moving with velocity $v_L + v_R$,  causing a phase difference $\sim M v_R v_L \delta T = \keff v_L \delta T$. In addition to the phase accrued by the atom as a result of time evolution, the atom also picks up the average phase of the laser during the atom-laser interaction. The $\frac{\pi}{2}$ and $\pi$ pulses consist of $N \sim 1000$ LMT pulses each of frequency $k$ ($\keff = 2 N k $). A timing error $\delta T$ changes the average laser phase of each LMT pulse by $k \delta T$ resulting in a total phase shift $\sim N k \delta T = \keff \delta T$. The net phase shift contributed by timing errors is then $\sim \keff \delta T + \keff v_L \delta T$. Differential measurement cancels the $\keff \delta T$ term and yields a phase shift $\keff \delta v_{L} \delta T$ where $\delta v_{L}$ is the difference between the launch velocities of the atom clouds. We assume the shutters that control the time between the pulses can be operated with picosecond precision. With $\delta T \sim 10^{-12} \text{~s}$, this background can be made smaller than shot noise by launching the atoms such that  $\delta v_{L} < 1 \text{ cm/s}$. 

Finally, variations in the overall interrogation time of the experiment cause a time varying phase shift $\sim \keff g T \delta T$ in each interferometer  resulting in a differential phase shift  $\keff \nabla g L T \delta T$ . With picosecond control over the shutters and the interrogation time of the experiment, this effect is $\sim 10^{-16} \frac{g}{\sqrt{\text{Hz}}} (\frac{L}{\text{1 km}})$.

\subsubsection{Effects of Rotation} 

For a laser fixed to the earth's surface, there is a differential Coriolis acceleration $\sim \omega_E \delta v$ between the two atom clouds where $\omega_E$ is the angular velocity of the earth and $\delta v$ is the difference between the transverse velocities of the clouds. $\delta v$ is caused by  thermal effects and transverse vibrations of the trap used to prepare the atom clouds. The statistical variation in the average thermal velocities of two atom clouds with $\sim 10^8$ atoms at $\sim 100$ picokelvin temperatures is  $\lessapprox 10^{-8} \text{ m/s}$.  Thermal effects will cause $\delta v$ to vary from shot to shot by $10^{-8}$ m/s. With $\omega_E \sim 10^{-4}$ rad/s, these thermal variations cause accelerations $\sim 10^{-13} g$ which is larger than shot noise. 

One way to control this problem is to servo the laser's axis so that the axis remains non-rotating in an inertial frame. In this case, the only residual backgrounds arise from errors $\delta \omega$ in the servoing mechanism. The  servoing apparatus can  operate with nanoradian precision. With $\delta\omega \sim 10^{-9} \frac{\text{rad/s}}{\sqrt{\text{Hz}}}$,  the variations in the  Coriolis acceleration due to thermal effects are  smaller than $\EarthShotNoise$ if $\delta v \lessapprox 10^{-5} \text{~m/s}$.   The wobble $\delta\omega$ of the laser's axis also causes a differential centrifugal acceleration  $L (\delta\omega)^2$ between the atom clouds. This acceleration is below  $\EarthShotNoise$ if  $\delta\omega < 10^{-9} \frac{\text{rad/s}}{\sqrt{\text{Hz}}} (\frac{\text{1 km}}{L})^{\frac{1}{2}}$.

The proposed experimental setup involves two atom interferometers vertically separated by a length  $L \sim 1$ km and run by a common laser. One interferometer will then be at a distance $L$ from the laser. The atoms in this interferometer will have a velocity $\sim L \omega \sim 10 \text{~cm/s} (\frac{L}{\text{1 km}})$  perpendicular  to the axis of the servoed laser due to the rotation of the Earth. Since these velocities have to be smaller than $10^{-5}$ m/s to suppress Coriolis accelerations from the jitter of the servoing apparatus, these atoms must be launched with a transverse kick that cancels the relative velocity between the laser's axis and the atom cloud to $10^{-5}$ m/s. These kicks could potentially be delivered using an appropriately positioned laser. The vertical vibrations of this laser will cause fluctuations in the launch velocity of the atom cloud leading to time varying accelerations as discussed in sub section  \ref{SubSec: EarthNewtonianGravity} and these vibrations must be appropriately damped. This transverse velocity could also be cancelled by locking both atom clouds in an optical lattice and rotating the lattice itself to counter the rotation of the Earth. 

The interferometer measures the component of $\vec{g}$ along the laser's axis. Jitters $\delta \omega$ in the laser's axis will cause differential accelerations $\sim \nabla g L (\omega_E  T)(\delta\omega T)$ and $\sim \nabla g L (\omega_E T) (\delta\omega_E T)$.  With nanoradian stability in $\delta\omega$, $L \sim 1\text{~km}$ and $T\sim 1\text{~s}$, the first term is smaller than $\EarthShotNoise$. The second term is also smaller than $\EarthShotNoise$ since at 1 Hz, $\delta\omega_E \ll 10^{-7}\frac{\text{rad/s}}{\sqrt{\text{Hz}}}$. 

The need to servo the lasers emerged from the demand to suppress Coriolis accelerations due to thermal fluctuations in the atom cloud. The experiment can be performed without servoing the lasers if the  interferometer is operated in the multiloop configurations described in sub section \ref{SubSec: EarthNewtonianGravity}. The Coriolis acceleration caused by a laser rotating with a constant angular velocity and an atom cloud moving with a constant transverse velocity is constant. If the interferometer is run in the multiloop configurations, the phase shift due to this acceleration can be completely cancelled  eliminating the need to servo the laser. In the multiloop configuration, the interferometer has a smaller bandwidth but has the same sensitivity to gravitational waves at its resonant frequency as the Mach Zender configuration. In these multiloop configurations, rotational backgrounds arise due to instabilities $\delta \omega$ in the rotation of the laser's axis leading to Coriolis and centrifugal accelerations $\sim \delta\omega \delta v + L (\delta\omega)^2$. These accelerations are smaller than $\EarthShotNoise$ if $\delta\omega \lessapprox 10^{-9} \frac{\text{rad/s}}{\sqrt{\text{Hz}}}$.


Due to unavoidable misalignments, the earth's gravitational field will have a component along the direction transverse to the laser's axis. This component will cause a differential velocity $\sim   \nabla g L \sin (\theta) T$ between the atom clouds where $\theta$ is the angle between the local gravitational field and the laser's axis. Jitter in the lasers' axis causes an acceleration $\sim \delta \omega  \nabla g L  \sin (\theta) T$. The interferometer needs to be operated with $\theta \sim 0.01 (\frac{\text{1 km}}{L})$ for this acceleration to be smaller than $\EarthShotNoise$.




\subsubsection{Effects of Magnetic Fields}

\label{SubSec: EarthMagneticFields}

A magnetic field $B$ changes the energy difference between the hyperfine ground states in the $m=0$ sublevel of the atom by an amount $\alpha_{\text{ZC}} B^{2}$ where  $\alpha_{\text{ZC}}$ is the Zeeman Clock shift of the atom. If the magnetic field varies by $\delta B$ during the course of the experiment, the energy difference between the atom states during the $\frac{\pi}{2} - \pi$ stage will be different from the energy difference during the $\pi - \frac{\pi}{2}$ stage of the experiment. This produces a phase shift $\sim \alpha_{\text{ZC}}\; B_0 \; \delta B \;T$ which must be smaller than the per shot phase sensitivity of the interferometer $\sim 10^{-5}$.  With a bias field $B_0 \sim 100$ nT and $\alpha_{\text{ZC}} \sim  1 \frac{\text{kHz}}{\text{G}^2}$ (for Rubidium), this phase shift can be made smaller than $10^{-5}$ rad if $\delta B$ is smaller than 1 $\frac{\text{nT}}{\rthz}$. Time varying magnetic fields in the interferometer are caused by time variations in the applied bias field and the Earth's magnetic field. The current source used to create the bias field may be made stable to 6 digits in the 1 Hz band, resulting in 1 Hz variations due to the 100 nT bias field smaller than 1 $\frac{\text{nT}}{\rthz}$.  Magnetic fields from external sources like the Earth can be shielded to the required 1 $\frac{\text{nT}}{\sqrt{\text{Hz}}}$ level by following the techniques of \cite{MagneticShielding}.

\section{Satellite Based Experiment}
\label{Sec: Space}

The search for gravitational waves in the sub-Hertz band on the Earth is impeded by time varying local gravitational fields due to seismic and atmospheric activity.  Additionally, the atom interferometer is maximally sensitive to a gravitational wave of frequency $\omega$ when the interrogation time $T$ of the experiment is such that $T \sim \frac{1}{\omega}$. Interrogation times larger than 10 s are difficult to achieve in a terrestrial interferometer, since the atoms are in free fall.  We are thus lead to consider satellite-based interferometer configurations to search for gravitational waves in the sub-Hertz band. 

\subsection{Experimental Setup}
\label{Sec: Space Setup}

\begin{figure}
\begin{center}
\includegraphics[width=\columnwidth]{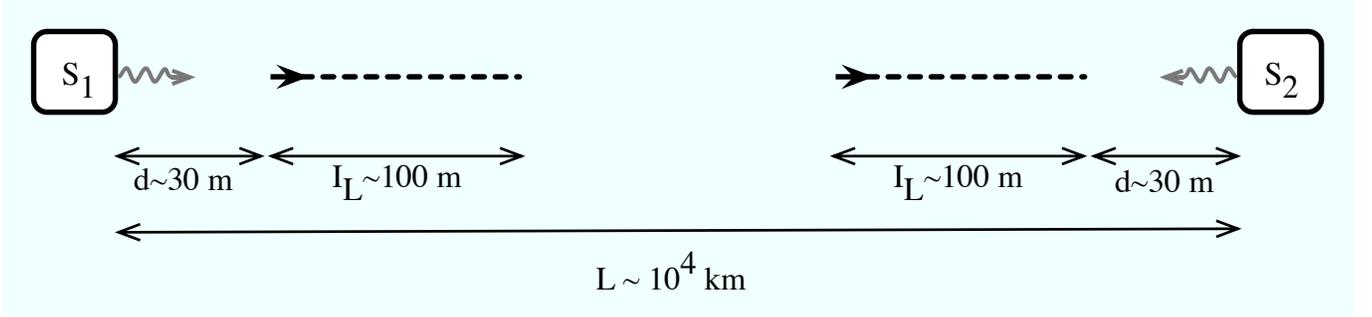}
\caption{ \label{Fig:space setup} The proposed setup for a satellite experiment. Two satellites $S_1$ and $S_2$ house the lasers and atom sources. The atoms are brought a distance $d \sim 30 \text{ m}$ from the satellites at the start of the interferometer sequence. The dashed lines represent the $I_L \sim 100 \text{ m}$ path travelled by the atoms during the interferometer sequence.  The gray lines represent the paths of the lasers along the axis between the satellites. In practice it is desirable to have a third satellite in a LISA-like constellation with such a pair of interferometers operated along each arm.}
\end{center}
\end{figure}

The atom interferometer configurations discussed in section \ref{Sec: GW Signal} can be realized in a satellite experiment by using two satellites ($S_1$ and $S_2$ in Figure \ref{Fig:space setup}) separated by a long baseline $L$. As we will show, the satellites can act as  base stations (housing lasers and atom sources) to operate  $I_L \sim 100 \text{ m}$ interferometers along the axis between the satellites. The atoms are brought a distance $d \sim 30 \text{ m}$ from the satellites at the start of the interferometer sequence. The atoms travel a distance $I_L \sim 100 \text{ m}$ during the course of the interferometry executed by the lasers in the satellites. As in the terrestrial interferometer, the experiment measures the differential phase shift $\sim \keff h L$ between two interferometers separated by $L$ and operated by common lasers in order to suppress backgrounds from low frequency vibrations and phase noise of the lasers. The satellites could be placed in heliocentric or geocentric orbits.

It is desirable to operate the atom interferometers outside the satellites.  As we will show, this suppresses backgrounds, resulting in significantly reduced satellite requirements.  Importantly, it ameliorates the control required over the position of the spacecraft. The gravitational force on the atom from the spacecraft will vary in time due to uncontrolled motion of the spacecraft and will mimic a gravitational wave signal. If the atoms are far from the spacecraft, the magnitude of this acceleration is reduced and makes the interferometer less sensitive to fluctuations in the position of the spacecraft.  Additionally this increases the available interferometer region, improving sensitivity by allowing longer interrogation times and larger recoil velocities.

The two atom interferometers can be constructed by initially placing atoms a distance $I_L + d$ from $S_2$ towards $S_1$ (Figure \ref{Fig:space setup}) using laser manipulations.  Similarly, a cloud from $S_1$ can be brought a distance $d$ towards $S_2$.  After the clouds are appropriately positioned, the same laser pulses can be used to operate both interferometers.  We will argue (see Section \ref{Sec:space LMT limit}) that it should be possible to run the interferometer outside the spacecraft to distances $\sim 100$ m.  With interrogation times $T \sim 100 \, \text{s}$, the length of the interferometer region limits $\keff \sim 10^9 \, \text{m}^{-1}$. 

In order to detect a gravitational wave it is only necessary to have one such baseline containing a pair of atom interferometers.  However, it may be desirable to have a third satellite in a LISA-like constellation with a pair of atom interferometers operated along each baseline. The addition of the third satellite provides another gravitational wave channel. As discussed in sub section \ref{Sec: Earth Backgrounds}, since the gravitational wave signal in each detector depends upon the orientation of the detector relative to the incident direction and the polarization of the gravitational wave, the cross-correlated sensitivity of the constellation will be set by the stochastic noise floor.  The addition of the third satellite can be useful in further suppressing laser phase noise (see Section \ref{Subsec:SpaceVibrations}).  Additionally, independent correlated measurements would increase the confidence of detection. Detecting a stochastic background of gravitational waves requires cross-correlating the output from two independent gravitational wave detectors. Furthemore, with three such single-axis gravitational wave detectors whose axes point in different directions, it is possible to determine information on the direction of the gravitational wave source.  



The signal in the interferometer is directly proportional to the size of the baseline $L$ and the effective momentum $\keff$ transferred by the atom optics. The transfer of a large momentum will impart a large recoil velocity to the atom. The operation of the interferometer with a large recoil velocity requires the interferometer region $I_L$ to be long and hence the limit on $I_L$ imposed by the satellite will limit $\keff$.   The detection of a time varying signal from a gravitational wave of frequency $f$ requires a data taking rate $f_d \gtrsim 2 f$. As discussed in sub section \ref{Sec: Earth Setup}, this requires the operation of concurrent atom interferometers along the common satellite axis. In the following, we examine the limits imposed on these quantities in a satellite experiment. 

\subsubsection{Baseline Limit from Atom Optics}
\label{Sec:space baseline limit}


Gravitational wave experiments benefit from long baselines since the signal increases linearly with the baseline. The most stringent limit on the baseline is imposed by the need to drive the atomic transitions that create the beamsplitter and mirror pulses using a laser that is a large distance away on the far satellite.  The laser field from the nearby satellite can be intense, but at large distances the laser field from the far satellite will necessarily spread and lose intensity.

To find this limit, consider an atom in the presence of the laser field from the nearby satellite with intensity $I_n$ and from the far satellite with intensity $I_f$.  The Rabi frequency $\Omega_\text{st}$ of stimulated 2-photon transitions in Fig. \ref{Fig:Raman} is \cite{Berman}
\begin{equation}
\label{Eqn: stimulated rabi}
\Omega_\text{st} = \left| \frac{ \langle e | {\bf d \cdot E_n} | 1 \rangle \langle e | {\bf d \cdot E_f} | 2 \rangle}{2 \Delta} \right| \approx \frac{\Gamma^2}{4 \Delta} \sqrt{\frac{I_n I_f}{\Isat^{(e1)} \Isat^{(e2)} }}
\end{equation}
where $\Delta$ is the detuning from the intermediate state $| e \rangle$ and $\Gamma$ is the decay rate of the excited state.  The saturation intensity, $\Isat$ is defined by $\frac{I}{\Isat^{(e1)}} = 2 \left( \frac{\Omega_{e \to 1}}{\Gamma} \right)^2$ where $\Omega_{e \to 1}$ is the resonant two-level Rabi frequency between the excited state and state $|1\rangle$ \cite{Berman}.  Note that $\Isat$ is an atomic property independent of laser intensity.  For simplicity we assume that the decay rate and dipole matrix element are independent of the choice of state $|1\rangle$ or $|2\rangle$, i.e. $\Isat^{(e1)} \approx \Isat^{(e2)}$. The intensity $I_f$ of the far satellite decreases as the baseline is increased, decreasing $\Omega_\text{st}$. The width of the 2-photon atomic transition is set by $\Omega_\text{st}$ and if $\Omega_\text{st}$ becomes too small, the transitions can become very velocity selective due to Doppler detuning. To avoid loss of SNR, the initial thermal velocity spread of the atoms must be smaller than the velocity selection imposed by $\Omega_\text{st}$.  Thus the lowest attainable temperature sets a lower limit on $\Omega_\text{st}$.  With cloud temperatures $\sim 0.1 \, \text{nK}$, the Rabi frequency  $\Omega_\text{st}$ must be $\gtrsim 2 \pi \left( 10^{2} \, \text{Hz} \right)$.  

To maximize sensitivity, a significant fraction of the atoms should not undergo spontaneous 2-photon transitions during the time the atom is in the presence of the light. A spontaneous 2-photon transition can occur when one laser field (in practice the more intense one) drives the atom up to the intermediate state and this state then decays due to spontaneous emission.  In this case the atom gets a momentum kick in an arbitrary direction and will be lost from the interferometer. The spontaneous transition rate due to the near laser is \cite{Metcalf, Steck}
\begin{equation}
R = \frac{ \frac{\Gamma}{2} \frac{I_n}{\Isat } }{1+ 4 \left( \frac{\Delta}{\Gamma} \right)^2 + \frac{I_n}{\Isat} }. 
\end{equation}
The detuning can be eliminated using Eqn. \eqref{Eqn: stimulated rabi}. The total time for which the the atom is in the presence of the light must be smaller than $\frac{1}{R}$. Since we wish to use LMT beamsplitters that deliver $N$ photon kicks to the atom, the atom will be in the presence of the transition light for time $\sim \frac{N}{\Omega_\text{st}}$. The need to suppress spontaneous emission yields $R \lesssim \frac{\Omega_\text{st}}{N}$. 

With stimulated Rabi frequency $\Omega_\text{st} \sim 2 \pi \left( 10^2 \, \text{Hz} \right)$ and spontaneous emission rate  $R \sim \left(10 \, \text{s}\right)^{-1}$, we find that baselines $L \sim 10^{3} \, \text{km}$ can be achieved with lasers of waist  $\sim 0.5 \, \text{m}$ and  power $\sim 1 \, \text{W}$. We have assumed atomic parameters (e.g. for Rb or Cs) $\lambda \approx 1 \, \mu\text{m}$, $\Isat \approx 1 \, \frac{\text{mW}}{\text{cm}^2}$, and $\Gamma  \approx 3 \times 10^7 \, \frac{\text{rad}}{\text{s}}$ \cite{Metcalf, Steck}, thus requiring a detuning $\Delta \approx 20 \text{~GHz}$ to reach this limit. This configuration would allow the interferometer to use $N\sim 100$ LMT beamsplitters. This is the main limitation on the distance between the satellites.  Improvements on this limit, either through higher laser power, an optimized choice of atom transitions or improved cooling techniques that can allow the transitions to proceed at smaller Rabi frequencies  should allow direct enhancements in the final sensitivity.



\subsubsection{Environmental Constraints on the Interferometer Region}
\label{Sec:space LMT limit}

The length of each interferometer must be at least $v_r T$ (where $v_r$ is the recoil velocity of the atom) since the two  arms of the interferometer will separate by $v_r T$ during the course of the experiment. The recoil velocity $v_r$ is  equal to $\left(\frac{\keff}{M_\text{a}}\right)$, where $\keff$ is the momentum transfered to the atom and $M_\text{a}$ is the mass of the atom.  For a fixed length $I_L$, an interrogation time $T$ requires $v_r < \frac{I_L}{T}$. Since the signal scales linearly with $\keff$, we would like to make $I_L$ as large as possible. If the atom trajectories are restricted to lie within the spacecraft, then $I_L$ has to be smaller than the typical dimensions of  spacecraft $\sim$ 1 m.   The atom interferometer requires the laser and the atom source be placed inside the satellite, near their power sources. However, the diffuse atom cloud trajectories that form the arms of the interferometer need not be inside the satellite. 


Collisions of the atoms with background gas ({\it e.g.} solar wind) and photons are the major problem posed by the environment to the atom interferometer. These collisions cause decoherence and atoms that undergo such interactions cannot engage in quantum interference. Collisions are therefore not a source of noise since they do not cause phase shifts. However, by knocking atoms away from the atom cloud, collisions reduce the number of atoms available to do the experiment, thereby decreasing the sensitivity of the instrument.

Solar photons are a source of decoherence.  The interaction cross section of the photon with the atom is appreciable only if the frequency of the photon is within the width of an atomic transition.  For example, for Rubidium the most important transition in this band is the 780 nm line with a width of a few MHz.  At a distance of 1 AU, the solar spectral intensity around 780 nm is $\sim 1 \, \frac{\text{W}}{\text{m}^2 \text{ nm}}$.  Thus the intensity within the atomic linewidth is $\sim 10^{-5} \, \frac{\text{W}}{\text{m}^2}$.  The number density of photons within this line is then $n \sim 10^5 \text{ m}^{-3}$.  The resonant photon-atom scattering cross section is $\sigma \sim \lambda^2 = \left( 780 \text{ nm} \right)^2$.  The mean photon scattering rate for an atom in the sun's light is then $n \sigma c \sim 10 \text{ s}^{-1}$.  This is a severe limit on the possible interrogation time for the atom interferometer and so must be avoided.  There are several possible solutions. 

Satellite experiments like the James Webb Space Telescope (JWST) mission rely on the use of large $\sim 200 \text{ m}^2$ ultra-light sunshields ($\sim 2.5 \text{ kg}$) to protect the satellite from solar radiation. These shields can reduce the solar intensity from $\sim  1000 \, \frac{\text{W}}{\text{m}^2}$ to $\sim  10 \, \frac{\text{mW}}{\text{m}^2}$ \cite{JWST1, JWST2}. The decoherence time scale due to the residual solar flux is $\sim 10^{-4} \text{ s}^{-1}$ which is significantly longer than the time scales of interest to this experiment.  In this case the size of the shield would be one limit on the size of the atom path. Another possibility is to place the satellites in a lunar or geocentric orbit. If the experiment is done in such an orbit, then the satellites will spend an order one fraction of the time on the dark side of the moon/earth where there is no problem (a similar estimate for a 300 K blackbody spectrum gives a very low scattering rate).  This leads to an order one loss of duty cycle and a small loss of statistical sensitivity.    It might also be possible to perform the experiment farther from the sun to reduce the solar intensity, though this would presumably increase the difficulty and expense while still limiting interrogation times.

The environment outside the satellite is dominated by the solar wind composed of protons and electrons with a number density $\sim 10^7 \, \frac{\text{particles}}{\text{m}^3}$ moving at velocities $\sim 500 \, \text{km/s}$. The typical interaction cross-section of these particles with the atom is $\sim 10^{-18} \, \text{m}^2$ leading to mean collision times $\gg 10^3 \, \text{s}$.  The local environment near the spacecraft is less pristine than the space vacuum due to emissions from the spacecraft. In order to make use of the local space environment to run the interferometer, the craft must be designed so that such emissions are not in the direction of the atom trajectories.


The atom will be subjected to the interplanetary magnetic field if the experiment is done outside the spacecraft. In order to suppress phase shifts from magnetic fields, the atom must be placed in a magnetically insensitive $m = 0$ state in the direction set by the field.  The Rabi frequency of the atomic transition is set by the internal state of the atom and the laser must be tuned to match this frequency. Since, in the random interplanetary magnetic field, the atom's spin precesses rapidly, the spin of the atom will change as the direction of the magnetic field changes.  If the direction of the magnetic field changes over the length of the interferometer, the atom will evolve away from the original $m = 0$ state. The laser's frequency is however tuned to the original $m = 0$ state.  The interaction between the laser and the atom will excite, for example, $m = \pm 1$ states along the new axis of quantization. Pollution into these states will cause phase shifts that are a background to the experiment. 

The interplanetary planetary magnetic field at 1 AU is $\sim$ 5 nT and has a correlation length $\sim$ 0.01 AU. The drift in the direction of the field is smaller than  $5 \degree$ over 10 minutes during an average time interval but can be as large as $10 \degree$ over 10 minutes during noisier times \cite{Burlaga}. The phase shift due to these direction changes in  the magnetic field are smaller than the shot noise requirements of this experiment (see sub section  \ref{SubSec: SpaceMagneticFields}).  In addition to this slow drift, the magnetic field also exhibits sharp discontinuities in its direction. These sharp directional discontinuities are separated by periods of an hour  \cite{Burlaga} and  are not a problem to an interferometer with interrogation time smaller than 100 s. The direction of the magnetic field in the interferometer region can be further stabilized by attaching a permanent magnet to the spacecraft, coaxial to the atomic trajectory. A bar magnet of size $ 1 \text { m } \times 10 \text{ cm } \times 10 \text{ cm }$ with magnetization $\sim 10^7$ A/m can provide magnetic fields $\sim$ 20 nT out to distances $\sim$ 100 m. This field is larger than the interplanetary magnetic field $\sim 5$ nT and can enhance the stability of the direction of the magnetic field in the interferometer. The experiment relies on differential measurement strategies which requires both interferometers to be operated by the same set of lasers.  The correlation length of the interplanetary magnetic field is significantly larger than the baseline $L \sim 10^3$ km of this experiment. With the addition of the $\sim 20$ nT bias field, the magnetic field direction in the two interferometers can be sufficiently aligned to enable the same lasers to operate both interferometers. 

The torque on the spacecraft from the action of the interplanetary magnetic field on the external magnet is $\sim 10^{-4}$ Nm, which is of the same order of magnitude as the torque produced by solar pressure on the satellite. Since the forces on the spacecraft due to the external magnet are comparable to the force from solar pressure,  the addition of the magnet will not significantly alter the dynamics of the spacecraft control system. With the addition of the bias field from the permanent magnet, the interferometer can be run over at least  $I_L \sim $ 100 m. Interferometer lengths longer than 100 m may be achievable when the interplanetary magnetic field is  quiet.

The hardware required to measure the phase shift in the  interferometer has to be housed in the spacecraft. A normalized measurement of the phase shift is done by counting the number of atoms in each final state of the atom after the interferometer pulse sequence. If the experiment is  performed outside the spacecraft, the counting must be performed with detectors located on the spacecraft. Remote detection of an atom in a given internal state can be done using absorption imaging wherein a light beam, whose frequency is tuned to an atomic resonance accessible from the internal state of interest, is pulsed from one spacecraft to the other. The atoms that are in the internal state of interest will absorb these photons. A photodetector on the other spacecraft will measure the change in the intensity of the initial beam measuring the number of absorbed photons and hence the number of atoms in the internal state of interest. 

The absorption detection technique must be sufficiently sensitive to detect the  required phase sensitivity $\sim 10^{-4}$ (see section \ref{Sec: Sensitivities}) of this experiment. With $N_a \sim 10^8$ atoms in the cloud, a phase sensitivity $\sim 10^{-4}$ requires the detection scheme to measure changes in the number as small as $\sqrt{N_a} \sim 10^4$ atoms. The absorption cross-section of the atom with the resonant laser light of wavelength $\lambda$ is $\sigma_\text{abs} \sim \lambda^2 \sim \left(1 \mu \text{m}\right)^{2}$.
If the detection is done over a period $\delta\tau$, then the total number of photons scattered by $\sqrt{N_a} \sim 10^4$ atoms  is $N_s \sim \sqrt{N_a} \left(\frac{I_f}{k}\right) \sigma_\text{abs} \delta\tau$ where $k \sim 10^{-19} \, \text{J}$ is the energy of the detection photon. This number must be larger than the photoelectron shot noise $\sim \sqrt{ \left(\frac{I_f}{k}\right) A \delta \tau}$ over the detection area $A$. The size  of the detection area (e.g. the size of a lens) must be as big as the typical size of the atom clouds used in this experiment $A \sim \left(10 \, \text{cm}\right)^2$. With these parameters, a satellite experiment with a baseline $L\sim 10^{3} \, \text{km}$ and  a detection laser with intensity $\sim 10^{-8} \, \frac{\text{W}}{\text{cm}^2}$ laser and a  $\sim 1 \text{ m}$ waist housed on the distant satellite can image the atom cloud with the necessary precision in a detection time $\delta \tau \sim 0.1 \, \text{s}$. Each atom undergoes $\left(\frac{I_f}{k}\right) \sigma_\text{abs} \delta \tau \sim 10^2$ absorptions during this imaging time. Since the atom undergoes rapid spontaneous emission upon excitation, absorption imaging must be performed between atomic states that have $\sim 100$ cycling transitions to prevent loss of atoms through spontaneous emission into other atomic states.

These arguments suggest that it should be possible to run the interferometer outside the spacecraft to distances $\sim 100$ m.

\subsubsection{Limit on Data Taking Rate}
\label{Sec:space fd limit}

The detection of a time varying signal from a gravitational wave of frequency $f$ requires a data taking rate $f_d \gtrsim 2 f$. As discussed in sub section \ref{Sec: Earth Setup}, this requires the operation of concurrent atom interferometers along the common satellite axis. Concurrent operation of atom interferometers requires that the laser fields that trigger interferometry in one interferometer not cause transitions in the other interferometers in the common beam axis. This can be achieved by launching the interferometers with different launch velocities so that the interferometers are all doppler detuned from one another. The width of a 2-photon transition is equal to the Rabi frequency $\Omega_\text{st}$ of the transition. Two interferometers are doppler detuned if the relative velocity between them is $\sim 10^{-4} \, \text{m/s} \left(\frac{\Omega_\text{st}}{2 \pi \, \left( 10^2 \, \text{Hz} \right) }\right)$.  While doppler detuning prevents unwanted stimulated transitions, the laser field can drive spontaneous 2-photon transitions, as discussed in subsection \ref{Sec:space baseline limit}.   Following the discussion in subsection \ref{Sec:space baseline limit},   the spontaneous emission rate is $R \sim  \left(10 \, \text{s}\right)^{-1}$ in a configuration with baseline $L \sim 10^{3} \, \text{km}$, with $\sim 1 \, \text{W}$ lasers and stimulated Rabi frequency  $\Omega_\text{st} \sim 2 \pi \left( 10^2 \, \text{Hz} \right)$. The operation of $q$ concurrent atom interferometers with transition times $\sim \frac{N}{\Omega_\text{st}}$ requires   $q  \frac{N}{\Omega_\text{st}} \lessapprox \frac{1}{R}$. With the beam parameters described above, we can operate $q \sim 10$ concurrent interferometers.

The interferometers will be operated outside the satellites with the phase shift in each interferometer measured through absorption detection. The process of measuring the phase shift in one interferometer through this technique should not affect the other interferometers operating in the beam line. This can be achieved by initially performing a velocity selective stimulated Raman transition that takes the atom state at the end of the interferometer (the `interferometer state') into another long lived ground state of the system (the `detection state'), detuned from the original interferometer state.  The phase shift can be measured by imaging the detection state. For example, in Rubidium, the hyperfine interaction splits the ground state into states separated by $\sim 2 \pi \left( 6.8 \, \text{GHz} \right)$. One of these states could be used to run the interferometer (the interferometer state) and velocity selective stimulated Raman transitions can be used to populate the other state  (the detection state) prior to detection.  A Rabi frequency $\sim 2 \pi \left( 10^2 \, \text{Hz} \right)$ for this stimulated Raman process can be achieved through laser and beam features described in the preceding paragraph.  The spontaneous emission rate induced by this process is then  $\sim  \left(10 \, \text{s}\right)^{-1}$ which is not a problem for the operation of the interferometers since this light will be on for only $\sim 0.01 \text{ s}$ during detection. The spontaneous emission rate for the atoms in the interferometer state due to the $\sim 10^{-8} \, \frac{\text{W}}{\text{cm}^2}$ detection light tuned to the detection state is $\sim \left(10^{5} \, \text{s}\right)^{-1}$ and is also not a problem for the experiment. 

With this configuration, the data taking rate $f_d$ can be $\lesssim 1 \, \text{Hz} \left(\frac{10 \, \text{s}}{T}\right)$ where $T$ is the interrogation time of the experiment, limited by spontaneous emission caused by the laser light used to operate the interferometers. This is the main limitation on the data taking rate of the experiment.  Improvements on this limit, either through higher laser power, an optimized choice of atom transitions or improved cooling techniques that can allow the transitions to proceed at smaller Rabi frequencies  should allow direct enhancements to this rate.

\subsection{Backgrounds}
\label{Sec: space backgrounds}

Our configuration consists of two satellites in orbit separated by $L \sim 10^3 $ km. The satellites act as base stations and run the atom interferometers along their axis using common laser pulses. With a stabilizing magnetic field $\gtrsim 20$ nT provided by a permanent magnet housed in the spacecraft, the satellite environment permits the operation of the atom interferometer out to distances $I_L \sim 100 \,\text{m}$  from the satellite and for interrogation times $\sim 100 \, \text{s}$.  Prior to launch, the atoms are positioned at distances $d \sim 30 \, \text{m}$ and $d + I_L$ from their base stations $S_1$ and $S_2$ respectively using laser manipulations (Figure \ref{Fig:space setup}). The atoms are then launched with a common launch velocity and the interferometry is performed using common laser pulses. 

The differential  acceleration caused by a gravitational wave of amplitude $h$ and frequency $\omega$ is $h L \omega^2$ causing a phase shift $\keff h L \omega^2 T^2$. A  $L \sim 10^3$ km long baseline interferometer can detect gravitational waves of amplitude  $h \sim 10^{-23}$ with $\sim 10^6$ s of integration time if it is sensitive to accelerations $\sim \spshotnoise$. The strain sensitivity of such a configuration would be $\sim \frac{10^{-20}}{\rthz}$. The proposed experiment could reach target sensitivity using $200 k$ LMT beamsplitters and  atom statistics phase sensitivity $\sqna \sim 10^{-4}$ using ensembles of $N_a \sim 10^8$ atoms and interrogation times  $T \sim \frac{1}{\omega}$. Phase shifts from noise sources must be made smaller than $10^{-4}$. In particular, acceleration backgrounds should be less than $\sim \spshotnoise$.  We will assume that a LISA-like three satellite atom interferomter constellation is placed in orbit. As discussed in sub section \ref{Sec: Space}, the cross-correlated sensitivity of the gravitational wave channels thus produced is limited by the stochastic noise floor.  Thus we can assume that the noise in the entire set of detectors is stochastic, even if certain noise sources have long correlation times in any individual detector.  In the following, we discuss these stochastic noise sources and strategies to suppress them to the level required to detect gravitational waves with strain sensitivity  $ \sim \frac{10^{-20}}{\rthz}$ in the $10^{-2} \text{ Hz - } 1 \text{ Hz}$ band.

\subsubsection{Vibrations and Laser Phase Noise}
\label{Subsec:SpaceVibrations}

Vibration and laser phase noise issues were discussed in Section \ref{Sec: Earth Backgrounds}. The solutions proposed to address these issues in the terrestrial interferometer can also be used for the space based experiment. Following the analysis in that section,  contributions from the vibration of the lasers to the phase shift are smaller than shot noise if these vibrations are smaller than  $10^{-5} \frac{\text{ m}}{\sqrt{\text{Hz}}} \left(\frac{10^{-2} \text{ Hz}}{\nu}\right)^{\frac{3}{2}} \left(\frac{10^3 \text{ km}}{L}\right) $ at frequencies $300 \text{ Hz}  \left(\frac{10^3 \text{km}}{L}\right) \gtrsim \nu \gtrsim 10^{-2} \text{ Hz}$. 

Additionally, in space, an alternate strategy to handle laser phase noise is to use the same passive laser to run interferometers along two non parallel baselines in a LISA-like three satellite configuration.  Each baseline consists of two interferometers. The interferometers along each baseline are operated by a common control laser. The passive laser is placed at the intersection of the two baselines with appropriate optical beamsplitters so that the beam from the passive laser is shared by both baselines. The same pulses from the passive laser can trigger transitions along the interferometers in both baselines if the control lasers along the two baselines are simultaneously triggered. The laser phase noise in the difference of the differential phase shift along each baseline is greatly suppressed since phase noise from the control laser is common to the interferometers along each baseline and the phase noise from the passive laser is common to the baselines. The gravitational wave signal is retained in this measurement strategy since the gravitational wave will have different components along the two non parallel baselines. 

The contribution to the differential phase shift along each baseline due to a drift $\delta k$ in the frequency of the laser is suppressed by the arm length of the baseline (see Section \ref{Sec: Earth Backgrounds}). The residual contribution of this frequency drift to the difference of the differential phase shift along each baseline is $\delta k \delta L$ where $\delta L$ is the difference in the length of the two baselines. The effect of this contribution can be cancelled to the extent to which the arm length difference $\delta L$ is known. With $\sim 1$ m knowledge of the arm lengths, these contributions are smaller than shot noise if the frequency drift $\delta k$ of the laser is controlled to better than $\sim 10^{4} \frac{\text{Hz}}{\rthz}$ at frequencies $\omega \sim 10^{-2}$ Hz. In addition to this effect, differences $\delta T$ between the timing of the control lasers that operate the interferometer will also change the phase of the passive laser that is imprinted along the interferometers in the two baselines. The phase shift due to this effect is $\sim \delta k \omega \delta T L$. With $\delta k \lesssim 10^{4} \frac{\text{Hz}}{\rthz}$ at frequencies $\omega \sim 10^{-2}$ Hz, this phase shift is smaller than the per shot phase sensitivity  $10^{-4}$ of this experiment if the two control lasers are synchronized with $\delta T \lesssim 100 \mu \text{s}$. 

In addition to classical sources of phase noise discussed above, the quantum nature of the laser field will contribute to noise in the imprinted phase.  This quantum noise was computed in \cite{Yamamoto} and was found to be $\sim \frac{1}{\sqrt{\text{N}_{\gamma}}}$ where $\text{N}_{\gamma}$ is the total number of photons that form the coherent state of the laser field. The interferometers in this experiment are operated with $\sim 1$ Watt lasers with transition times $\sim 10^{-2}\text{ s}$  leading to $N_{\gamma} \sim 10^{17}$. Phase noise in the interferometer from the quantum nature of light is negligibly small.

\subsubsection{ Newtonian Gravity Backgrounds }
\label{Sec: Space Gravity Backgrounds}

The gravitational field of the satellite will cause a phase shift in the interferometer. Since the gravitational field of the spacecraft changes by $\mathcal{O}(1)$ over the length of the interferometer, the spacecraft is a local mass anomaly of mass $M$ at a distance $d_I \lessapprox v_R T$ from the interferometer (subsection \ref{SubSec: EarthNewtonianGravity}) . The phase shift in the interferometer due to the spacecraft is given by \cite{GR Atom}

\begin{equation}
\Delta \phi \sim  \keff \left( \frac{G M}{ (d_I \;  v_R T)} \right) \left(1   -    \left(\frac{v_L T}{d_I}\right)\right)T^2 + \dots
\end{equation}
when the launch velocity $v_L$ of the atom cloud satisfies $v_L T \ll d_I$ and the recoil velocity $v_R$ is such that $d_I \lessapprox v_R T$.

The relative distance between the spacecraft and the atom will change due to random motions of the spacecraft. Additionally, the average initial position of the atom clouds with respect to the spacecraft will also change from shot to shot due to thermal variations in the atom clouds and vibrations  of the trap. A variation $\delta R$ in this distance due to these effects will cause an  acceleration $\sim \frac{G M}{d_I \; v_R T} \frac{\delta R}{d_I}$. This acceleration is smaller than $\sim \spshotnoise$ if $\delta R (\omega) \lessapprox 10 \frac{\mu\text{m}}{\rthz} (\frac{\omega}{10^{-2}\text{ Hz}})^{\frac{3}{2}} (\frac{d_I}{\text{30 m}})^2 (\frac{\text{1000 kg}}{M})$.  With $N_a \sim 10^8$ atoms, shot to shot variations in the central position of the atom clouds due to thermal fluctuations can be made smaller than  $10 \frac{\mu \text{m}}{\rthz}$ by confining the atoms within traps of size  $\sim 1 \text{ cm} \sqnasubs$.  The atom trap and the spacecraft must be engineered so that their vibrations at frequency $\omega$ are smaller than  $10 \frac{\mu \text{m}}{\rthz} (\frac{\omega}{10^{-2}\text{ Hz}})^{\frac{3}{2}} (\frac{d_I}{\text{30 m}})^2 (\frac{\text{1000 \text{kg}}}{M})$. 

The average launch velocity $v_L$ of the atom cloud will change from shot to shot due to thermal variations in the atom clouds. These variations $\delta v_L$ will change the trajectory of the atoms in the gravitational field of the spacecraft. Due to the non-zero gradient of this field, these trajectories will experience different gravitational fields resulting in  time varying accelerations $\sim \frac{G M}{d_I \; v_R T} \frac{\delta v_L T}{d_I}$. This  acceleration is smaller than $\spshotnoise$ if  $\delta v_L (\omega) \lessapprox 100 \frac{\text{ nm/s}}{\rthz} (\frac{\omega}{10^{-2}\text{ Hz}})^{\frac{5}{2}} (\frac{d_I}{\text{30 m}})^2 (\frac{\text{1000 \text{kg}}}{M})$. The atom cloud used in this experiment will contain $N_a \sim 10^8$ atoms. Thermal fluctuations in the average velocity of this cloud are smaller than 100 $\frac{\text{ nm/s}}{\rthz}$ if the cloud is cooled to temperatures $\sim 100  \text{ pK} \sqnasubs$. Thus the thermal velocity of the atoms  do not limit the detection of gravitational waves in the frequency band and sensitivities of interest in this paper.


 We note that the control over the position and velocity of the spacecraft required by this experiment are weaker than the requirements of the LISA mission. LISA's inertial masses need to be placed inside the spacecraft since these masses must be shielded from the external environment. This increases the gravitational force of the spacecraft on the inertial masses making the inertial masses more sensitive to fluctuations in the position of the spacecraft. In the atom interferometer, the inertial atoms do not require the protection of the spacecraft and the experiment can be performed at distances $d_I \sim 30 \text{ m} $ from the spacecraft thereby decreasing the gravitational acceleration of the  atoms by a factor of $10^4$ relative to LISA \cite{LISAFTR}. The decreased gravitational acceleration makes the interferometer less sensitive to vibrations of the spacecraft. 
 
\subsubsection{Timing Errors}

The effect of asymmetries in the time between the $\frac{\pi}{2} - \pi$ and $\pi-\frac{\pi}{2}$ pulses were discussed earlier under backgrounds for the terrestrial interferometer. A timing error $\delta T$ causes a differential phase shift   $\sim \keff \Delta v_L \delta T$ where $\Delta v_L$ is the relative launch velocity between the atom clouds. This phase shift must be smaller than the per shot phase sensitivity of the instrument $\sim 10^{-4}$. With picosecond control over $\delta T$, this background is smaller than shot noise if the atoms are launched such that  $\Delta v_L < 10 \text{ cm/s}$. If the spacecrafts are in solar orbits separated by a distance  $L \sim 10^3$ km,  then the relative velocity between the spacecrafts is $\sim 10 \text{ cm/s} (\frac{L}{10^3 \text{ km}})$. But, this velocity is transverse to the interferometer baselines  and hence the atoms can be launched with relative velocities  smaller than 10 cm/s along the baseline.

\subsubsection{Effects of Rotation} 
\label{SubSec: SpaceRotations}

The  angular velocity of one spacecraft relative to the other is equal to its orbital angular velocity $\omega_S \sim 10^{-7}$ rads/s around the sun at $\sim 1$ AU.  The atom clouds are also in orbits around the sun and will therefore rotate around the passive laser housed in the spacecraft $S_2$ with the same angular velocity $\omega_S$.  The laser axis will always be kept along the line between the satellites.  If this axis rotates with angular velocity $\omega_S$, transverse velocities $v_T$ of the atom cloud  result in Coriolis accelerations $\sim \omega_S v_T$. In an atom cloud with $N_a \sim 10^8$ atoms cooled to $100 \text{ pK} \sqnasubs$ temperatures, the average transverse velocity of the clouds will change from shot to shot by $\sim 10^{-8} \text{ m/s}$. These thermal variations cause accelerations $\sim 10^{-16} g$ which are higher than the required $\sim \spshotnoise$ acceleration tolerance of this experiment. 

This problem can be tackled by fixing the direction of the laser's axis with respect to an inertial reference. The Coriolis acceleration due to the thermal velocity $v_T \sim 10^{-8}$ m/s of the atom cloud is smaller than the shot noise $\spshotnoise$ of this experiment if the residual rotational velocity  $\delta \omega$ of the laser axis is  smaller than $10^{-10} $ rads/s. Control over the rotation axis at the level of $10^{-14}$ rads/s has been achieved \cite{GPB}. However, if the laser axis is inertial, the satellite at distance $L$ away from it will have a transverse velocity $L \omega_S \sim 10 \text{ cm/s} \left(\frac{L}{10^3 \text{ km}}\right)$ with respect to the laser axis. The residual rotational velocity $\delta \omega$ of the laser axis couples to this transverse velocity and causes a Coriolis acceleration $\sim L \omega_S \delta \omega $ which is smaller than $\spshotnoise$ if $\delta \omega$ is smaller than $10^{-17} \text{ rads/s } \left(\frac{L}{10^3 \text{ km}}\right)$ . The control required over the rotation axis can however be relaxed by applying forces on one satellite while using the other as an inertial reference to cancel the relative rotation between them. The gravitational tidal force on the satellites due to the Sun is $\sim 10^{-4} \text{ N} \left(\frac{M}{\text{1000 kg}}\right) \left(\frac{L}{10^3 \text{ km}}\right)$ while the force on the satellites due to solar radiation pressure $\sim 10^{-5} $ N. These forces are small enough to be compensated by FEEP and colloid thrusters \cite{LISAFTR}.  The  application of these forces cancels the relative transverse velocity between the laser's axis and the distant satellite.  The residual transverse velocity $v_T$ of the atom clouds due to their thermal velocity and vibrations of the atom trap can also cause Coriolis accelerations.  If the atoms are cooled to $\sim 100 \text{ pK} \sqnasubs$ temperatures, their thermal velocities are smaller than $10^{-8}$ m/s. The Coriolis acceleration is then smaller than $\spshotnoise$ if $\delta \omega$ is controlled better than $10^{-10}$ rad/s and the transverse vibrations of the atom clouds are smaller than $ 10 \text{ nm/s } (\frac{\omega}{10^{-2} \text{ Hz}})^2$.

In addition to the Coriolis acceleration, any instability $\delta\omega$ in the laser's angular velocity (e.g. in the rotation servoing mechanism) causes a differential centrifugal acceleration  $\sim L (\delta \omega)^2$. This acceleration is smaller than $\spshotnoise$ if $\delta \omega \lessapprox 10^{-11} \frac{\text{ rad/s }}{\rthz}  \sqrt{(\frac{10^3 \text{ km}}{L})} (\frac{\omega}{10^{-2}\text{ Hz}})^{\frac{1}{2}}$ at frequency $\omega$. The control over the rotation of the laser's axis can be potentially further relaxed by tuning the radius of curvature of the laser beam. Since the atom senses the local phase of the laser beam, the atoms will not sense rotations of the laser's axis if the phase fronts are appropriately curved.  If the radius of curvature $R$ of the beam is equal to the distance $L$ between the atom and the laser then the  atom is insensitive to centrifugal accelerations $\sim L (\delta \omega)^2$. The control over the rotation of the laser's axis can be relaxed to the extent to which the radius of curvature of the beam at the distant interferometer can be tuned to equal the distance between that interferometer and the laser. The differential setup proposed in this experiment requires one interferometer to be close to the laser at a distance $I_L \sim 100$ m while the other is at a distance $L \sim 10^3 $ km. The centrifugal acceleration of the atoms near the laser will produce accelerations $\sim I_L \left(\delta\omega \right)^2$. These accelerations set the minimal control required over the laser's rotation to $\delta\omega \lessapprox  10^{-9} \frac{\text{ rad/s }}{\rthz}  \sqrt{(\frac{\text{100 m}}{I_L})} (\frac{\omega}{10^{-2}\text{ Hz}})^{\frac{1}{2}}$. 

In this configuration, due to the finite radius of curvature of the laser beam, the interferometer is sensitive to transverse vibrations of the lasers. The effects of these vibrations on the two interferometers will not be entirely common if the radii of curvature of the laser beams that interact with the two interferometers are not equal. The phase shift from a transverse vibration $\delta y$ to a single interferometer is $\sim \keff \frac{\delta y^2}{R}$ where R is the radius of curvature of the beam. This phase shift can be made smaller than shot noise even without relying on common mode cancellation by damping the transverse vibrations of the laser  below $100 \mu m \sqrt{\frac{R}{10^3 \text{km}}}$ over the frequencies of interest. 


The need to reference the axis of the laser to an inertial reference emerged from the demand to suppress Coriolis accelerations due to the thermal velocity of the atom cloud. Another way to deal with this problem is to operate the interferometer  in the multiloop configurations described in sub section  \ref{SubSec: EarthNewtonianGravity}. The Coriolis acceleration caused by a laser rotating with a constant angular velocity and an atom cloud moving with a constant transverse velocity is constant. The phase shift due to such a constant acceleration is completely cancelled in these multiloop configurations.  In this configuration, the interferometer has a smaller bandwidth but has the same sensitivity to gravitational waves at its resonant frequency as the Mach Zender configuration.

Rotational backgrounds in this multi-loop setup can be controlled by servoing the laser to track the rotation of the satellites. An instability $\delta \omega$ in the angular velocity of the axis will cause a centrifugal acceleration $L \left(\delta\omega \right)^2$. Moreover, the transverse velocity $v_T$ of the atom cloud  caused by the thermal velocity of the atom and vibrations of the trap used to confine the atoms will cause accelerations $\delta \omega v_T  $.  These backgrounds can be made smaller than $\spshotnoise$ by making $\delta \omega $ smaller than $10^{-10}$ rads/s as discussed earlier in this section.
 

\subsubsection{Effects of Magnetic Fields}
\label{SubSec: SpaceMagneticFields}

A time variation $\delta B$ in the magnetic field $B_0$ produces a phase shift $\sim \alpha_{\text{ZC}} B_0 \delta B T$ in the interferometer as discussed in sub section \ref{SubSec: EarthMagneticFields}.  This phase shift must be smaller than $10^{-4}$. Time variations in the interplanetary magnetic field at $\sim 1$ AU have been measured to be $\sim 0.1 \frac{\text{nT}}{\sqrt{\text{Hz}}} (\frac{10^{-2} \text{ Hz}}{\omega})$ \cite{SpaceMagneticFields}. The applied bias magnetic field $B_0$  is $\sim 100$ nT over the interferometer region $I_L$ and $\alpha_{\text{ZC}} \sim 1 \frac{\text{kHz}}{\text{G}^2}$ (for Rubidium). With these values, $\alpha_{\text{ZC}} B_0 \delta B T \sim 10^{-5}$ for $T \lessapprox 100$ s. 

Note this is true only if the atom interferometer is operated using Raman transitions, so the atom is in different internal levels during the course of the interferometer. This phase shift will be absent if the interferometer is operated using Bragg transitions, since the phase accrued along each arm is the same. However, there will still be a phase shift that goes like  $\sim \alpha_{\text{ZC}} \nabla\left( B_0 \delta B \right) \left(v_R T\right) T$.

The atoms are in magnetically insensitive ($m = 0$) states and they move through a non-uniform magnetic field. The gradient $\nabla B$ of the magnetic field causes a force  $\sim \alpha_{\text{ZC}} B \nabla B$ on the atom due to the second order Zeeman effect.  The atom experiences a gradient $\nabla B \sim \frac{100 \text{ nT}}{30 \text{ m}}$ from the external bias magnet in the configuration considered in this experiment. With this gradient, time variations $\delta B \sim 0.1 \, \frac{\text{nT}}{\sqrt{\text{Hz}}} (\frac{10^{-2} \text{ Hz}}{\omega})$ \cite{SpaceMagneticFields} in the interplanetary magnetic field cause accelerations $\sim 10^{-19} g$ which is equal to the shot noise requirement of the experiment.  The time varying acceleration caused by fluctuations in the position of the bias magnet are smaller than $\sim 10^{-19} g$ if these fluctuations are smaller than $\sim 1 \,  \frac{\text{mm}}{\sqrt{\text{Hz}}}$ in the $10^{-2} \text{ Hz}$ band. 

The atom is placed in a $m = 0$ state with respect to the external magnetic field at the start of the interferometer to minimize the effects of accelerations from time dependent magnetic fields.  The Rabi frequency of the atomic transition is set by the internal state of the atom and the laser is tuned to match this frequency. Changes to the direction of the external magnetic field during the interrogation time of the experiment are adiabatic compared to the rapid precession rate of the atom's spin. If the direction of the magnetic field changes, the quantization axis of the atom's spin will track this direction change.  Since the laser is tuned to the original $m = 0$ state, the atom-laser interaction will excite  $m = \pm$ 1 states along the new axis of quantization. The phase shift from these states is a background. 

The $m = \pm$1 components developed by the atom as a result of a misalignment by an angle $\theta$ between the magnetic field and the quantization axis are proportional to $\sin(\theta)$. The probabilities induced by this mixing are therefore proportional $\sin^2(\theta)$. The contributions of this mixing to the phase shift in the interferometer are smaller than $10^{-4}$ when $\theta \lessapprox  10^{-2}$. The direction of the interplanetary magnetic field was characterized by \cite{Burlaga}. During an average time, the drift in the direction of the magnetic field was found to be smaller than $5 \degree$ over 10 minutes. In the presence of a $\sim 100$ nT bias field over the interferometer region,  these angular variations of the $\sim$ 5 nT interplanetary magnetic field will change the overall direction of the magnetic field in the interferometer by less than $10^{-2}$ in 100 seconds. 

The above arguments indicate that the effects of time varying interplanetary electromagnetic fields on the atom interferometer are naturally small and close to the shot noise floor of the experiment.  The effects of these fields can be additionally suppressed to the extent to which these fields can be measured. The response of the atom interferometer to a given electromagnetic field is determined by known quantities like the magnetic moment of the atom and its polarizability. Since these quantities are known to several digits, a measurement of the electromagnetic fields will enable us to predict the effect of these fields on the atom interferometer. These effects can then be subtracted out from the measured phase shift. 

We note that the effects of electromagnetic forces on the atom interferometer are significantly suppressed compared to their effects on LISA's inertial test masses. Spurious electromagnetic forces on the test masses due to charge transfer between the test masses and the satellite environment is a major background for LISA. The test mass acquires a random charge from its environment and its response to time varying electromagnetic fields cannot be predicted even if the electromagnetic fields themselves are measured. Since the atom interferometer is operated using magnetically insensitive atomic states, electromagnetic forces on the atom are greatly diminished.  The response of the atom interferometer to electromagnetic fields can be predicted to the extent to which these fields are measured providing additional immunity to the atom interferometer from time varying electromagnetic fields. 

\subsubsection{The Radius of Curvature of the Beam}

The temperature of the atom cloud will cause the atom to have thermal velocities along the direction transverse to the laser fields propagating along the interferometer axis. This velocity will cause the atoms to move in a direction transverse to the laser beam. Owing to the finite radius of curvature $R$ of the beam, an atom that is slightly off-axis by $\delta y$ from the center of the beam will see an additional phase $\keff (\frac{\delta y^2}{R})$. With $N_a \sim 10^8$ atoms in the cloud, shot to shot variations in this phase are $\sim \sqna \keff (\frac{\delta y^2}{R})$ and these must be smaller than $\sim 10^{-4}$. With thermal velocities $\sim 100 \mu \text{m/s}$, the maximum transverse distance travelled by the clouds is  $\delta y \sim 1 \text{cm}$  over an interrogation time $T \sim 100$ s. The phase shift $ \sqna \keff (\frac{\delta y^2}{R})$  is then smaller than $10^{-4}$ if the radius of curvature $R$ of the beam is greater than $\sim 100 \text{ km} \left(\frac{10^9 \text{ m}^{-1}}{\keff} \right)$.



\subsubsection{Blackbody Clock Shift}

Black body radiation shifts the hyperfine transition frequency of the atom by $  \sim 10^{-4} \text{ Hz} \left(\frac{\tau}{\text{300 K}}\right)^4$ \cite{BlackBodyShift}.   The ambient temperature $\tau$ at 1 AU is $ \sim \text{300 K}$. Time variations $\delta \tau$ in the temperature of the interferometer region during the interrogation time $T$ of the experiment will  change the hyperfine transition frequency by $\delta \nu \sim 4 \times 10^{-4} \text{ Hz} \left(\frac{\tau}{\text{300 K}}\right)^4 \left(\frac{\delta\tau}{\tau}\right)$ causing a phase shift $\delta \nu \; T \sim 10^{-2}  \left(\frac{\tau}{\text{300 K}}\right)^4 \left(\frac{\delta\tau}{\tau}\right) \left(\frac{\omega}{10^{-2} \text{ Hz}}\right)$.  This phase shift  is smaller than $10^{-4}$ if the temperature fluctuations $\delta \tau$ in the frequency band $\omega$ are smaller than $\sim 1 \text{ K} \left(\frac{\omega}{10^{-2} \text{ Hz}}\right)$. Time dependence in the temperature of the interferometer is caused by variations in the solar output and fluctuations of the spacecraft temperature. Time variations of the solar output typically occur over the time scale of a day at distances $\sim 1$ AU\cite{Rottman}. The solar output changes by $\sim$ 1 Watt during this period leading to a temperature change $\sim 0.05$ K in the time scale of a day. These variations are therefore not a problem for the interferometer. 

The effects of the thermal variation of the satellite on the interferometer are suppressed since the interferometer is operated at a distance $d_I \sim 30 \text{ m}$ away from the satellite. Temperature variations of the satellite at frequency $\omega$ have to be larger than $\sim 10 \text{ K} \left(\frac{\omega}{10^{-2} \text{ Hz}}\right) \left(\frac{d_I}{\text{30 m}}\right)^{\frac{1}{2}}$ in order to change the temperature of the interferometer region by $1$ K. The spacecraft receives heat from the Sun and the solar wind. As discussed above, variations in the solar output are small over the time scale of interest. The solar wind is composed of 2 keV protons and electrons with density $\sim \frac{5}{\text{cm}^3} $ moving at speeds $\sim 400 \text{ km/s}$. The change in temperature of the satellite from an order one change in the flux of the solar wind is $\sim 1 \text{ mK}$. The environment of the satellite will therefore not cause its temperature to fluctuate at levels of interest to this experiment. 

The satellite will also establish a spatial thermal gradient over the interferometer region due to its shadow. This spatial gradient will contribute to the phase shift in the interferometer.  The natural time scale for the variation of this phase shift is equal to the orbital period of the satellite $\sim 1$ year and is therefore not a problem for the current experiment. Time variations of this spatial gradient are also created by random motions of the satellite during the interrogation time of the experiment. However, these motions need to be well controlled to suppress Newtonian gravity backgrounds which are much larger than the small phase shift produced by the spatial thermal gradient.  The variations in this phase shift due to the residual random motions of the spacecraft will therefore be smaller than shot noise.

\subsection{Comparison with LISA}
\label{Sec: LISA Comparison}

The detection of gravitational waves requires techniques that are sensitive to the miniscule effects of gravitational waves and can simultaneously  suppress noise in the measurement bandwidth to permit the extraction of the signal. The atom interferometer configurations discussed in this paper can probe the same frequency spectrum as satellite based light interferometers like LISA  with comparable sensitivity (see Section \ref{Sec: Sensitivities}). However, as discussed in subsection \ref{Sec: space backgrounds}, these configurations may naturally permit significant suppression of several serious backgrounds faced by LISA, see Table \ref{Tab: LISA comparison}.

LISA aims to detect gravitational waves by measuring the relative distance between two inertial proof masses separated by an arm length $\sim 5$ million kilometers. Position noise of these masses is a background for LISA.  The significant gravitational coupling between random motions of the satellite and the proof mass is a dominant cause of this position noise. In order to sufficiently suppress this noise, LISA requires satellite position control at $\sim 1 \frac{\text{nm}}{\rthz}$ in its measurement bandwidth \cite{LISAFTR}. However, as argued in subsections \ref{Sec:space LMT limit} and \ref{Sec: Space Gravity Backgrounds}, since the atom interferometer can be operated outside the satellite over a $\sim 100$ m region from the satellite, the effects of position noise of the satellite on the interferometer are significantly suppressed. For gravitational wave sensitivity similar to LISA, our atom interferometer setup would require satellite position control at $\sim 10 \frac{\mu \text{m}}{\rthz}$ in the measurement bandwidth.

 In addition to random motions of the satellite, spurious electromagnetic forces on the LISA proof mass also contribute to its position noise. These forces are caused by direct collisions between the proof mass and the background gas and due to charge accumulation on the proof mass from interactions with cosmic rays and the solar wind. The test mass acquires a random charge from its environment and its response to time varying electromagnetic fields cannot be predicted even if the electromagnetic fields themselves are measured. Since the atoms are neutral and the atom interferometer is operated using magnetically insensitive ($m=0$) states, electromagnetic forces on the atom clouds are naturally small.  The response of the atom interferometer to electromagnetic fields can be predicted to the level at which these fields are measured.  This provides additional immunity from time varying electromagnetic fields. Collisions of the atoms with background particles from the solar wind or cosmic rays lead to particle deletion from the cloud and not charging of the cloud. These deletions result in a minor reduction in the sensitivity (for interrogation times  $\lesssim 1000 \text{~s}$) but do not cause phase shifts to the remaining atoms and hence are not a background for this experiment.

Laser phase noise is another major background for gravitational wave detectors. This noise can be suppressed by the simultaneous operation of  interferometers along the arms of a three-satellite constellation. In this configuration, the effects of laser phase noise are cancelled up to knowledge of the arm lengths (see subsection \ref{Subsec:SpaceVibrations}). Both LISA and the atom interferometer can benefit by exploiting this idea. However, due its long ($\sim 5$ million km) arm length, LISA faces unique challenges in determining the absolute distance between its satellites \cite{Shaddock:PRD68, Cornish}. Owing to these difficulties, LISA requires control over the frequency drift of its lasers at $\sim 1 \frac{\text{ Hz}}{\rthz}$ at $10^{-2} \text{ Hz}$. The atom interferometer setup considered in this paper can reach sensitivities comparable to LISA with significantly smaller arm lengths $\sim 10^3$ km. The compactness of this baseline might allow for the determination of the arm lengths of the atom interferometer constellation with better precision than LISA. If these arm lengths are known to within $\sim 1$ m, our experiment can reach sensitivities similar to LISA with control over laser frequency $\sim 10^{4} \frac{\text{ Hz}}{\rthz}$ at $10^{-2} \text{ Hz}$. 

The atom interferometer setup discussed in this paper might significantly relax the requirements on several major backgrounds faced by light interferometers like LISA while achieving comparable sensitivity. We have attempted to consider the relevant backgrounds introduced by the atom interferometer setup in section \ref{Sec: space backgrounds} and show that they could be controlled with practical technology in a realistic setup. Since many of these backgrounds require careful engineering, further study is necessary.  However, the experiment appears to be feasible and exciting enough to merit more serious consideration.

\begin{table}
\begin{center}
\begin{math}
\begin{array}{|c|c|c|}
\hline
\text{Attribute} & \text{AGIS} & \text{LISA}  \\
\hline
\hline
\text{baseline} & 10^3 \text{~km} & 5 \times 10^6 \text{~km} \\
\hline
\text{satellite control (at} \sim 10^{-2} \text{ Hz)} & 10^4 \frac{\text{nm}}{\sqrt{\text{Hz}}} & 1 \frac{\text{nm}}{\sqrt{\text{Hz}}} \\
\hline
\text{laser frequency control (at} \sim 10^{-2} \text{ Hz)} & 10^4 \frac{\text{Hz}}{\sqrt{\text{Hz}}} & 1 \frac{\text{Hz}}{\sqrt{\text{Hz}}} \\
\hline
\text{rotational control (at} \sim 10^{-2} \text{ Hz)} & 10^{-2} \frac{\text{nrad}}{\sqrt{\text{Hz}}} & 1 \frac{\text{nrad}}{\sqrt{\text{Hz}}} \\
\hline
\text{electromagnetic forces} & \text{atoms neutral, EM forces naturally small,  } & \text{cosmic ray charging of proof mass} \\
\text{ } & \text{predictable response to measured EM field } &  \\
\hline
\text{collisions with background gas} & \text{delete atoms, not a noise source} & \text{cause acceleration noise} \\
\hline
\end{array}
\end{math}
\caption{\label{Tab: LISA comparison} A comparison between specifications for a three satellite AGIS configuration that could potentially allow comparable sensitivity to LISA, and the LISA requirements.  There are many caveats and details that cannot be captured in a table and are discussed in Sections \ref{Sec: space backgrounds} and \ref{Sec: LISA Comparison} and in the LISA papers (see e.g. \cite{LISAFTR, LISAPPA}).}
\end{center}
\end{table}

	

\section{Gravitational Wave Sources}
\label{Sec: Sources}


There are many known and potential sources for gravitational waves from astrophysics and cosmology.  Here we will discuss only a few, including the well-known compact object binaries, which give a coherent oscillatory gravitational wave signal, and more speculative cosmological sources, which give a stochastic background of gravitational waves.  There are many reviews of this subject that discuss other sources including gamma-ray bursts, supernovae, and spinning neutron stars (see, for example, \cite{Cutler:2002me, Maggiore:1999vm}).

\subsection{Compact Object Binaries}
One of the most promising sources of observable gravitational waves is a binary star where both components are compact objects such as white dwarfs, neutron stars, or black holes \cite{Cutler:2002me}.  These compact binaries emit strongly in gravitational waves because they contain large mass stars relatively close to each other.  The amplitude of the gravitational waves emitted is
\begin{eqnarray}
h \sim G \mu \frac{(G M \Omega)^{\frac{2}{3}} }{r} \sim \frac{(G M_1) (G M_2)}{r R}
\end{eqnarray}
where $M_{1,2}$ are the masses of the components, $M = M_1 + M_2$ and $\mu = \frac{M_1 M_2}{M}$ are the total and reduced masses, $R$ is the radius of the binary, $\Omega$ is the orbital frequency of the binary, and $r$ is the distance from the binary at which the wave is observed.  As neutron stars and white dwarfs are both roughly 1 solar mass, $M_\odot$, we will primarily be interested in compact binaries composed of two $1 \, M_\odot$ mass components as sources.  The amplitude of the emitted gravitational waves then depends only on the orbital period and the distance to the star.  For a binary with $\Omega = 1 \text{ s}$ in our galaxy we expect $h \sim 10^{-18}$, in our local cluster $h \sim 10^{-21}$, and in a Hubble volume (i.e. out to redshifts $z \sim 1$) $h \sim 10^{-23}$.

The main frequency component of the emitted gravitational wave is at twice the binary's orbital frequency, $\omega \propto 2 \Omega$ \cite{BlandfordThorne}.  This is clear for equal mass stars, and can also be seen for unequal masses from the fact that gravitational radiation arises from the second time derivative of the quadrupole moment of the binary.

Near the end of its life, the dominant energy loss mechanism for a compact binary is gravitational radiation.  As a compact binary loses energy, the stars spiral inward, increasing the orbital frequency.  This can bring the emitted gravitational waves into the observable part of the spectrum for gravitational wave detectors.  This process ends  when the two compact objects collide.  Thus, the highest gravitational wave frequency emitted depends on the radii of the compact objects.  A neutron star binary can reach frequencies of over 100 Hz while a white dwarf binary can only reach roughly 0.5 Hz before collision.  The power emitted in gravitational waves is $P \sim M_\text{planck}^2 h^2$.  Because this power depends mainly on a few variables like the masses and orbital period of the binary, the inspiral of a compact binary near the end of its life is consistent and predictable and therefore so is the waveform of the emitted gravitational waves.  Using the power emitted in gravitational waves, the remaining lifetime of a compact binary is \cite{BlandfordThorne}
\begin{eqnarray}
\tau \sim \frac{1}{50 G \mu (G M)^\frac{2}{3} \Omega^\frac{8}{3}}.
\end{eqnarray}
As the orbital frequency increases, the rate of energy loss increases and the remaining lifetime decreases rapidly.  This means that at any given $\Omega$ most of the remaining life of the binary will occur near that frequency.

There are thus two main advantages to being able to observe gravitational waves at lower frequencies.  First, the population of binary stars that are potentially observable is increased, both because new classes of stars such as white dwarf or high-mass black hole binaries are observable and because a greater fraction of any given class, such as neutron star binaries, is at lower frequencies than at higher ones.  Indeed, for a gravitational wave detector such as LISA which can observe waves with frequencies as low as $10^{-3} \text{ Hz}$, the large number of white dwarf binaries creates a stochastic background of gravitational waves for the detector in this frequency band \cite{Farmer:2003pa}.  Second, a lower frequency binary has a longer time left to live which increases the observation time and thus the sensitivity of the detector for this source.

\subsection{Stochastic Sources}

In addition to a large number of white dwarf binaries, several potential cosmological sources can produce a stochastic background of gravitational waves including inflation and reheating, a network of cosmic strings, or phase transitions in the early universe.

A period  of inflation can produce a fairly flat (scale-invariant) stochastic gravitational wave background \cite{Starobinsky:1979ty}.  This could be as high as $\Omega_\text{GW} (f) \approx 10^{-13}$, as limited by the COBE bound \cite{Allen:1994xz}, though slow-roll inflation models probably give a smaller value and a tilted spectrum \cite{Maggiore:1999vm}.  This is fairly difficult for planned experiments to detect, but reheating after inflation can give a more peaked spectrum of gravitational waves with a higher value of $\Omega_\text{GW}$.  For example, reheating after hybrid inflation \cite{GarciaBellido:2007af} or preheating \cite{Easther:2006vd, Dufaux:2007pt}, can give a spectrum of gravitational waves with $\Omega_\text{GW}$ several orders of magnitude higher than that from the period of inflation itself.  The frequency of the peak is model-dependent, proportional to the scale of reheating.  It probably lies within a range from roughly $1 \text{ Hz}$ to $10^9 \text{ Hz}$.  There is then a possibility that this enhanced strength of gravitational waves from reheating will allow a detection by interferometers.  There are also other possibilities such as pre-big bang \cite{Gasperini:2002bn} or extended \cite{Turner:1990rc} inflation that can lead to much higher values of $\Omega_\text{GW} (f)$ in the phenomenologically interesting frequency range for interferometers.

A first-order phase transition in the early universe can produce gravitational waves through bubble nucleation and turbulence \cite{Kamionkowski:1993fg, Caprini:2007xq}.  The frequency of the gravitational waves today is given by redshifting the frequency at which they were produced, which is proportional to the Hubble scale at the phase transition.  There are, however, significant uncertainties in the calculations of these frequencies (see \cite{Maggiore:1999vm}).  The best expectation is that a phase transition at the electroweak scale is likely to produce gravitational waves with a frequency today in a range near $10^{-3} \text{ Hz}$.  Earlier phase transitions at higher temperatures produce gravitational waves with proportionally higher frequencies.  In some models with new physics at the weak scale \cite{Grojean:2006bp}, including some supersymmetric \cite{Apreda:2001us} and warped extra-dimensional \cite{Randall:2006py} models, the electroweak phase transition can produce gravitational waves with very large $\Omega_\text{GW}$, well above the threshold for detection by atom interferometers.

A network of cosmic strings produces a stochastic background of gravitational waves from vibrations of the strings.  Cusps and kinks in the strings produce bursts of gravitational waves which could be seen individually or as a stochastic background.  Unfortunately, even in the simplest models there are large uncertainties in the calculation of the formation and subsequent gravitational radiation of such string networks.  Thus, it is very difficult to get a precise prediction from theory about the strength of gravitational waves coming from a network of cosmic strings.  Using the current understanding of cosmic string networks, the sensitivities of atom interferometers on earth and in space to a stochastic gravitational wave background (see Figs. \ref{Fig:earth stochastic sensitivity} and \ref{Fig:space stochastic sensitivity}) could allow detection of cosmic strings with $G \mu \sim 10^{-8}$ to $10^{-11}$ ($\mu$ is the string tension) or lower, depending on the sensitivity achieved and the uncertainties in the cosmic string calculations.  For a recent review of this subject see for example \cite{Polchinski:2007qc}.

There are many other possible sources for gravitational waves from fundamental physics in the early universe including Goldstone modes of scalar fields \cite{Hogan:1998dv}, or radion modes and fluctuations of our brane in an extra dimensional scenario \cite{Hogan:2000is, Hogan:2000aa}.  There are also other astrophysical sources that may lead to an interesting stochastic gravitational wave background (for a review see \cite{Maggiore:1999vm}).

The possibility of accessing these cosmological and astrophysical sources makes gravitational waves a very interesting avenue for exploring the universe and probing fundamental physics.  Indeed, observing gravitational waves could be one of our only ways of getting information about the universe before the last scattering surface.

\section{Sensitivities}
\label{Sec: Sensitivities}


In this Section we find projected sensitivity curves for the terrestrial and satellite experiments.  There is always significant uncertainty in projecting the sensitivity of a proposed experiment.  We have attempted to give a range of sensitivities to show more conservative and more aggressive assumptions about what may be experimentally achievable.  There is also some uncertainty in these sensitivity curves because we have not attempted to perform a careful statistical study of the exact sensitivity for a particular configuration.  Especially in the case of the stochastic gravitational wave background, this can make important differences that have been worked out carefully be many authors for laser interferometers.  We leave such considerations to future work.


\subsection{Binary Sources}

\begin{figure}
\begin{center}
\includegraphics[width=\columnwidth]{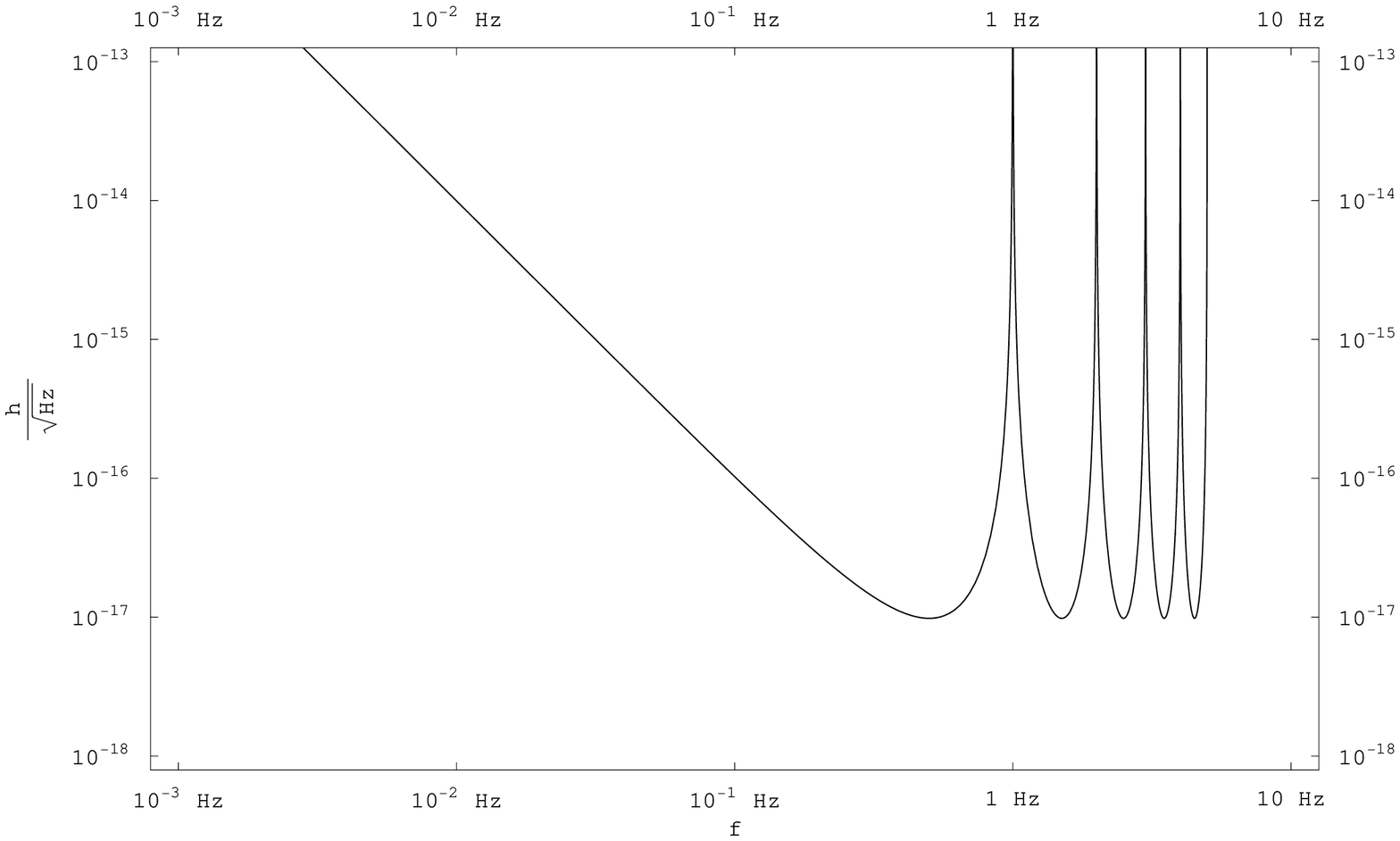}
\caption{ \label{Fig:demon binary sensitivity} An example sensitivity curve to a gravitational wave of frequency $f$.  It is a shot noise power spectrum in the response of the atom interferometer to a gravitational wave of amplitude $h$.  Here we have taken the two atom interferometers to be a distance $L=1 \text{ km}$ apart, with interrogation time $T = 1 \text{ s}$, $100 \hbar k$ LMT beamsplitters, a per shot phase sensitivity of $10^{-4} \text{ rad}$, and a data-taking rate of $10 \text{ Hz}$.}
\end{center}
\end{figure}

The inherent limit on the sensitivity to a gravitational wave due to shot noise can be found from Eqn. \eqref{Eqn:GW phase shift}.  This limit, equivalently the power spectrum of the shot noise in the experiment $h_n (f)$, is shown in Figure \ref{Fig:demon binary sensitivity} for an example configuration, as described in Sections \ref{Sec: Space Setup} or \ref{Sec: Earth Setup}.  Here we have taken the two atom interferometers to be a distance $L=1 \text{ km}$ apart, with interrogation time $T = 1 \text{ s}$, $100 \hbar k$ LMT beamsplitters, and a per shot phase sensitivity of $10^{-5} \text{ rad}$.  We have also assumed a data-taking rate of $10 \text{ Hz}$.  The plot is cutoff at the Nyquist frequency of $5 \text{ Hz}$.  There would in actuality be some sensitivity to higher frequencies but they will be aliased to look like lower frequencies potentially also leading to confusion with backgrounds.

At low frequencies the sensitivity rises as $f^{-2}$ as is clear from Eqn. \eqref{Eqn:GW phase expand}.  For higher frequencies, the sensitivity flattens out because a longer interrogation time does not increase the response of the interferometer once it is longer than the period of the gravitational wave $T > f^{-1}$.  The sensitivity then reaches it's maximum when $\sin^2 \left( \frac{\omega T}{2} \right) = 1$, i.e. when there are an odd number of periods of the gravitational wave in the entire time $2 T$ of the interferometer.  This agrees with the intuition that the interferometer is sensitive to changes in the relative timing of the laser pulses caused by the stretching of the metric and therefore maximally sensitive when there is the greatest change in the distance to the laser (the clock) between each successive laser pulse.

The singularities in the sensitivity curve in Figure \ref{Fig:demon binary sensitivity} come at frequencies which are integral multiples of $f = T^{-1}$, when an integral number of periods of the gravitational wave fit in the interrogation time $T$.  Roughly the periods when the gravitational wave is causing a `stretch' exactly equal the periods when it is causing a `squeeze'.  The net integrated effect of the gravitational wave is then zero and the phase shift response of the atom interferometer goes to zero.  Thus the atom interferometer has no sensitivity to such frequencies.

Note that the best sensitivities in Figure \ref{Fig:demon binary sensitivity} come at frequencies halfway between the singularities, when there are an odd, integral number of periods of the gravitational wave in the entire atom interferometer (a time of $2 T$).  This can be understood since the atom interferometer is essentially taking the difference between the phases accrued by the atom in the first and second halves of the sequence.  The maximal difference arises when the `stretch' part of the gravitational wave (which is when the coefficient of the $dx^2$ term in Eqn \eqref{Eqn: GW metric TT gauge} is greater than 1) occurs during one of the halves and the `squeeze' during the other.  These are the frequencies to which the atom interferometer responds maximally.

\begin{figure}
\begin{center}
\includegraphics[width=\columnwidth]{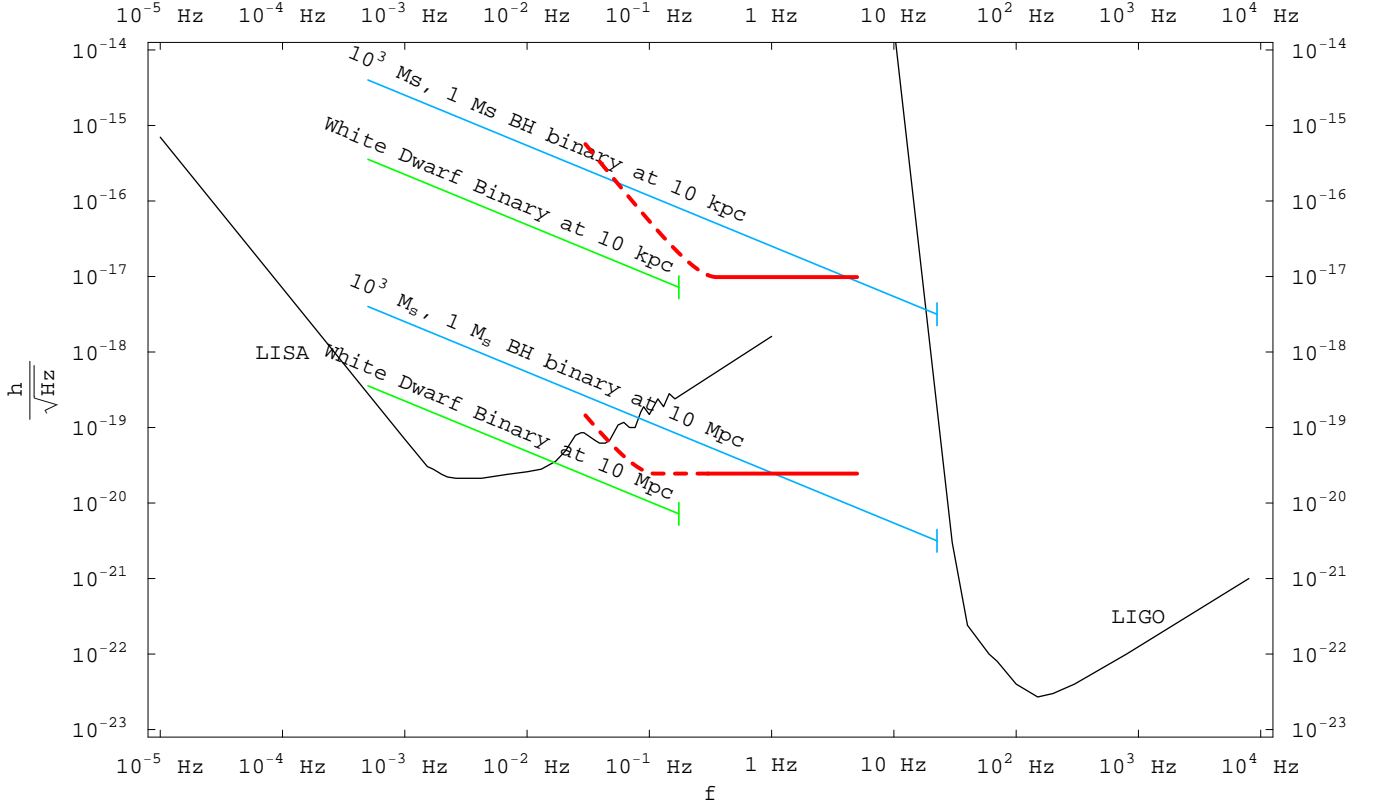}
\caption{\label{Fig:earth binary sensitivity} (Color online) The thick (red) curves show projected sensitivities of our proposed terrestrial experiments to a gravitational wave of frequency $f$.  The choices of experimental parameters for these two configurations are shown in Table \ref{Tab: sensitivities}.  These are only projected shot noise power spectra in the response to a gravitational wave of amplitude $h$.  They do not include other backgrounds, since as we have argued, these may be reduced below shot noise.  Possible sources are shown.  Expected noise curves are shown for initial LIGO and LISA. }
\end{center}
\end{figure}

A longer interrogation time for the experiment does not actually improve the peak sensitivity in the sense of lowering the curve in Figure \ref{Fig:demon binary sensitivity}.  Instead, it slides the curve left, lowering the frequency at which the maximum sensitivity is reached.  Of course, a larger length $L$ or higher momentum beamsplitters improves the entire sensitivity curve linearly.  We have cut off the sensitivity curve above the Nyquist frequency.  In reality there will be a slightly more gradual loss of sensitivity before this frequency and even some sensitivity to higher frequencies, although they will be aliased.  Assuming a constant number of atoms per second that can be cooled and run through the interferometer, a faster data-taking rate does not improve sensitivity.  It would merely improve the high frequency cutoff.  Thus it seems unnecessary to strive for a data-taking rate faster than $\OO(10 \text{ Hz})$.  The sensitivity would also decrease as the frequency of the gravitational wave approached the light travel time (or really the gravitational wave travel time) across the whole experiment, namely $L$.  However this frequency is much higher than the frequency of maximal sensitivity for an atom interferometer.  This would not be true for a light interferometer where the light travel time across the device is also the time length of a `shot', the analogue of the interrogation time.  For example, this explains why LISA loses sensitivity above $\sim 0.05 \text{ Hz}$ while the atomic interferometer's sensitivity curve remains flat (see, for example, Figure \ref{Fig:earth binary sensitivity}).

\begin{figure}
\begin{center}
\includegraphics[width=\columnwidth]{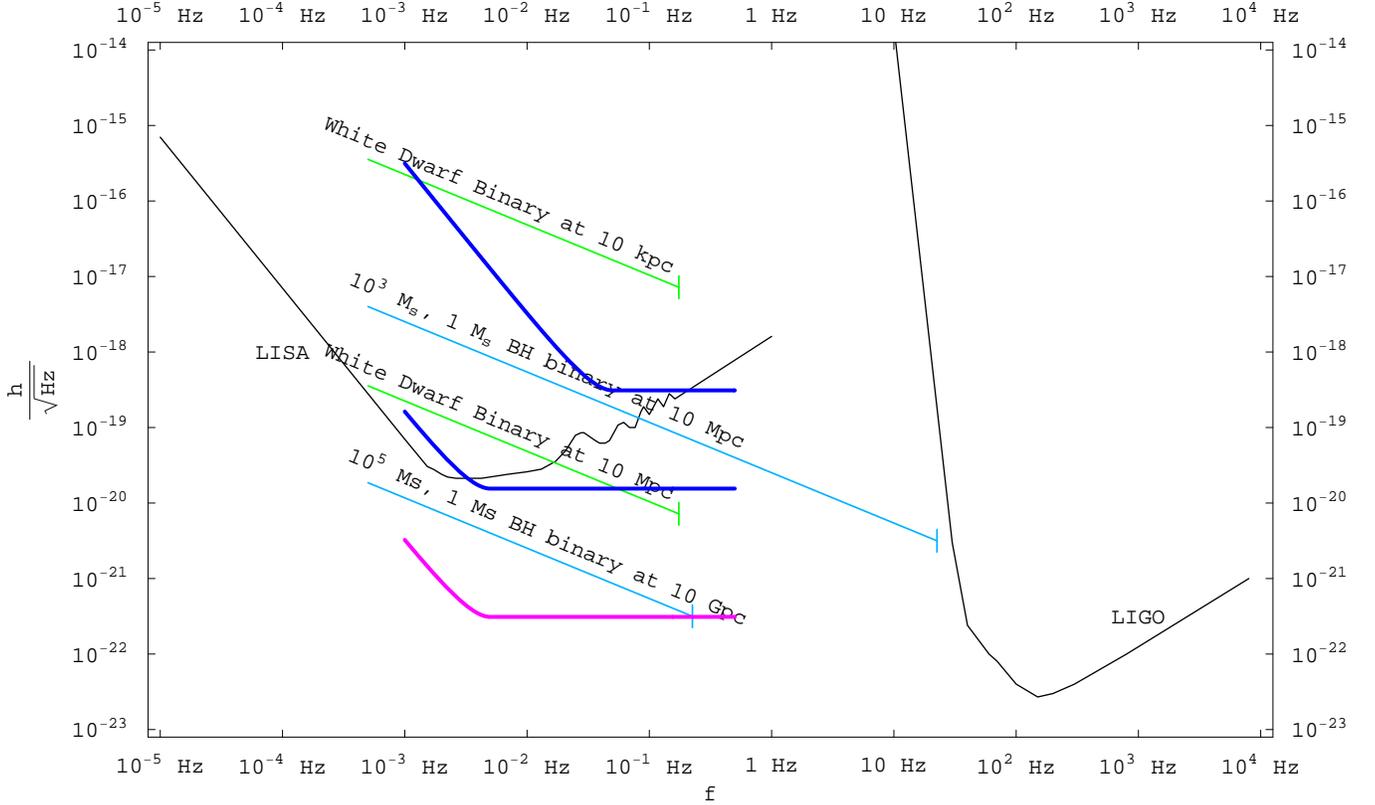}
\caption{\label{Fig:space binary sensitivity} (Color online) The thick (blue and purple) curves show the projected sensitivities of our proposed satellite experiments to a gravitational wave of frequency $f$.  The choices of experimental parameters for these three configurations are shown in Table \ref{Tab: sensitivities}.  The lowest (purple) curve, satellite 3 in Table \ref{Tab: sensitivities}, is an aggressive possibility that might be realizable in the future.  These are just projected shot noise power spectra in the response to a gravitational wave of amplitude $h$.  Possible sources are shown.  Expected noise curves are shown for initial LIGO and LISA.}
\end{center}
\end{figure}

\begin{table}
\begin{center}
\begin{math}
\begin{array}{|c|c|c|c|c|c|c|}
\hline
\text{Setup} & L & \keff & T & I_L & \text{Phase Sensitivity} & f_d  \\
\hline
\text{Terrestrial~} 1 & 1 \text{~km} & 1.6 \times 10^9 \text{~m}^{-1} & 1.4 \text{~s} & 10 \text{~m} & 10^{-4} \text{~rad} & 10 \text{~Hz} \\
\text{Terrestrial~} 2 & 4 \text{~km} & 1.6 \times 10^{10} \text{~m}^{-1} & 4.5 \text{~s} & 100 \text{~m} & 10^{-5} \text{~rad} & 10 \text{~Hz} \\
\text{Satellite~} 1 & 100 \text{~km} & 1.6 \times 10^{9} \text{~m}^{-1} & 10 \text{~s} & 100 \text{~m} & 10^{-4} \text{~rad} & 1 \text{~Hz} \\
\text{Satellite~} 2 & 10^3 \text{~km} & 3.2 \times 10^{9} \text{~m}^{-1} & 100 \text{~s} & 200 \text{~m} & 10^{-4} \text{~rad} & 1 \text{~Hz} \\
\text{Satellite~} 3 & 10^4 \text{~km} & 1.6 \times 10^{9} \text{~m}^{-1} & 100 \text{~s} & 100 \text{~m} & 10^{-5} \text{~rad} & 1 \text{~Hz} \\
\hline
\end{array}
\end{math}
\caption{\label{Tab: sensitivities} The experimental parameters chosen for the benchmark sensitivity curves in Figures \ref{Fig:earth binary sensitivity} and \ref{Fig:space binary sensitivity}.  The phase sensitivity is the per shot sensitivity.  $L$ is the length of the baseline, $f_d$ is the data-taking or shot repetition rate, $\keff$ is the effective momentum transfer of the beamsplitters, $T$ is the interrogation time of each shot, $I_L$ is the length of each interferometer region.}
\end{center}
\end{table}

The projected sensitivities for two possible configurations of the proposed earth-based experiments are shown in Figure \ref{Fig:earth binary sensitivity}.  The choice of experimental parameters for these two configurations, shown in Table \ref{Tab: sensitivities}, is meant to illustrate the range of possible sensitivities that could be achievable.  These are the envelopes of curves similar to the one in Figure \ref{Fig:demon binary sensitivity}.  We have chosen to remove the singularities that appear in Figure \ref{Fig:demon binary sensitivity} to emphasize the frequency scaling for a general AI detector.  In an actual experiment, the entire area of the envelope curve can be swept out by increasing the interrogation time $T$ by a factor of roughly two.  The sensitivities plotted are only the inherent sensitivity of the atom interferometer, i.e. the power spectra of the expected shot noise.  We have argued in Section \ref{Sec: Earth Backgrounds} that other backgrounds are smaller than this level.  The one exception is time-varying gravity gradient noise and so the sensitivity curves are shown dashed below the frequency at which we expect gravity gradient noise to become the dominant noise source (see Figures \ref{Fig:timevargrav10Km} and \ref{Fig:timevargrav1Km}).  The upper sensitivity curve assumes a $L=1 \text{ km}$ distance between two $10 \text{ m}$ atom interferometers, with, therefore, an interrogation time of $T = 1.4 \text{ s}$.  Each interferometer has $100 \hbar k$ LMT beamsplitters, a per shot phase sensitivity of $10^{-4} \text{ rad}$, and a data-taking rate of $10 \text{ Hz}$.  The more aggressive curve assumes $L=10 \text{ km}$, $1000 \hbar k$ LMT beamsplitters, $100 \text{ m}$ interferometers with $T = 4.5 \text{ s}$, a per shot phase sensitivity of $10^{-5} \text{ rad}$ and the same data-taking rate.  The curves are cut off at the Nyquist frequency.  The sensitivity of initial LIGO \cite{LIGO} and the projected sensitivity of LISA \cite{Larson:1999we} are also shown.

Figure \ref{Fig:space binary sensitivity} shows the projected sensitivities for three possible configurations of the proposed satellite experiment, with parameters shown in Table \ref{Tab: sensitivities}.  The most conservative curve assumes $L=100 \text{ km}$, $100 \hbar k$ LMT beamsplitters, $T = 10 \text{ s}$, per-shot phase sensitivity of $10^{-4} \text{ rad}$ and data-taking rate of $10 \text{ Hz}$.  The middle curve is the same except it assumes $L=10^4 \text{ km}$, $100 \hbar k$ LMT beamsplitters, and  $T = 100 \text{ s}$.  The most aggressive curve assumes the same length, beamsplitters, and interrogation time as the middle curve but assumes an extra factor of 10 in the per shot phase sensitivity, either from a larger number of atoms or from squeezed states.

\subsection{Stochastic Gravitational Wave Backgrounds}

\begin{figure}
\begin{center}
\includegraphics[width=\columnwidth]{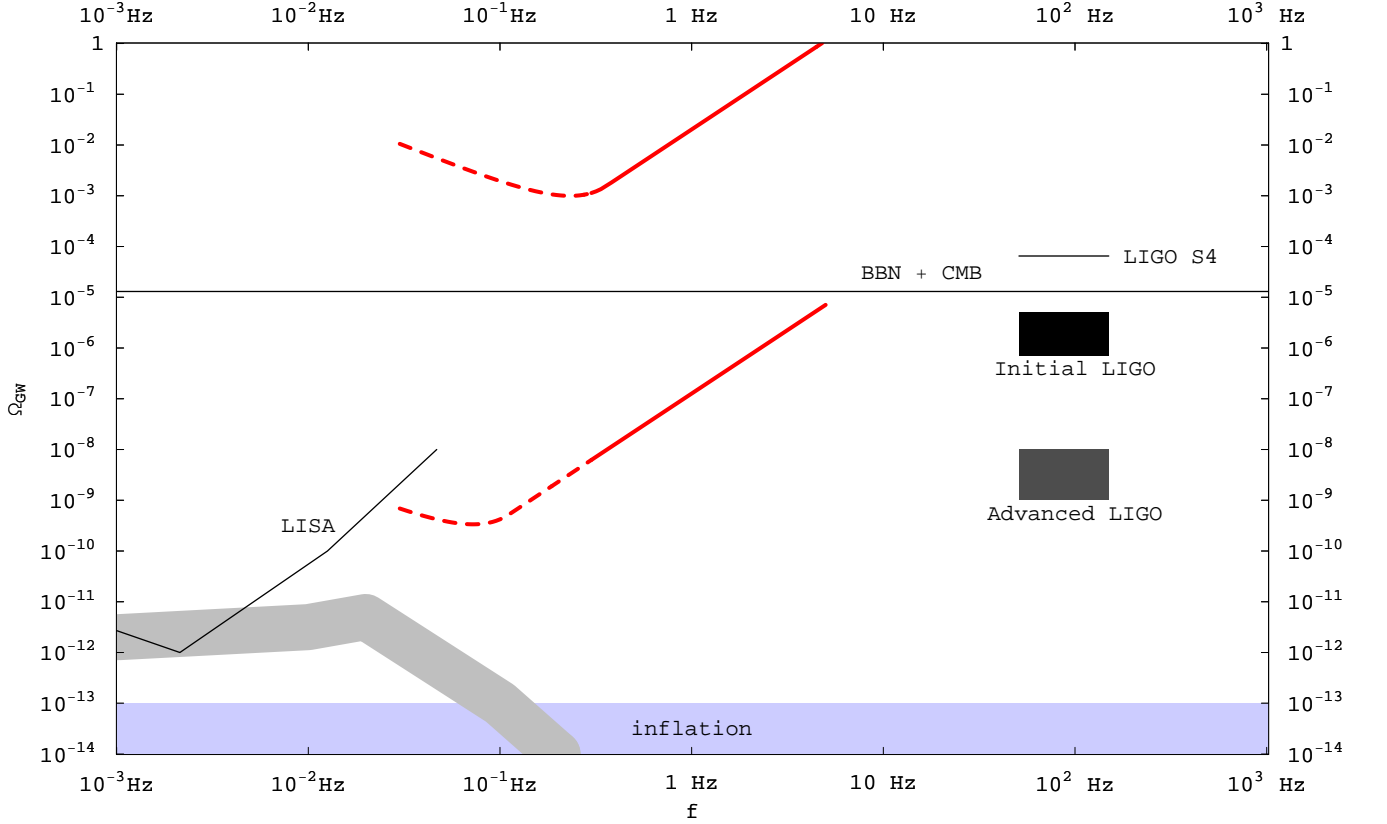}
\caption{ \label{Fig:earth stochastic sensitivity} (Color online) The projected sensitivity in $\Omega_\text{GW}$ of our proposed Earth based experiments, thick (red) curves, to a stochastic background of gravitational waves.  The parameter choices are as in Fig. \ref{Fig:earth binary sensitivity}.  These curves only take into account shot noise.  The limit from LIGO Science Run 4 and the projected limits from initial and advanced LIGO are shown \cite{Abbott:2006zx}.  The limits from BBN \cite{Allen:1996vm} and the CMB \cite{Smith:2006nka} apply to the integral of the stochastic gravitational wave background over frequency.  The possible region of gravitational waves produced by a period of inflation (not including reheating) is shown.  The upper limit on this region is set by the COBE bound \cite{Allen:1994xz}.  The gray band shows a prediction for the stochastic gravitational wave background from extragalactic white dwarf binaries; its width shows an expected error \cite{Farmer:2003pa}.}
\end{center}
\end{figure}

\begin{figure}
\begin{center}
\includegraphics[width=\columnwidth]{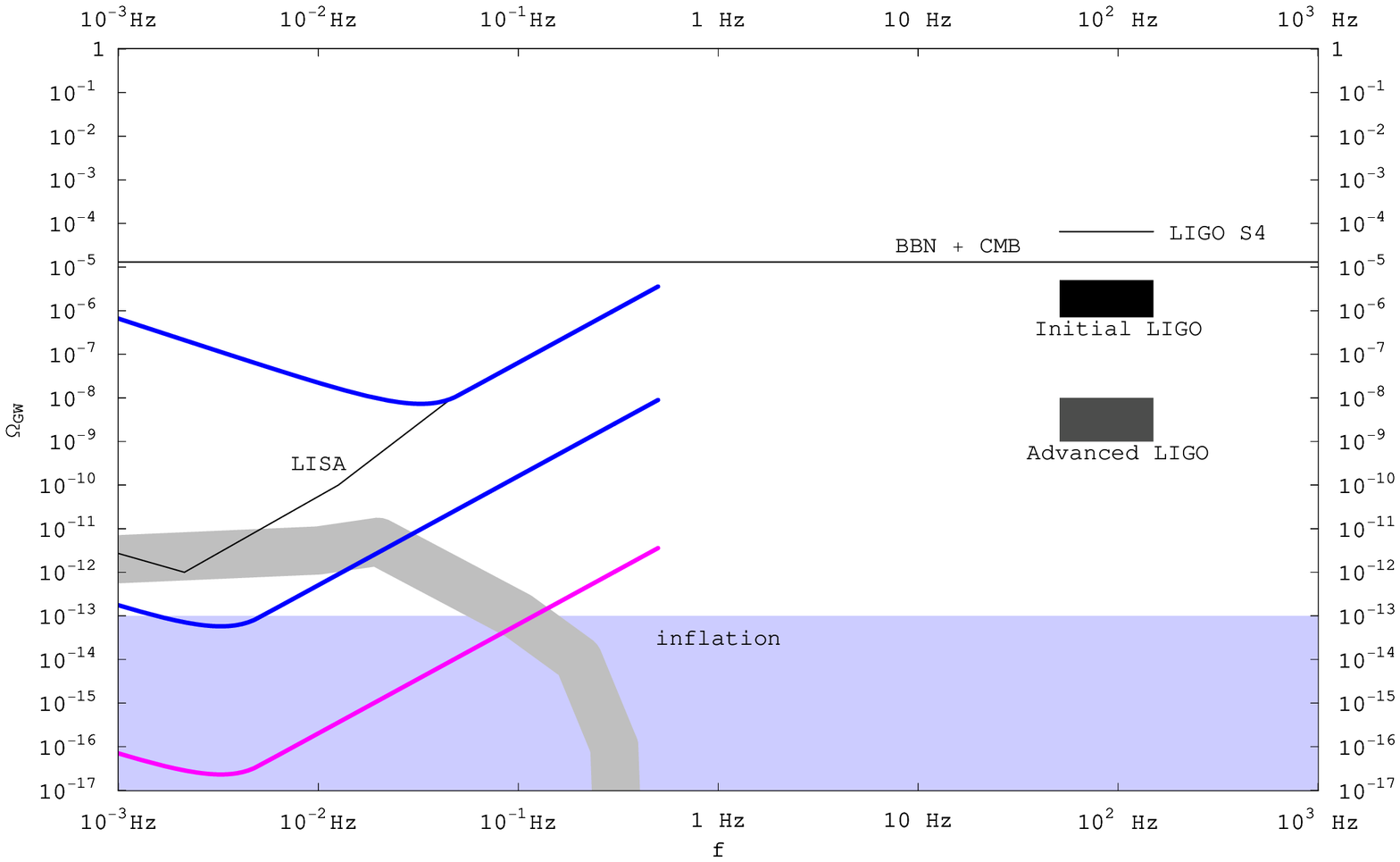}
\caption{ \label{Fig:space stochastic sensitivity} (Color online) The projected sensitivity in $\Omega_\text{GW}$ of our proposed satellite experiments, thick (blue and purple) curves, to a stochastic background of gravitational waves.  The parameter choices are as in Fig. \ref{Fig:space binary sensitivity}.  These curves only take into account shot noise.  The limit from LIGO Science Run 4 and the projected limits from initial and advanced LIGO are shown \cite{Abbott:2006zx}.  The limits from BBN \cite{Allen:1996vm} and the CMB \cite{Smith:2006nka} apply to the integral of the stochastic gravitational wave background over frequency.  The possible region of gravitational waves produced by a period of inflation (not including reheating) is shown.  The upper limit on this region is set by the COBE bound \cite{Allen:1994xz}.  The gray band shows a prediction for the stochastic gravitational wave background from extragalactic white dwarf binaries; its width shows an expected error \cite{Farmer:2003pa}.}
\end{center}
\end{figure}

A stochastic background of gravitational waves would be undistinguishable from any other background noise in a single gravitational wave detector.  A single detector means for example one of the LIGO sites or one AGIS configuration (i.e. two atom interferometers with a long laser baseline in between as in Figures \ref{Fig: Earth diff} or \ref{Fig:space setup}).  With two gravitational wave detectors it is possible to cross-correlate the measurements and obtain sensitivity to a stochastic background of gravitational waves.  It is preferable if these two detectors are far apart either on the earth or in space since a stochastic background of gravitational waves coming from astrophysical or cosmological sources would be common to both detectors, but other sources of noise (nearby motions of the earth for example) would not.  Thus a single gravitational wave detector can never detect a stochastic background of gravitational waves (or at least can never prove that is what is being detected) but more than one detector allows sensitivity to a stochastic background of gravitational waves.  This standard strategy is also employed by LIGO and is described for example in \cite{Abbott:2006zx}.

As is standard, the sensitivity to such gravitational waves is shown in Figures \ref{Fig:earth stochastic sensitivity} and \ref{Fig:space stochastic sensitivity}, plotted in the variable
\begin{equation}
\Omega_\text{GW} (f) = \frac{f}{\rho_c} \frac{d \rho_\text{GW}}{df}
\end{equation}
where $\rho_c$ is the critical energy density of the universe and $\rho_\text{GW}$ is the local energy density in gravitational waves.  These curves follow from the standard analysis, so we plot the $95 \%$ confidence limit on the spectrum of stochastic gravitational waves.  Following \cite{Christensen:1992wi} (but see also \cite{Flanagan:1993ix, Allen:1997ad}) we estimate this limit by
\begin{equation}
\Omega_\text{GW} (f) = \frac{\pi c^2 f^3}{\rho_c G | \gamma \left( \vec{x}_1, \vec{x}_2, f \right) |} \sqrt{\frac{2}{\tau_\text{int} \Delta f}} (1.645) h_n^2 (f)
\end{equation}
where $\tau_\text{int}$ is the total time length of the experiment, $\gamma$ is a geometric factor taking into account the positions of the two detectors which we take equal to its maximum value $\frac{8 \pi}{5}$ (it will probably be slightly smaller in a real configuration), and $h_n$ is the power spectrum of the noise in the gravitational wave detector as plotted for example in Figure \ref{Fig:demon binary sensitivity}.  To produce the curves in Figures \ref{Fig:earth stochastic sensitivity} and \ref{Fig:space stochastic sensitivity} we use the $h_n$ from Figures \ref{Fig:earth binary sensitivity} and \ref{Fig:space binary sensitivity}, respectively.  As is standard, we assume a $\tau_\text{int} \sim 1 \text{ yr}$ integration time for the experiment.  This is only a benefit if two detectors can be cross-correlated.  Otherwise, the sensitivity to a stochastic background is no better than the noise on each shot and it is only possible to place limits on and not to detect such a background.

It is advantageous to measure at lower frequencies to gain sensitivity in the variable $\Omega_\text{GW}$ because it scales favorably with low $f$.  Further, there is a cutoff in $\gamma$ and thus the sensitivity when the two gravitational wave detectors are father apart than the wavelength of the gravitational waves, $f^{-1}$.  For frequencies below $\OO(10 \text{ Hz})$ this is not a problem for our detectors, but for LIGO this is an issue.  The sensitivity of LIGO to stochastic gravitational waves is reduced because of the large distance between their two detectors, $\sim 3000 \text{~km}$ \cite{Flanagan:1993ix}.


There is predicted to be a stochastic background of gravitational waves from the large number of galactic and extragalactic close binaries, mainly white dwarf binaries.  There are significant uncertainties in the calculation of the spectrum from this source due to uncertainty in stellar population models.  Figures \ref{Fig:earth stochastic sensitivity} and \ref{Fig:space stochastic sensitivity} show one prediction \cite{Farmer:2003pa} for this background with the approximate uncertainty represented by the size of the band.  This background is reduced to some extent by the ability to measure and subtract known binary sources.  It can limit the ability of gravitational wave detectors to see other, cosmological sources of gravitational waves in this low frequency band.

\begin{figure}
\begin{center}
\includegraphics[width=\columnwidth]{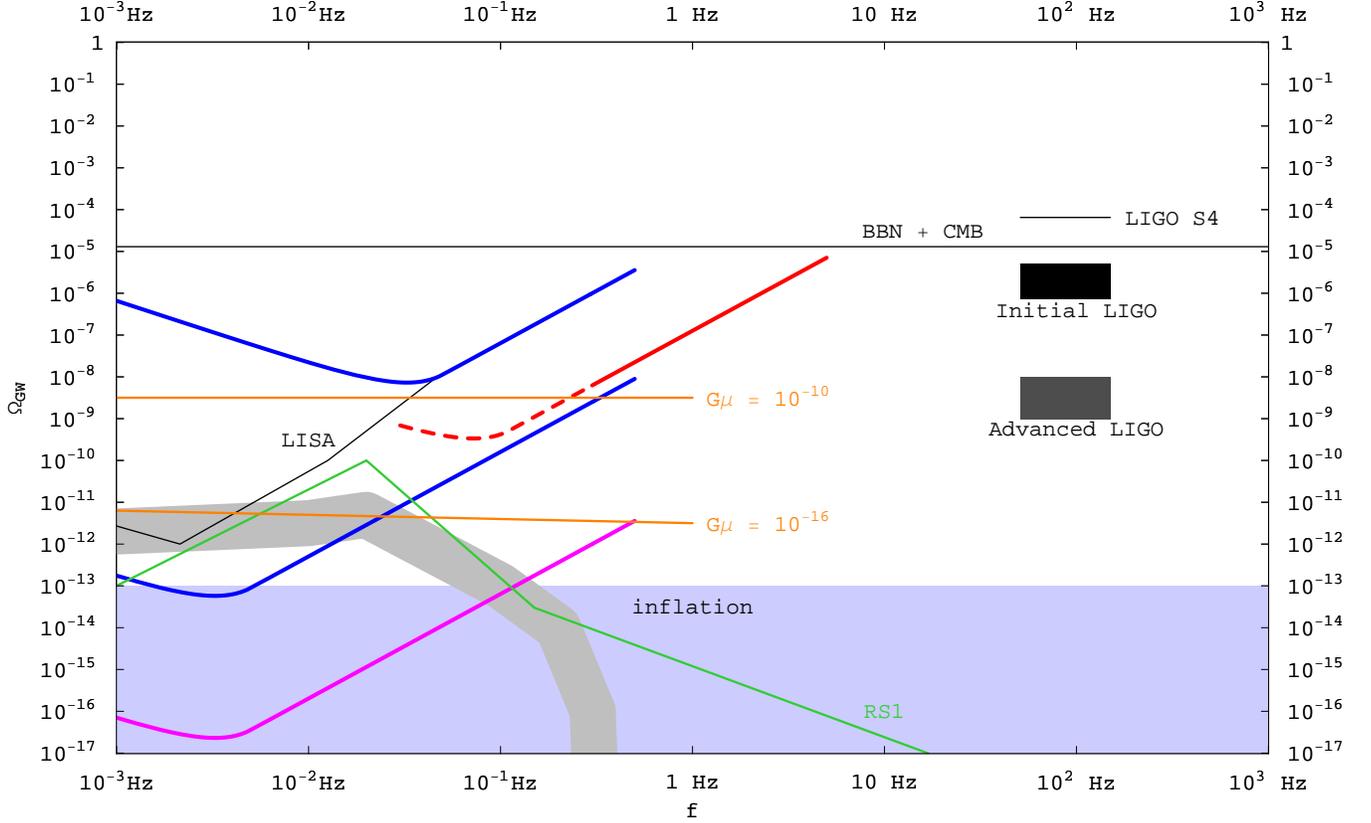}
\caption{ \label{Fig:space stochastic sources} (Color online) The same plot as in Figures \ref{Fig:earth stochastic sensitivity}, \ref{Fig:space stochastic sensitivity} with possible new physics sources of stochastic gravitational waves.  The (green) curve labeled RS1 corresponds to an example spectrum of gravity waves from a TeV scale phase transition, in this case in RS1 \cite{Randall:2006py}.  The (orange) lines labeled with $G \mu$ correspond to one prediction for a network of cosmic strings with tensions $G \mu = 10^{-10}$ and $10^{-16}$ (with $\alpha = 0.1$ and $\gamma = 50$) \cite{DePies:2007bm}.  Note that these cosmic string estimates have large uncertainties and may be optimistic assumptions.}
\end{center}
\end{figure}

\subsubsection{New Physics Signals}
Figure \ref{Fig:space stochastic sources} shows several possible new physics sources of gravitational waves.  An example spectrum from the TeV scale phase transition in RS1 taken from \cite{Randall:2006py} is shown to illustrate roughly what the spectrum from an electroweak scale phase transition might look like.  It shows a peak at frequencies around $10^{-2}$ Hz and can certainly be strong enough to be visible in these interferometric detectors.  Of course there is much model dependence in this spectrum; for example, only a first order weak scale phase transition will produce gravitational waves at all.

Two example spectra are shown for cosmic strings with tensions $G \mu = 10^{-10}$ and $10^{-16}$ from \cite{DePies:2007bm}.  It is important to note that not only is there model dependence in the spectrum from cosmic strings, but there is also much uncertainty in the calculation and so these should probably be considered to be upper limits on the spectrum of gravitational waves from such cosmic string networks.  However, given these optimistic assumptions, it may be possible to detect a network of cosmic strings with tension as low as $G \mu = 10^{-16}$ using these interferometers.  This is becoming limited by the white dwarf background, whose calculation itself has large uncertainties.

The region labeled `inflation' in the figure is really the upper limit on the possible inflation spectrum assuming it is perfectly flat from the low frequency CMB bound.  However, realistic models of inflation give $\Omega \lesssim 10^{-15}$ in our frequency band with the highest values of $\Omega$ from the highest scale models of inflation.  Low scale inflation models will not directly give observable gravitational wave spectra, but could give observable gravitational waves from reheating (see e.g.  \cite{Dufaux:2007pt}).


\section{Conclusion}
\label{Sec: Conclusion}

\subsection{Comparison with Previous Work}

Previous studies on the role of atom interferometers in gravitational wave detection concluded that they would be of limited use in probing the gravitational wave spectrum. Our proposal differs significantly from these efforts owing to the central role played by light pulse interferometry in our setup.  

The work of \cite{Chiao+Speliotopoulos}, \cite{Foffa:2004up} and \cite{Roura:2004se} used material mirrors like diffraction gratings to execute the interferometer. The gravitational wave signal in the configurations considered in these papers is  $\sim k h d$ where $k$ is the momentum of the atom, $h$ the amplitude of the gravitational wave and $d$ the distance between the mirrors. It is experimentally difficult to make the distance between these mirrors bigger than $\sim 1$ m. Even if the distance between the mirrors were to be increased, the experiment would still be difficult since the separation between the two arms of the atom's wave function must also be equally scaled. These considerations forced the authors to conclude that  an unrealistic atom flux would be needed to see a gravitational wave. The use of material mirrors suffers from the additional drawback that the mirrors would be subject to vibration noise. The mirrors would have to be placed on vibration isolation stacks so this interferometer would be subject to the same limitations as light based interferometers like LIGO.

The work of \cite{Delva:2006qa} and \cite{Tino:2007hs} described atom interferometers which used light pulse interferometry. However, these authors did not consider the effect of the gravitational wave on the light pulses used to execute the interferometry. Without this effect, the phase shift in the interferometer is $\sim k h d$ where $d$ is the separation between the two arms of the interferometer (see discussion in Section \ref{Sec: Signal Results}). Since the separation between the two arms of the interferometer cannot be easily scaled, these authors were also forced to consider unrealistic atom fluxes. Moreover, these papers did not discuss strategies to handle crucial backgrounds to gravitational wave detection like vibration and laser phase noise.  

In this paper, we point out that the effect of the gravitational wave on the light pulses used to execute the interferometer is crucial and can be easily scaled to increase the signal. When the interferometer is operated by a laser at a distance $L$, a gravitational wave of amplitude $h$ causes a phase shift $\sim k h L$. This signal increases as long as $L$ is smaller than the wavelength of the gravitational wave.  Unlike the separation between the two arms of the atom's wave function, the distance between the atom and the laser can be easily scaled.   With $L \sim 10$ km, the signal in this interferometer is $10^4$ larger than the signal in the configurations previously considered. In addition to boosting the signal, the configuration considered in this paper offers an effective way to deal with vibration and laser phase noise. By using the same laser to run two widely separated interferometers and measuring the differential phase shift between the two interferometers, this setup drastically suppresses the effects of vibrations and laser phase noise. Our setup thus achieves a large, scaleable enhancement in the signal while simultaneously suppressing backgrounds thereby making it possible to search for gravitational waves with current technology. 

The SAGAS \cite{SAGAS} project that uses atom  interferometry and ion clock techniques to explore gravity in the outer solar system was proposed. SAGAS will improve current bounds on stochastic gravitational waves in the frequency band $10^{-5} \text{ Hz} - 10^{-3} \text{ Hz}$ but is not expected to be sensitive to known sources of gravitational radiation. In contrast, our proposal  will search for gravitational waves in the $10^{-3} \text{ Hz} - 10 \text{ Hz}$ band at sensitivities that can detect gravitational waves from expected sources. 

\subsection{Summary}

We have proposed two experiments, terrestrial and satellite-based, to observe gravitational waves using atom interferometry.  Both experiments rely on similar underlying ideas to achieve a large, scaleable enhancement to the gravitational wave signal while naturally suppressing many backgrounds.  A differential measurement is performed between two atom interferometers run simultaneously using the same laser pulses.  The lasers provide a common `ruler' for comparison of the two interferometers.  The distance between the interferometers can be large because only the light travels over this distance, not the atoms.  The signal still scales with this distance and so can be competitive with light interferometers.  In a sense, the atom interferometers are the analogue of the mirrors in a light interferometer and it is the distance between them that determines the size of the signal.

Further, many backgrounds are naturally suppressed by this method.  Laser phase noise, which must be cancelled between the two arms of a light interferometer, is here cancelled by the differential measurement between the two atom interferometers.  Since this subtraction is between two interferometers along the same laser axis with only vacuum in between, vibrations of the lasers (and any optics) are cancelled as well.  The atoms themselves, the analogues of the mirrors in a light interferometer, are in free fall and are unaffected by vibrations.  This removes one of the major backgrounds which prohibits terrestrial light interferometers from achieving sensitivity to lower frequencies.  For example, Advanced LIGO will lose sensitivity below $\sim 10$ Hz due to direct (non-gravitational) coupling to vibrations (see e.g. \cite{Hughes:1998pe}).  Similarly, in the satellite-based experiment the atoms can be far from the satellite, greatly reducing the engineering requirements on the control of the satellites.  Satellite position control and laser noise are two of the major hurdles for an experiment such as LISA.  For similar gravitational wave sensitivity, these requirements are significantly reduced for our atom interferometer proposal.

Of course, new backgrounds may enter in an atomic experiment.  We have attempted to consider all the relevant backgrounds and show that they are controllable with practical technology in a realistic setup.  Many backgrounds will require careful engineering, just as for any gravitational wave detector.  We are certainly not experts in every relevant area of expertise necessary for such experiments, but this experiment seems possible and exciting enough to merit more serious consideration.

An interesting consequence of having a differential measurement between two interferometers along the same baseline is that this setup would have sensitivity to scalar-type perturbations, that would, for example, change the length of the perimeter of the LISA triangle \footnote{Thanks to Bob Wagoner for pointing this out.}.  LIGO lacks sensitivity to these signals since two perpendicular laser arms are used to remove backgrounds including laser phase noise.  Our setup confers sensitivity to overall changes in the length of a single arm since each arm is a laser phase noise free combination.  In LISA these Sagnac channel events would be vetoed.  One interesting signal of this type would arise from large mass dark matter particles passing near the detector \cite{Adams:2004pk, Seto:2004zu}.

There are many avenues for improvement of these proposals in the future.  The atom statistics may be improved with improved cooling techniques, ultimately limited only by the limit on the density of the cloud from atom-atom interactions and on the total number of atoms from opacity of the cloud.  The use of squeezed atom states may also allow significant improvements in atom statistics beyond the standard quantum limit.  Improved sensitivities could also come from better classical and atom optics including multi-photon LMT beamsplitters, higher laser powers and larger laser waist sizes.  There may also be clever ideas for improved atomic systems, for example which suppress the spontaneous 2-photon transition rate without suppressing the stimulated rate.  It is difficult to predict what advances will be made in the future.  Nevertheless, the rapid advance of atom interferometry motivates us to consider a range of sensitivity curves that illustrate the possibilities not just for current but also near future technology.

The proposed gravitational wave detectors may allow the observation of low frequency sources in the band $10^{-3} - 10 \text{~Hz}$.  This is a very exciting range for astrophysical and cosmological sources.  Compact binaries including black holes, neutron stars, and white dwarfs live for a long period in this band.  Such low frequencies also allow enhanced sensitivity to a stochastic background of gravitational waves, assuming at least two such AGIS detectors are built.  Many cosmological sources arising from physics beyond the Standard Model could be present in this range including inflation and reheating, early universe phase transitions, or cosmic strings.  The observation of gravitational waves has the potential to reveal significant information about new physics at both the shortest and longest length scales.

\section*{Acknowledgments}
We would like to thank Allan Adams, Mustafa Amin, Asimina Arvanitaki, Roger Blandford, Seth Foreman, David Johnson, Lev Kofman, Vuk Mandic, Holger Mueller, Aaron Pierce, Stephen Shenker, Jay Wacker,  Robert Wagoner, Neal Weiner, and Yoshihisa Yamamoto for valuable discussions.
PWG acknowledges the support of the Mellam Family Graduate Fellowship during a portion of this work.

\end{document}